\documentclass[11pt,letterpaper]{article}
\pdfoutput=1
\usepackage{jcappub}
\usepackage{bbm}
\usepackage{mathrsfs}
\usepackage{slashed}
\usepackage{caption}
\usepackage{epstopdf}
\usepackage[normalem]{ulem}
\usepackage[bottom]{footmisc}
\usepackage{subcaption}
\usepackage{bbold}
\usepackage{titlesec}
\usepackage{threeparttable}
\usepackage{booktabs}
\usepackage{changepage}
\usepackage[utf8]{inputenc}
\usepackage{dsfont} 
\usepackage{grffile}
\usepackage{graphicx}  
\usepackage{dcolumn}   
\usepackage{bm}        
\usepackage{amssymb}   
\usepackage{setspace}
\usepackage{amsmath, amssymb, setspace}
\usepackage{array}
\usepackage{booktabs}
\usepackage{caption}
\usepackage{float}
\usepackage{lmodern}
\usepackage{multirow}
\usepackage{soul}
\usepackage[normalem]{ulem}
\usepackage{braket}
\usepackage{comment}
\usepackage[draft]{pgf}
\usepackage{adjustbox} 
\usepackage{xspace} 
\usepackage{url}
\usepackage{xcolor}

\newcommand{\DoBox}[1]{\begin{center}
\color{red}\fbox{
\begin{minipage}{0.9\textwidth}
\end{minipage}}
\end{center}}

\newcommand{\be}{\begin{equation}}
\newcommand{\ee}{\end{equation}}
\newcommand{\ba}{\begin{eqnarray}}
\newcommand{\ea}{\end{eqnarray}}
\newcommand{\nn}{{\nonumber}}

\newlength{\myimageoversize}
\newsavebox{\myimage}

\usepackage[most]{tcolorbox}
\usepackage{empheq}

\tcbset{colback=white!10!white, colframe=black!50!black, 
        highlight math style= {enhanced, 
            colframe=black,colback=white!10!white,boxsep=0pt}
        }

\begin{document}
\title{\huge{Birefringence Tomography for Axion Cloud}}

\author{Yifan Chen,$^{a,b}$}
\author{Chunlong Li,$^{a}$}
\author{Yosuke Mizuno,$^{c,d}$}
\author{Jing Shu,$^{e,f}$}
\author{Xiao Xue,$^{g,h}$}
\author{Qiang Yuan,$^{i,j}$}
\author{Yue Zhao,$^k$}
\author{and Zihan Zhou$^l$}
\emailAdd{yifan.chen@nbi.ku.dk}
\emailAdd{chunlong@itp.ac.cn}
\emailAdd{mizuno@sjtu.edu.cn}
\emailAdd{jshu@pku.edu.cn}
\emailAdd{xiao.xue@desy.de}
\emailAdd{yuanq@pmo.ac.cn}
\emailAdd{zhaoyue@physics.utah.edu}
\emailAdd{zihanz@princeton.edu}

\affiliation{
{$^a$CAS Key Laboratory of Theoretical Physics, Institute of Theoretical
Physics, Chinese Academy of Sciences, Beijing 100190, China\\
$^b$Niels Bohr International Academy, Niels Bohr Institute, Blegdamsvej 17, 2100 Copenhagen, Denmark\\
$^c$Tsung-Dao Lee Institute and School of Physics and Astronomy, Shanghai Jiao Tong University, Shanghai, 200240, China\\
$^d$Institute for Theoretical Physics, Goethe University Frankfurt, Frankfurt am Main, 60438, Germany\\
$^e$School of Physics and State Key Laboratory of Nuclear Physics and Technology, Peking University, Beijing 100871, China\\
$^f$Center for High Energy Physics, Peking University, Beijing 100871, China\\
$^g$II. Institute of Theoretical Physics, Universit\"{a}t  Hamburg, 22761, Hamburg, Germany\\
$^h$Deutsches Elektronen-Synchrotron DESY, Notkestr. 85, 22607, Hamburg, Germany\\
$^i$Key Laboratory of Dark Matter and Space Astronomy, Purple Mountain
Observatory, Chinese Academy of Sciences, Nanjing 210023, China\\
$^j$School of Astronomy and Space Science, University of Science and
Technology of China, Hefei 230026, China\\
$^k$Department of Physics and Astronomy, University of Utah, Salt Lake City, UT 84112, USA\\
$^l$Department of Physics, Princeton University, Princeton, NJ 08544, USA
}}
\date{\today}

\abstract{
An axion cloud surrounding a supermassive black hole can be naturally produced through the superradiance process. Its existence can be examined by the axion induced birefringence effect. It predicts an oscillation of the electric vector position angle of linearly polarized radiations. Stringent constraints of the existence of the axion in a particular mass window has been obtained based on the recent Event Horizon Telescope measurement on M87$^\star$.
The future Very-Long-Baseline Interferometry (VLBI) observations will be able to measure the vicinity of many supermassive black holes, thus it opens the possibility to search for the existence of axions in a wide mass regime.
In this paper, we study how different black hole properties and accretion flows influence the signatures of the axion induced birefringence. We include the impacts of black hole inclination angles, spins, magnetic fields, plasma velocity distributions, the thickness of the accretion flows. 
We pay special attention to characterize the washout effects induced by the finite thickness of the accretion flows and the lensed photons.
Based on this study, we give prospects on how to optimize the axion search using future VLBI observations, such as the next-generation Event Horizon Telescope, to further increase the sensitivity.
}

\maketitle

\section{Introduction}
Taking advantage of the Very Long Baseline Interferometer (VLBI) technology, the Event Horizon Telescope (EHT) opens a new era of probing the physics under extreme conditions near the horizon of a supermassive black hole (SMBH) \cite{Akiyama:2019cqa,Akiyama:2019bqs,Akiyama:2019eap,Akiyama:2019fyp}. This allows us to test general relativity in the strong gravity region around the black hole and study the accretion flow around it. Beyond constructing the intensity image of the accretion flow of the SMBH M87$^\star$, the EHT recently performed a polarimetric measurement on the radiation from its vicinity, with a high spatial resolution. From the astrophysical point of view, it helps us to understand the magnetic structure of the accretion flow \cite{EHTP,EHTM}.

Besides the applications to study astrophysics, such horizon-scale measurements also provide us opportunities to test particle physics, especially ultralight bosons. With a proper mass, ultralight bosonic particles can be spontaneously accumulated around a Kerr black hole through the superradiance mechanism \cite{Penrose:1971uk,ZS,Press:1972zz,Damour:1976kh,Zouros:1979iw,Detweiler:1980uk,Strafuss:2004qc,Dolan:2007mj,Brito:2015oca}.
Among various choices of ultralight bosons beyond the standard model, the most promising candidate is the QCD axion or axion-like particles \cite{Peccei:1977hh,Peccei:1977ur,Weinberg:1977ma,Wilczek:1977pj}. They generically appear in theories with extra dimensions \cite{Arvanitaki:2009fg}, and they can be good cold dark matter candidates \cite{Preskill:1982cy,Abbott:1982af,Dine:1982ah}. In \cite{Chen:2019fsq}, using the axion induced birefringence effects \cite{Carroll:1989vb,Harari:1992ea} is proposed to search for the existence of the axion cloud around a SMBH. If it exists, the coherently oscillating axion field will lead to a periodic variation to the electric vector position angles (EVPAs) of linearly polarized radiations from the accretion flow. Based on this theoretical proposal, the signatures of the axion cloud are further investigated using the recent EHT polarimetric measurement on M87$^\star$ \cite{Chen:2021lvo} and stringent constraints on the axion parameter space are achieved.

The future VLBI observations, such as the next-generation EHT (ngEHT) \cite{Raymond_2021,Lngeht} and space VLBI \cite{Gurvits:2022wgm}, with more observed frequencies and potentially longer baseline in the space, can further increase the spatial resolution and perform detailed measurements on the horizons of a large number of SMBHs \cite{Pesce:2021adg}. Since the axion cloud can only be produced when the axion Compton wavelength is comparable to the black hole size, by observing black holes with various masses, the future {VLBI experiments} open the opportunities to study the existence of axion in a large mass regime. 

Given the potential information of a large landscape of SMBHs with various properties, such as spins, inclination angles, types of accretion flows etc, it is necessary to construct the foundation of the axion search at future VLBIs. In this paper, we perform a comprehensive study on the polarimetric signals caused by the axion induced birefringence, with various properties of SMBHs.

The layout of the paper is as follows. In Sec.\,\ref{ASMBH}, we review the production of the axion cloud around a Kerr black hole through the superradiance mechanism. In Sec.\,\ref{BRT}, we review the axion-induced birefringence in a curved space-time. We show how to embed the axion-photon coupling into the polarized radiative transfer equations. In Sec.\,\ref{TDRT}, we focus on the thin accretion disk model.  With different choices on the inclination angle and the spin of the black hole, we show how ray tracing influences the birefringence signals from the axion cloud. We further define a new observable, the Fourier decomposition of the differential EVPA along the azimuthal direction, that can be generally applied to nearly face-on black holes, such as M87$^\star$.
In Sec.\,\ref{RW}, we consider more realistic accretion disk models, characterized by Radiative Inefficient Accretion Flows (RIAFs). Particularly, we study two washout effects in our signal, the sum of the linear polarization along line of sight through the accretion flow, and the incoherent sum from the lensed photons. 
Finally we present the prospects for the future axion search in Sec.\,\ref{Pn}.

{Throughout this study, we work in units where $G = \hbar = c = 1$, and  adopt the metric convention $(-,+,+,+)$}.

\section{Axion Cloud from Black Hole Superradiance}\label{ASMBH}
According to the superradiance mechanism, a rapidly spinning black hole can generate an exponentially growing axion cloud, when the axion's reduced Compton wavelength is comparable to the gravitational radius of a Kerr black hole ~\cite{Penrose:1971uk,ZS,Press:1972zz,Damour:1976kh,Zouros:1979iw,Detweiler:1980uk,Strafuss:2004qc,Dolan:2007mj}, for a review see~\cite{Brito:2015oca}. 

The reduced Compton wavelength is related to the axion mass as $\lambda_c \equiv 1/\mu$, and the gravitational radius is determined by the black hole mass as $r_g \equiv M$.  Specifically, ignoring the axion self-interaction, the Klein–Gordon equation of axion field in a curved spacetime takes the form
\begin{equation}
    (\nabla^\mu \nabla_\mu - \mu^2) a = 0.
    \label{kge}
\end{equation}
In the following discussion, we take the covariant derivative $\nabla_{\mu}$ in terms of the Kerr metric of rotating black holes, with the mass $M$ and the angular momentum $J$ in Boyer-Lindquist (BL) coordinates $x^{\mu}=[t, r, \theta, \phi]$.
Under the Kerr background, the variables in the solution of Eq.\,(\ref{kge}) are separable \cite{Brill:1972xj,Carter:1968ks}, and we take the ansatz as
\begin{equation}
    a(t, \mathbf{r})=e^{-i\omega t+im\phi}R_{nlm}(r)S_{lm}(\theta),
\end{equation}
where $R_{nlm}(r)$ is the radial function, $S_{lm}(\theta)$ is the spheroidal harmonics which simplifies to the spherical harmonics $Y_{lm}$ in the non-rotating limit of the black hole or non-relativistic limit of the axion cloud. In addition, $\omega_{nlm}$ is the eigen-frequency of the corresponding eigenstate, and the number $n$, $l$ and $m$ satisfy $n \geq l+1, l \geq 0$ and $l \geq|m|$. One further imposes the ingoing bound condition at the Kerr black hole's outer horizon, and the wavefunction goes to zero at infinity. This makes the eigen-frequencies $\omega$ generally take a complex form
\begin{equation}
    \omega_{nlm}=\omega_{nlm}^r+i\omega_{nlm}^i.
\end{equation}
We first consider small values of $\alpha$ satisfying $\alpha \ll 0.1$. In this limit, the real part $\omega^r_{nlm}$ and the imaginary part $\omega_{nlm}^i$ can be written as \cite{Ternov:1978gq,Detweiler:1980uk}
\begin{align}
    &\omega_{nlm}^r=\mu\left(1-\frac{\alpha^2}{2n^2}+\mathcal{O}(\alpha^4)\right), \\
    &\omega_{nlm}^i\propto\alpha^{4l+5}\left(m\Omega_H-\omega_{nlm}^r\right)\left(1+\mathcal{O}(\alpha)\right).
    \label{omegai}
\end{align}
The dependence on the number $l$ and $m$ is included in the higher order terms of $\alpha$ whose expressions can be found in \cite{Baumann:2019eav}. Here $\Omega_{\rm H}\equiv a_J/(2r_+)$, with the radius of the outer horizon as $r_+\equiv M+M\sqrt{1-a_J^2}$ and the dimensionless spin as $a_J \equiv J/M^2$. When the superradiance condition is met, \begin{equation}
    \Omega_{H} > \frac{\omega^r_{nlm}}{m},
    \label{src}
\end{equation}
$\omega_{nlm}^i$ becomes positive. This leads to an exponential growth with the timescale as $\tau_{SR}= 1/\omega_{nlm}^i$. 

The radial profile of the axion cloud peaks at
\begin{equation}
    r_{{\rm max},n} \approx \Big(\frac{n^2}{2\alpha^2}\Big)r_g ~.
    \label{scalerealtion}
\end{equation}
This relation gives a simple scaling relation between the peak radius $r_{\rm max}$ and the gravitational fine-structure constant $\alpha$.

\begin{figure*}[thb]
    \centering
    \includegraphics[width=0.45\textwidth]{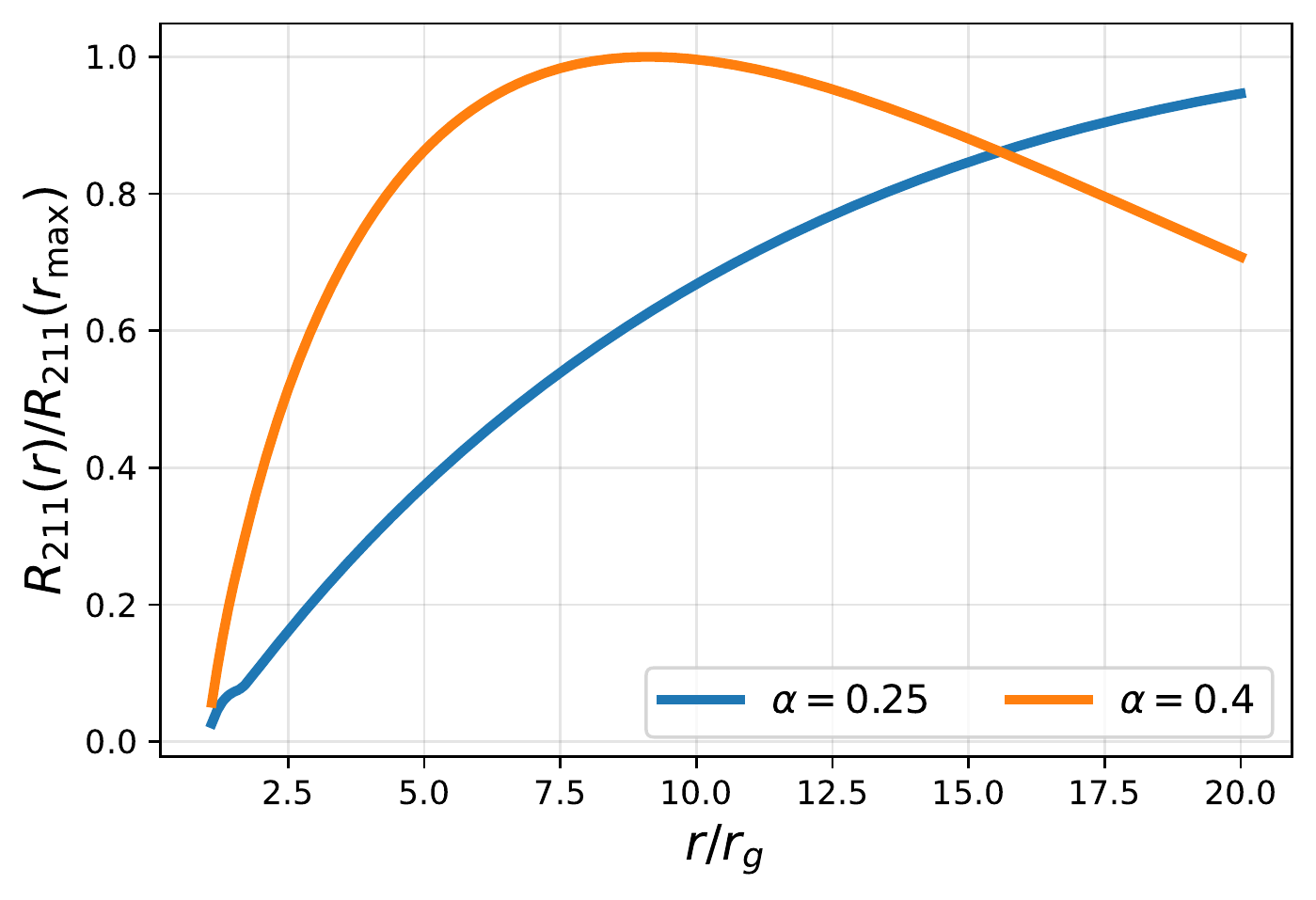}
    \includegraphics[width=0.45\textwidth]{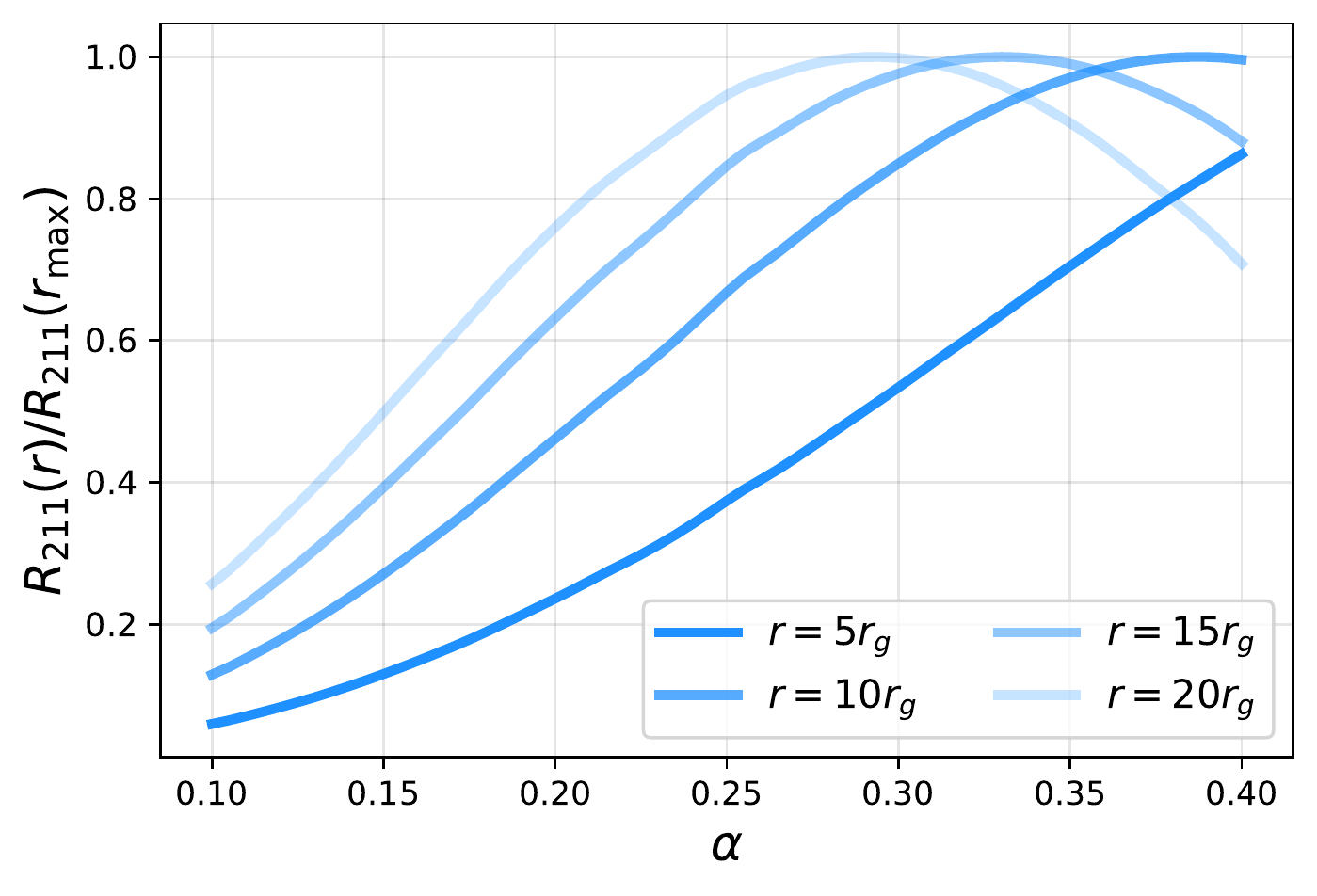}
    \caption{\footnotesize{Axion radial profile $R_{211}(r)$ with $a_J = 0.99$ by using the method of \cite{Dolan:2007mj}. Left: Profile with respect to distance $r$ for fixed $\alpha= 0.25,0.4$. Right: Profile with respect to gravitational fine-structure constant $\alpha$ for fixed $r=5,10,15,20\,r_g$.}}
    \label{RWFaf}
\end{figure*}

As for larger values of $\alpha$, one can perform the numerical calculation to obtain the solution of the axion field. According to the numerical study in \cite{Dolan:2007mj}, the state with $n=2, l = 1, m = 1$ has the highest superradiant rate. This is the lowest energy state among the ones which satisfy the superradiance condition. The axion wavefunction of such a state peaks at the equatorial plane ($\theta=90^\circ$) of the black hole. In Fig.\,\ref{RWFaf}, the radial function of this state, i.e., $R_{211}(r)$, is displayed for $a_J=0.99$. We emphasize that the axion cloud, with the Compton wavelength satisfying $\alpha = 0.4$, peaks close to $r \approx 5\,r_g$ \cite{Chen:2019fsq}. This is in a good agreement with the result presented in Eq.\,(\ref{scalerealtion}), as in the limit of $\alpha \ll 0.1$. 
For a bigger value of the angular momentum number $l$, $r_\textrm{max}$ becomes larger, and the axion cloud takes a much longer time to build up according to Eq.\,(\ref{omegai}). Thus in this study we only focus on the state with $l = m = 1$. 

Ignoring the $\alpha$'s higher order terms in Eq.\,(\ref{src}), for a fixed azimuthal mode $m$ and black hole spin $a_J$, the superradiance condition imposes an upper limit on $\alpha$
\be \alpha \lesssim \frac{a_J\ m}{2\ \left(1 + \sqrt{1-a_J^2} \right)}.\label{SRC}\ee
Choosing $m = 1$, $\alpha$ can be at most $0.5$ for an extreme Kerr black hole and $0.25$ for $a_J = 0.8$. Once the superradiance condition is satisfied, the axion cloud profile is only slightly influenced by the value of $a_J$ \cite{Amorim:2019hwp}.
In this study, we focus on the axion mass region satisfying $\alpha>0.1$ so that the superradiant timescale is much shorter than the age of the universe, i.e., within the range of $10^9$ years \cite{Dolan:2007mj}. The black hole spin can be as low as $a_J=0.5$ in order to satisfy the superradiance condition for $\alpha=0.1$.
As shown in Eq.\,(\ref{SRC}), the specific range of $\alpha$ which satisfies the superradiance condition is sensitive to the black hole spin $a_J$. Though the spin of M87$^\star$ is still uncertain \cite{EHTP}, Refs. \cite{Tamburini:2019vrf,Feng:2017vba} claim M87$^\star$ to be a nearly extreme Kerr black hole. In this study, we take the black hole spin $a_J$ to be 0.99 and 0.8 as two benchmarks, since these might be good representatives of the M87$^\star$ spin. 

Finally, one may question whether the axion cloud produced by the superradiance process is stable. For specific astrophysical systems, the stability of the axion cloud is discussed in \cite{Arvanitaki:2010sy}, where several potential perturbations which may destroy the axion cloud are discussed. Particularly, the presence of accreting matter and the tidal force from a companion star turn out to be negligible. For the parameter region we are interested in, the metric is always dominated by the SMBH. One may be concerned about the possibility of a merger with another SMBH in the past. However, we mainly focus on the axion mass which triggers a relatively short timescale for the superradiance. Even such a drastic merger happened once, the axion cloud should generically have enough time to build up again after the merger. Thus we neglect this possibility in our study. At last, the superradiance can be terminated by the axion self-interaction. Indeed, with the growth of the axion cloud, the axion field value in certain region of the cloud gets close to its decay constant $f_a$. The axion self interaction, described by $V(a) = \mu^2 f_a^2 \left( 1 - {\cos [ a/f_a ]} \right)$, leads to a correction to the potential energy. Due to 
the nontrivial self-interaction, the axion cloud can enter a violent bosenova or a saturating phase \cite{Yoshino:2012kn,Yoshino:2013ofa,Yoshino:2015nsa,Baryakhtar:2020gao,Omiya:2020vji}. Interestingly, both the numerical simulation \cite{Yoshino:2012kn,Yoshino:2013ofa,Yoshino:2015nsa} and the analytic estimation \cite{Baryakhtar:2020gao} indicate that the maximum of the field value $a_{\rm max}$ remains close to $f_a$ in either case, as long as the nonlinear regime is ever reached.

\section{Axion Induced Birefringence and Radiative Transfer}\label{BRT}

In this section, we first review the birefringence effects induced by the axion-photon coupling, based on the geometric optics approximation. The axion field background leads to modified Maxwell equations with different dispersion relations for the left and right circular polarized photons, which consequently causes a variation in electric vector position angles (EVPAs) of linearly polarized photons. Without medium effects, this birefringence effect is achromatic and topological since the shift of the EVPAs only depends on axion field values at emission and observation points \cite{Carroll:1989vb,Harari:1992ea,Plascencia:2017kca,Ivanov:2018byi,Fujita:2018zaj,Liu:2019brz,Fedderke:2019ajk,Caputo:2019tms,Yuan:2020xui}. This property also holds in curved space-time \cite{Schwarz:2020jjh}. 

Further we need to properly characterize the axion induced effects when photons propagate in the medium.  We show that they can be properly taken into account by a simple modification of the Faraday rotation terms in the covariant radiative transfer equations. 
Such additional terms are proportional to the gradient of axion field along the geodesics, which can be easily included into a numerical radiative transfer code like \texttt{IPOLE} \cite{Moscibrodzka:2017lcu, Noble:2007zx}.

\subsection {Axion-Photon Coupling and Birefringence}
\label{APCB}
We start with the photon propagation in a curved space-time with the axion background field, without including medium effects. In this case, the Lagrangian can be written as
\begin{align}
\mathcal{L}=&-\frac{1}{4} F_{\mu \nu} F^{\mu \nu}-\frac{1}{2} g_{a\gamma\gamma} a F_{\mu \nu} \tilde{F}^{\mu \nu}+\frac{1}{2} \nabla^{\mu} a \nabla_{\mu} a-V(a).
\label{overall_la}
\end{align}
Here {$g_{a\gamma\gamma}$ is the axion-photon coupling constant (not to be confused with the spacetime metric tensor)}, $\tilde{F}^{\mu\nu}=\epsilon^{\mu\nu\alpha\beta}F_{\alpha\beta}/2$ is the dual tensor of the electromagnetic field strength tensor, and $V(a)$ is the axion potential. In the Lorenz gauge $\nabla_\mu A^{\mu}=0$, the equation of motion for the electromagnetic field is
\begin{align}
    \nabla_\mu \nabla^{\mu}A^{\nu}-R_{\nu}{}^{\mu}A_{\mu} = -g_{a\gamma\gamma}(\nabla_\mu a)\tilde F^{\mu\nu}.
    \label{ame}
\end{align}
With a good accuracy, we follow \cite{Schwarz:2020jjh} and apply the geometric optics approximation, which is valid for photons with frequency much larger than the variation scale of the background metric and the axion field. This allows us to take the ansatz
\begin{align}
    A_{\mu}(x)=\bar{A}_{\mu}(x)\exp\left(\frac{i}{\epsilon}S(x)\right),
    \label{geoticex}
\end{align}
with the four dimensional wave-vector $k_\mu$ identified as
\begin{align}
    k_{\mu}\equiv\frac{1}{\epsilon}\partial_{\mu}S(x).
\end{align}
We take $\epsilon$ as a small number characterizing the geometric optics approximation. Our following calculations will be based on the perturbation on $\epsilon$. 
By substituting Eq.\,(\ref{geoticex}) into the Eq.\,(\ref{ame}), we find that the leading order term, i.e., $\mathcal{O}$($1/\epsilon^2$), gives
\begin{align}
    k^{\mu}k_{\mu}=0.
    \label{sdr}
\end{align}
We require this condition to hold along the path of photons. It indicates that the derivative of $k^{\mu}k_{\mu}$ with respect to the affine parameter equals to zero. This gives $k^{\mu}\nabla_{\mu}k_{\alpha}=0$, which means that photons follow null geodesics. 
The next order, i.e., $\mathcal{O}$($1/\epsilon$), expansion in Eq.\,(\ref{ame}) gives
\begin{align}
    k^{\mu}\nabla_{\mu}\bar{A}^{\nu}+\frac12\bar{A}^{\nu}\nabla_{\mu}k^{\mu}+g_{a\gamma\gamma}\epsilon^{\mu\nu\rho\sigma}\bar{A}_{\sigma}k_{\rho}\nabla_{\mu}a=0.
    \label{ptewos}
\end{align}
The Lorenz gauge $\nabla_{\mu}A^{\mu}=0$ under the geometrical optics approximation becomes $\bar{A}^{\mu}k_{\mu}=0$. 

To further simplify the calculation, we introduce the normalised space-like polarization vector $\xi^{\mu}$, and the vector potential can be written as $ \bar{A}^{\mu}=\bar{A} \xi^{\mu}$. The polarization vector satisfies $\xi^{\mu}\xi_{\mu}^*=1$ and $\xi_{\mu}k^{\mu}=0$. In this case, Eq.\,(\ref{ptewos}) can be decomposed into equations of motion for the amplitude $\bar{A}$ and the polarization vector $\xi^{\mu}$ \cite{Schwarz:2020jjh} respectively,
\ba
k^{\mu}\nabla_{\mu}\bar{A}+\frac12\bar{A}\nabla_{\mu}k^{\mu} &=&0,\label{eomI}\\
k^{\mu}\nabla_{\mu}\xi^{\sigma}+g_{a\gamma\gamma}\epsilon^{\mu\nu\rho\sigma}k_{\mu}\xi_{\nu}\nabla_{\rho}a &=& 0.\label{eomxi}
\ea
The equation of motion for $\bar{A}$, i.e., Eq.\,(\ref{eomI}), does not contain the axion field. This means that the axion field does not affect the observed intensity of the light. The first term in Eq.\,(\ref{eomxi}) describes the parallel transport of the polarization vector $\xi^{\mu}$. The second term contains the axion effect, which is the birefringent effect that we are focused on.

In order to see the evolution of the polarization direction, one needs to project Eq.\,(\ref{eomxi}) to the reference frame of an observer. Such a reference frame can be properly characterized by an orthonormal basis of vectors $e^{\mu}_{(a)}$. These base vectors satisfy $e^{\mu}_{(a)} e_{\mu (b)} = \eta_{(a)(b)}$, where $a$ or $b = 0,1,2,3$. Particularly, $e^{\mu}_{(0)}$ is the time-like 4-velocity of the observer, which will be specified later, and $e_{(3)}^{\mu}\equiv(k^{\mu}-\omega e^{\mu}_{(0)})/\omega$ is a space-like vector with $\omega\equiv-k_{\mu} e^{\mu}_{(0)}$. Furthermore, $e^{\mu}_{(1)}$ and $e^{\mu}_{(2)}$ are space-like vectors which span the transverse plane orthogonal to both  $e^{\mu}_{(0)}$ and $e^{\mu}_{(3)}$.  The residual gauge freedom allows us to set $\xi^{\mu}e^{(0)}_{\mu} = \xi^{\mu}e^{(3)}_{\mu}= 0$ and thus $|\xi^{(1)}|^2 + |\xi^{(2)}|^2 = 1$. 
By parallel transporting the basis $e^{\mu}_{(a)}$ with the condition $k^{\mu}\nabla_{\mu}e^{\nu}_{(a)}=0$, we project Eq.\,(\ref{eomxi}) into the vector fields $e^{\mu}_{(a)}$ and obtain
\be
\partial_s\xi^{(j)}+g_{a\gamma\gamma}\partial_s a\epsilon^{(0)(i)(3)(j)}\xi_{(i)}=0.
\label{Aeq}
\ee
Here $s$ is the affine parameter of the photon trajectory, and $i$ or $j$ takes a value of $1$ or $2$. Writing the polarization vectors in the basis of circular polarization, we have $\xi_{L,R} \equiv \xi_{(1)}\pm i \xi_{(2)}$. The Eq.\,(\ref{Aeq}) can be easily solved as
\be
\xi_{L,R}(x^{\mu}_o)=\exp \left(\pm i\Delta\chi\right)\ \xi_{L,R}(x^{\mu}_e), \label{bi-eff}
\ee
where $\Delta\chi\equiv g_{a\gamma\gamma}\left[a(x^{\mu}_o)-a(x^{\mu}_e)\right]$. It only depends on the difference of the axion field values at the emission and the observation points, i.e., $x^{\mu}_e$ and $x^{\mu}_o$, respectively \cite{Carroll:1989vb,Harari:1992ea,Plascencia:2017kca,Ivanov:2018byi,Fujita:2018zaj,Liu:2019brz,Fedderke:2019ajk,Caputo:2019tms,Chen:2019fsq,Yuan:2020xui,Schwarz:2020jjh}. The linear polarization is a superposition of left and right circular polarization, thus $\Delta\chi$ represents the shift of EVPA for the linear polarization.
Interestingly, the ordinary birefringence, i.e., the Faraday rotation in the plasma with a magnetic field, has a nontrivial frequency dependence. On the other hand, the axion-induced birefringence is achromatic, as long as the geometric optics approximation is valid.

\subsection{Radiative Transfer}
The photon propagation nearby a SMBH is properly described by the covariant radiative transfer equations where both the plasma and the curved space-time are taken into account. In this subsection, we follow the formalism developed in \cite{Gammie_2012} and demonstrate how to include the axion effects. Comparing with the photon propagation equation without the medium, i.e., Eq.\,(\ref{ptewos}), the plasma effect leads to additional terms and the corresponding equation can written as
\be
    2 i k^{\mu} \nabla_\mu \bar{A}^{\nu} + i \bar{A}^{\nu} \nabla_{\mu} k^{\mu}  + 2i g_{a\gamma\gamma} (\nabla_\mu a)\epsilon^{\mu\nu\alpha\beta}k_{\alpha} \bar{A}_{\beta} = \Pi^{\nu}_{\mu} \bar{A}^{\mu} + j^{\nu}.
    \label{te}
\ee
Here the first term on the right hand side is the induced current from plasma, with $\Pi^{\sigma}_{\mu}$ being the linear response tensor.
Further $j^{\nu}$ is the external current describing the plasma emission.
To describe the propagation of the incoherent superposition of a large number of electromagnetic waves, one introduces macroscopic polarization tensor $N^{\mu\nu}$ \cite{Gammie_2012}
\begin{align}
    N^{\mu \nu} \equiv\left\langle \bar{A}^{\mu} \bar{A}^{* \nu}\right\rangle,\label{Nmn}
\end{align}
where $\left\langle\cdots\right\rangle$ denotes ensemble average. Using Eq.\,(\ref{Nmn}), one can rewrite Eq.\,(\ref{te}) into a compact form 
\begin{align}
k^{\mu} \nabla_{\mu} N^{\alpha \beta}=J^{\alpha \beta}+\tilde{H}^{\alpha \beta \kappa \lambda} N_{\kappa \lambda}.
\label{mrt}
\end{align}
Here $\tilde{H}^{\alpha \beta \kappa \lambda}$ is defined as
\begin{align}
\tilde{H}^{\alpha \beta \kappa \lambda} \equiv- i\left(g^{\beta \lambda} \tilde{\Pi}^{\alpha \kappa}-g^{\alpha \kappa} \tilde{\Pi}^{* \beta \lambda}\right),
\label{Pi1}
\end{align}
and the modified response tensor $\tilde{\Pi}^{\nu \mu}$ is
\begin{align}
\tilde{\Pi}^{\nu \mu} \equiv \Pi^{\nu \mu}-2i g_{a\gamma\gamma}\left(\partial_{\lambda} a\right) \epsilon^{\lambda \nu \rho \mu} k_{\rho}.
\label{Pi2}
\end{align}
The emissivity tensor $J^{\alpha \beta}$ is related to the external current as
\be
J^{\alpha \beta}=-i\left(\langle j^{\alpha}\bar{A}^{\beta *}\rangle-\langle j^{\beta *}\bar{A}^{\alpha}\rangle\right).
\ee
In addition to the axion-photon coupling , $\tilde{H}^{\alpha \beta \kappa \lambda}$
in Eq.\,(\ref{mrt}) contains various plasma effects, whose coefficients can be calculated conveniently if one chooses a comoving Cartesian frame with respect to the plasma. In such a frame, $e^{\mu}_{(0)}$ points along the plasma 4-velocity, and we further choose the other three base vectors as the same way described in Sec.\,\ref{APCB}. We now project Eq.\,(\ref{mrt}) using these base vectors, after applying the parallel transport condition $k^{\alpha} \nabla_{\alpha} e_{(a)}^{\nu}=0$, we obtain
\begin{align}
\frac{d N^{(a)(b)}}{d s}=J^{(a)(b)}+\tilde{H}^{(a)(b)(c)(d)} N_{(c)(d)},
\label{prcoratr}
\end{align}
with
\begin{align}
\tilde{H}^{(a)(b)(c)(d)} = H^{(a)(b)(c)(d)} -  g_{a\gamma\gamma} k_{(f)}\nabla_{(e)} a\Big[\eta^{(b)(d)} \epsilon^{(e)(a)(f)(c)}+\eta^{(a)(c)} \epsilon^{(e)(b)(f)(d)}\Big].
\label{tildeH}
\end{align}
Here $\tilde H^{(a)(b)(c)(d)}$ contains the ordinary plasma effect, i.e., $H^{(a)(b)(c)(d)}$, and the axion contribution. In this local tangent space, the total intensity and the polarization intensities can be parameterized by $4$ Stokes parameters as
\be I^S \equiv m^S_{(a)(b)} N^{(a)(b)},\ee
where $I^S \equiv (I,Q,U,V)$ are locally Lorentz invariant Stokes parameters.  They contain the total intensity $I$, the linear polarization intensities at two different directions, $Q$ and $U$, and circular polarization intensity $V$, respectively. The projection matrix $m^S_{(a)(b)}$ is defined as 
\begin{align}
    m^{I}\equiv\left(\begin{array}{llll}0 & 0 & 0 & 0 \\ 0 & 1 & 0 & 0 \\ 0 & 0 & 1 & 0 \\ 0 & 0 & 0 & 0\end{array}\right),\ \qquad m^{Q}\equiv\left(\begin{array}{cccc}0 & 0 & 0 & 0 \\ 0 & 1 & 0 & 0 \\ 0 & 0 & -1 & 0 \\ 0 & 0 & 0 & 0\end{array}\right), \nn\\
    m^{U}\equiv\left(\begin{array}{llll}0 & 0 & 0 & 0 \\ 0 & 0 & 1 & 0 \\ 0 & 1 & 0 & 0 \\ 0 & 0 & 0 & 0\end{array}\right),\ \qquad m^{V}\equiv\left(\begin{array}{cccc}0 & 0 & 0 & 0 \\ 0 & 0 & -i & 0 \\ 0 & i & 0 & 0 \\ 0 & 0 & 0 & 0\end{array}\right).
\end{align}

Similarly, the four Stokes emissivities $j^S \equiv (j_I,j_Q,j_U,j_V)$ are obtained through
\be j^S \equiv m^S_{(a)(b)} J^{(a)(b)}.\ee
Contracting $\tilde{H}^{(a)(b)(c)(d)}$ with the projection matrices, we define
\begin{align}
M^{S T} \equiv -\frac{1}{2} m_{(a)(b)}^{S*} \tilde{H}^{(a)(b)(c)(d)} m_{(c)(d)}^T.
\label{Minverse}
\end{align}
Splitting the contributions from the plasma effects and the axion-photon coupling, Eq.\,(\ref{Minverse}) can be decomposed as
\be M^{S T}=M_{\rm plasma}^{S T}+M_{\rm axion}^{S T}, \ee
where the first term is exactly the Muller Matrix in the ordinary radiative transfer equations,
\begin{align}
    M_{\rm plasma}^{S T} \equiv\left(\begin{array}{cccc}\alpha_{I} & \alpha_{Q} & \alpha_{U} & \alpha_{V} \\ \alpha_{Q} & \alpha_{I} & \rho_{V} & -\rho_{U} \\ \alpha_{U} & -\rho_{V} & \alpha_{I} & \rho_{Q} \\ \alpha_{V} & \rho_{U} & -\rho_{Q} & \alpha_{I}\end{array}\right).
    \label{Mueller Matrix}
\end{align}
Here $\alpha_I$, $\alpha_Q$, $\alpha_U$, $\alpha_V$ are the absorption coefficients, while $\rho_V$, $\rho_U$, $\rho_Q$ are the Faraday rotation and conversion coefficients.
{For example, in \texttt{IPOLE} \cite{Moscibrodzka:2017lcu, Noble:2007zx}, the Stokes $U$ is taken to align with the magnetic field so that $j_U = \alpha_U = \rho_U = 0$.}

Further, the axion contribution is simply characterized as
\begin{align}
    M_{\rm axion}^{S T}=
\left(\begin{array}{cccc}0 & 0 & 0 & 0 \\ 0 & 0 & -2g_{a\gamma\gamma} \frac{d a}{d s} & 0 \\ 0 & 2g_{a\gamma\gamma} \frac{d a}{d s} & 0 & 0 \\ 0 & 0 & 0 & 0\end{array}\right).
\end{align}
Therefore the modified radiative transfer equation in a local tangent space can be written as
\be
\frac{d}{d s}\left(\begin{array}{l}I \\ Q \\ U \\ V\end{array}\right)=\left(\begin{array}{l}j_{I} \\ j_{Q} \\ j_{U} \\ j_{V}\end{array}\right)-\left(\begin{array}{llll}\alpha_{I} & \alpha_{Q} & \alpha_{U} & \alpha_{V} \\ \alpha_{Q} & \alpha_{I} & \rho^\prime_{V} & \rho_{U} \\ \alpha_{U} & -\rho^\prime_{V} & \alpha_{I} & \rho_{Q} \\ \alpha_{V} & -\rho_{U} & -\rho_{Q} & \alpha_{I}\end{array}\right)\left(\begin{array}{l}I \\ Q \\ U \\ V\end{array}\right),
\label{finalrte}
\ee
with
\be\label{ipole-mod} \rho^\prime_{V} = \rho_{V} - 2g_{a\gamma\gamma} \frac{d a}{d s}.\ee
The axion contributions are simply included by a  change in the ordinary Faraday rotation coefficient.
It is clear from Eq.\,(\ref{finalrte}) that $\rho^\prime_V$ plays the role of changing the EVPA, defined as 
\begin{align}
\chi \equiv \frac{1}{2} \arg (Q+i U).
\end{align}
In the absence of the emissivities and the plasma effects, Eq.\,(\ref{finalrte})
leads to consistent results as in Eq.\,(\ref{bi-eff}).

\section{Birefringence from Axion Cloud -- Thin Disk and Ray Tracing}\label{TDRT}

The following sections study the birefringent signals induced from the superradiant axion cloud accumulated around the SMBH, with various astrophysical conditions. We only consider the cases in which the radiations are emitted from the accretion flow, rather than the jet. 
We first consider the geometrically thin and optically thick disk. Then we will further discuss the RIAF models, whose geometric thickness is an input parameter. Both cases are expected to be explored at horizon scale by the future VLBI observations  \cite{Raymond_2021,Lngeht,Gurvits:2022wgm}.

For the thin disk, after photons are emitted, they propagate in the vacuum without the plasma. Consequently, the EVPA shift of the linear polarized photon can be simply described by Eq.\,(\ref{bi-eff}).  For the frequency regime that we consider here, a geometrically thin disk is opticially thick, thus the contribution from lensed photons can be safely ignored. For each point on the sky plane, we can trace back along the line of sight, and the emission only comes from the point of its first intersection with the disk.
Neglecting the axion field value near the Earth, the EVPA shift $\Delta\chi$ in Eq.\,(\ref{bi-eff}) becomes
\ba 
\Delta \chi (t, \rho, \varphi) = -  \frac{b \ c\  R_{211} (r_E) \cos{ [\omega t_E - m \phi_E]}} {2 \pi R_{211} (r_{\textrm{max}})}.\label{deltachi}
\ea
The ratio $R_{211} (r_E) / R_{211} (r_{\textrm{max}})$ is shown in Fig.\,\ref{RWFaf}. 
The peak value of the axion cloud is parametrized by
\be b \equiv \frac{a_{\textrm{max}}}{f_a},\ee  which can be  $\mathcal{O} (1)$ \cite{Yoshino:2012kn,Yoshino:2013ofa,Yoshino:2015nsa,Baryakhtar:2020gao} as mentioned above. $f_a$ is required to be below $10^{16}$ GeV so that the extraction of black bole rotation energy is negligible, thus complementary to black hole spin measurements \cite{Arvanitaki:2010sy,Arvanitaki:2014wva,Brito:2014wla,Davoudiasl:2019nlo,Stott:2020gjj,Unal:2020jiy,Saha:2022hcd} and direct shadow observations \cite{Roy:2019esk,Cunha:2019ikd,Creci:2020mfg,Roy:2021uye,Chen:2022nbb}.

One also defines the fundamental constant
\be c \equiv 2\pi g_{a\gamma\gamma} f_a,\label{defC}\ee
that translates the axion-photon coupling $g_{a\gamma\gamma}$ to a dimensionless quantity in the unit of the decay constant $f_a$. Here $c$ is the fundamental constant that we aim to constrain in our study \cite{Chen:2019fsq,Chen:2021lvo}.
In Eq.\,(\ref{deltachi}), there are two sets of coordinates on the two sides of the equation. First, $(t, \rho, \varphi)$ are the time of observation and the polar coordinates on the sky plane. Further, $(t_E, r_E, \phi_E)$ label the emission time and the polar coordinates of the black hole equatorial plane. These two sets of coordinates are related to each other by ray tracing, following photons' geodesics. Both the inclination angle, $i$, of the Kerr black hole with respect to the sky plane and the magnitude of the black hole spin, $a_J$, have impacts. 

The EVPA measurements performed by the EHT are presented as a function of the azimuthal angle on the skype plane \cite{EHTP, EHTM}. Without loss of the generality, we use the following ansatz to parametrize the EVPA variations 
\be 
\Delta \chi (t, \varphi, \rho)  = \mathcal{A} (\varphi, \rho) \cos{ [\omega t \pm \varphi + \delta(\varphi, \rho)]}.
\label{ansatz}
\ee
Here $\mathcal{A}$ and $\delta$ characterize the amplitude and the relative phase of the EVPA oscillation respectively. The $\pm \varphi$ term in the bracket comes from the angular dependence of the axion cloud since $|m|=1$. The sign is for two possibilities of the black hole spin orientation, either opposite to us ( $i > 90^\circ$ ) or towards us ( $i < 90^\circ$ ) respectively. In the following analysis, we normalize the amplitude $\mathcal{A}$ in terms of the maximum value of the axion field 
\be g_{a\gamma\gamma} a_{\textrm{max}} \equiv \frac{b\,c}{2\pi},\ee
according to Eq.\,(\ref{deltachi}).

We note that Eq.\,(\ref{ansatz}) has not taken into account of the intrinsic variations of the accretion flows. 
To reduce the nontrivial uncertainties from the time-dependent astrophysical background, we introduced differential EVPAs in the time domain \cite{Chen:2021lvo}. In this case, we extract the axion signal by comparing the EVPA observations at two different times $t_i$ and $t_j$
\be \Delta \chi (t_i, \varphi, \rho) - \Delta \chi (t_j, \varphi, \rho) = 2 \sin{[\omega t_{\textrm{int}}/2]}\,
\mathcal{A} (\varphi, \rho)\, \sin{ \left[\omega (t_i+t_j)/2 \pm \varphi + \delta(\varphi, \rho)\right]},\ee
where the interval time between two sequential observations is $t_{\textrm{int}} \equiv t_j - t_i$.
As far as $t_{\textrm{int}}$ is shorter than the timescale of the accretion flow dynamics, the astrophysical uncertainties can be suppressed. 
On the other hand, one pays the price for the suppression factor $2 \sin{[\omega t_{\textrm{int}}/2]}$ if the axion oscillation period is much longer than $t_{\textrm{int}}$.
More details about the optimized analysis method will be given in the later discussion.

\subsection{Ray Tracing from Novikov Thorne Thin Disk}

In this subsection, we start from the thin disk model to study the properties of the axion-induced birefringence signals, with different choices on the inclination angle $i$ and the black hole spin $a_J$. Shakura and Sunyaev first developed this kind of model in \cite{Shakura:1972te}, and later it was generalized to a fully general relativistic version by Novikov and Thorne \cite{Novikov:1973kta} (NT model). The NT model is an axisymmetric and stationary solution, with an optically thick and geometrically thin disk on the equatorial plane. All photons we receive come directly from it without the contribution of lensed photons, and one can safely neglect the thickness of the accretion disk. The fluid in the disk has a nearly Keplerian orbit. 
The polarization of the radiation in this model is calculated from the electron scatterings in a semi-infinite atmosphere \cite{RTChandra}.
The spectrum of the thin disk is approximately thermal. The geometrically thin disk model can be applicable for some classes of active galactic nuclei (AGN) with mass accretion rate being nearly Eddington mass accretion rate such as quasars, which future VLBI observations have the potential to measure.

\begin{figure*}[htb] 
\centering 
\includegraphics[width=0.45\textwidth]{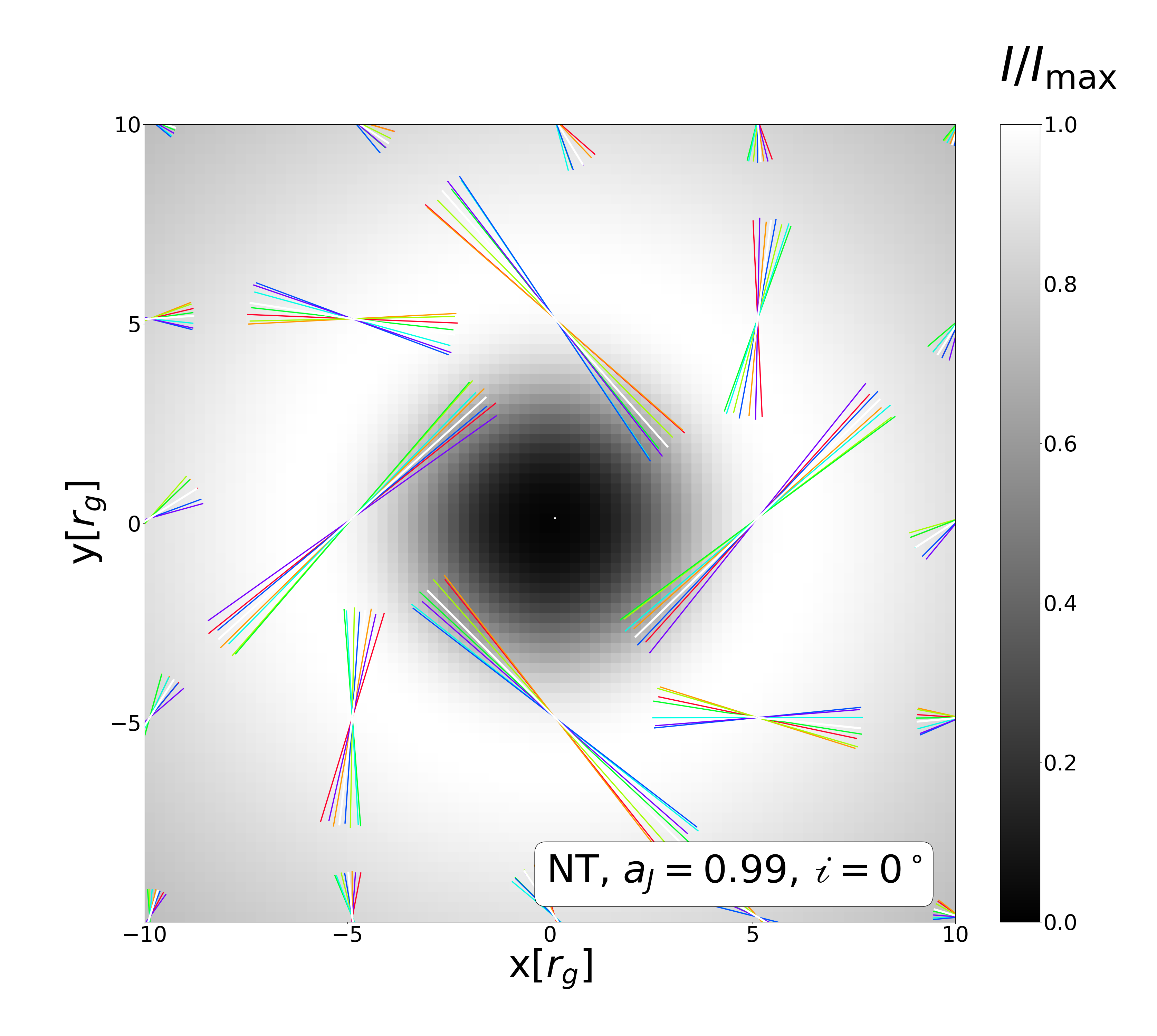}
\includegraphics[width=0.45\textwidth]{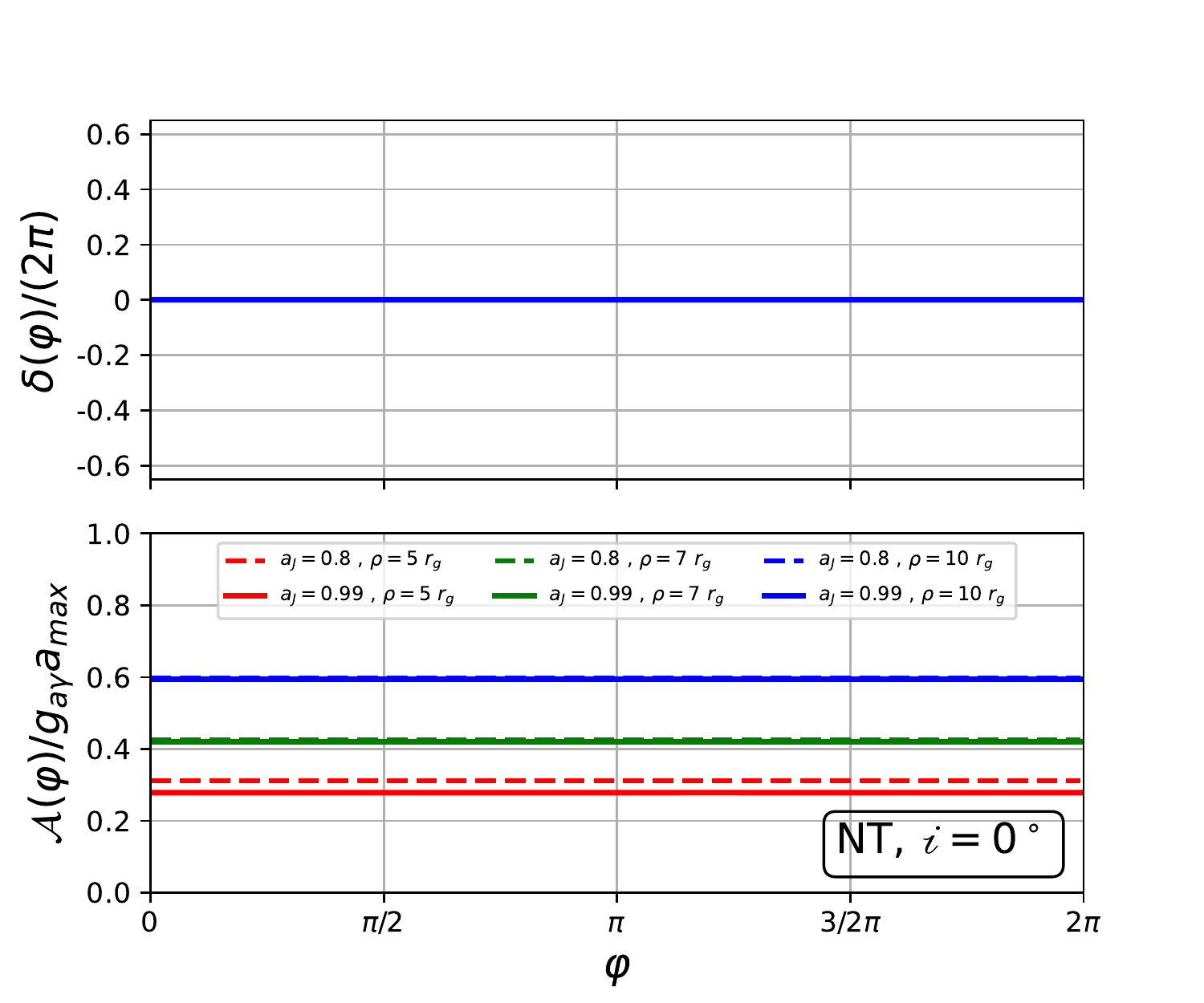}
\includegraphics[width=0.45\textwidth]{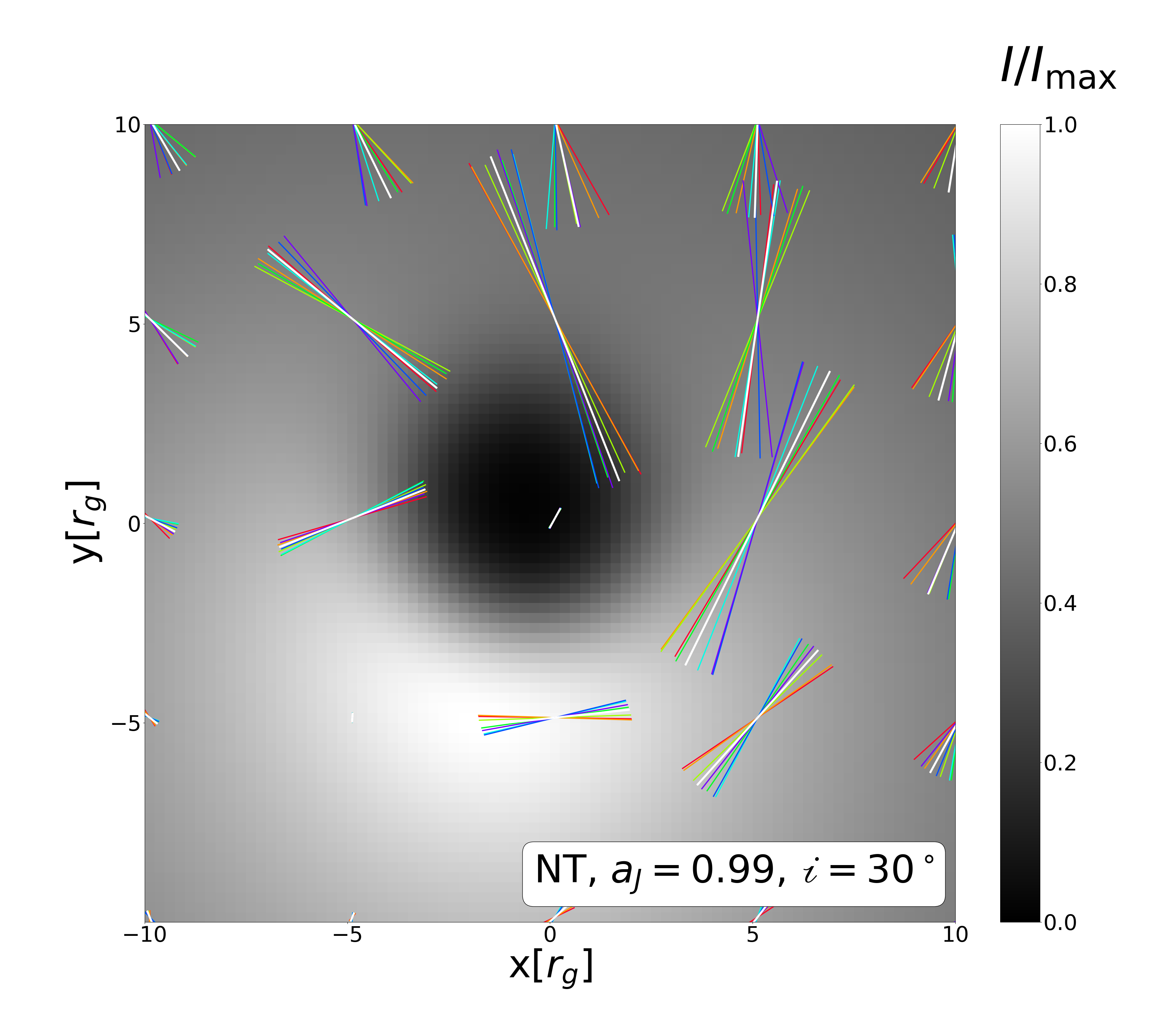}
\includegraphics[width=0.45\textwidth]{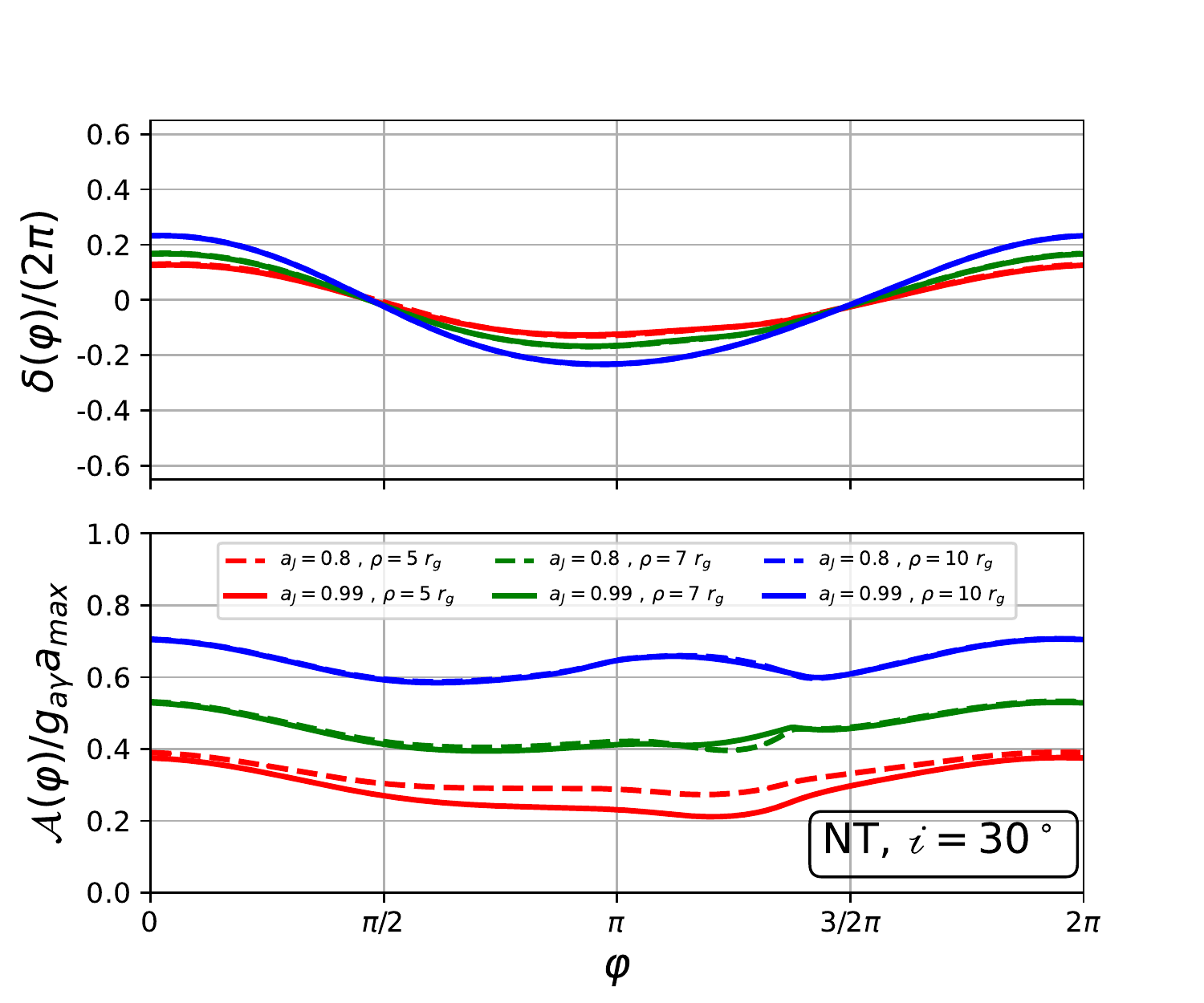}
\includegraphics[width=0.45\textwidth]{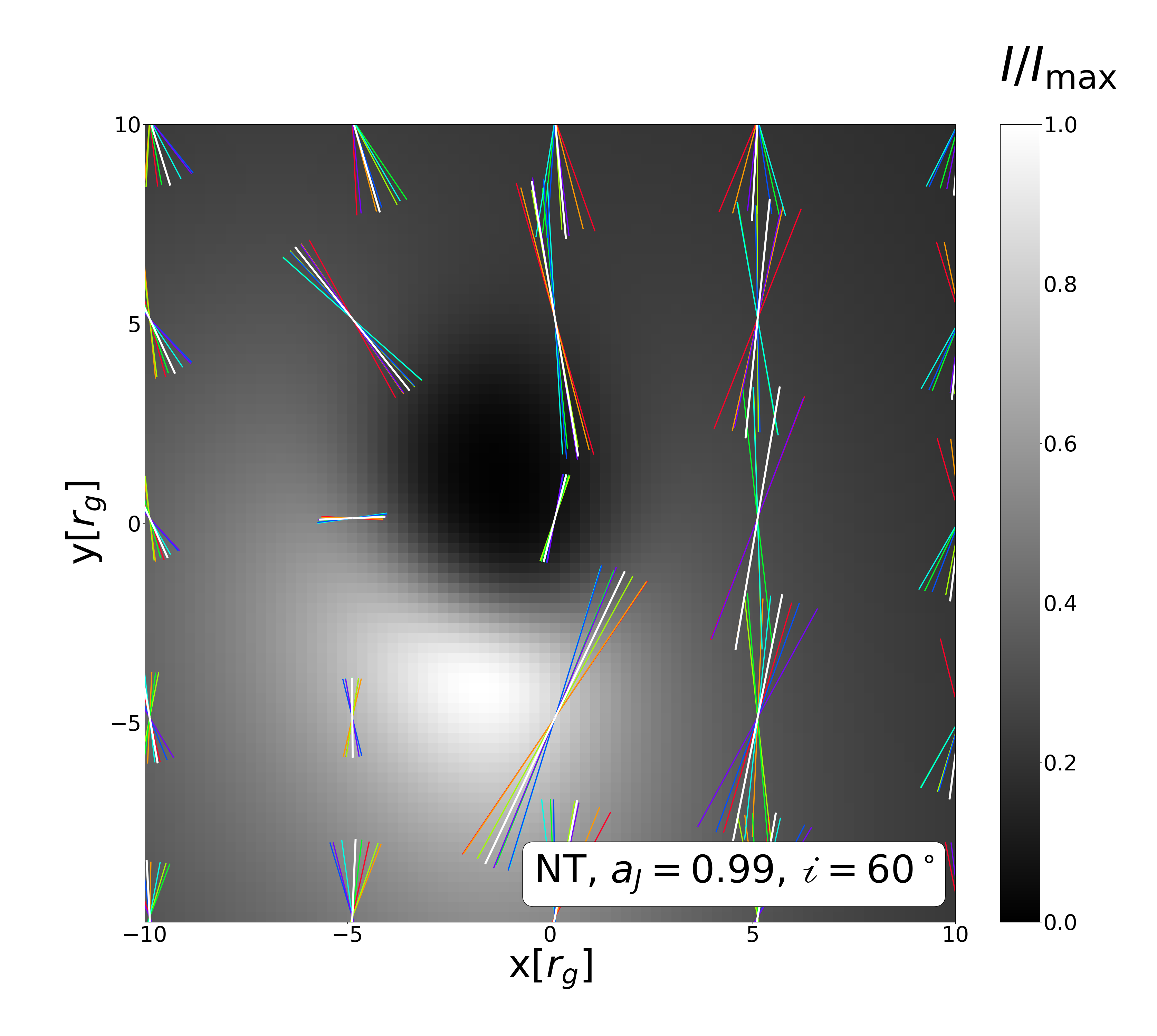}
\includegraphics[width=0.45\textwidth]{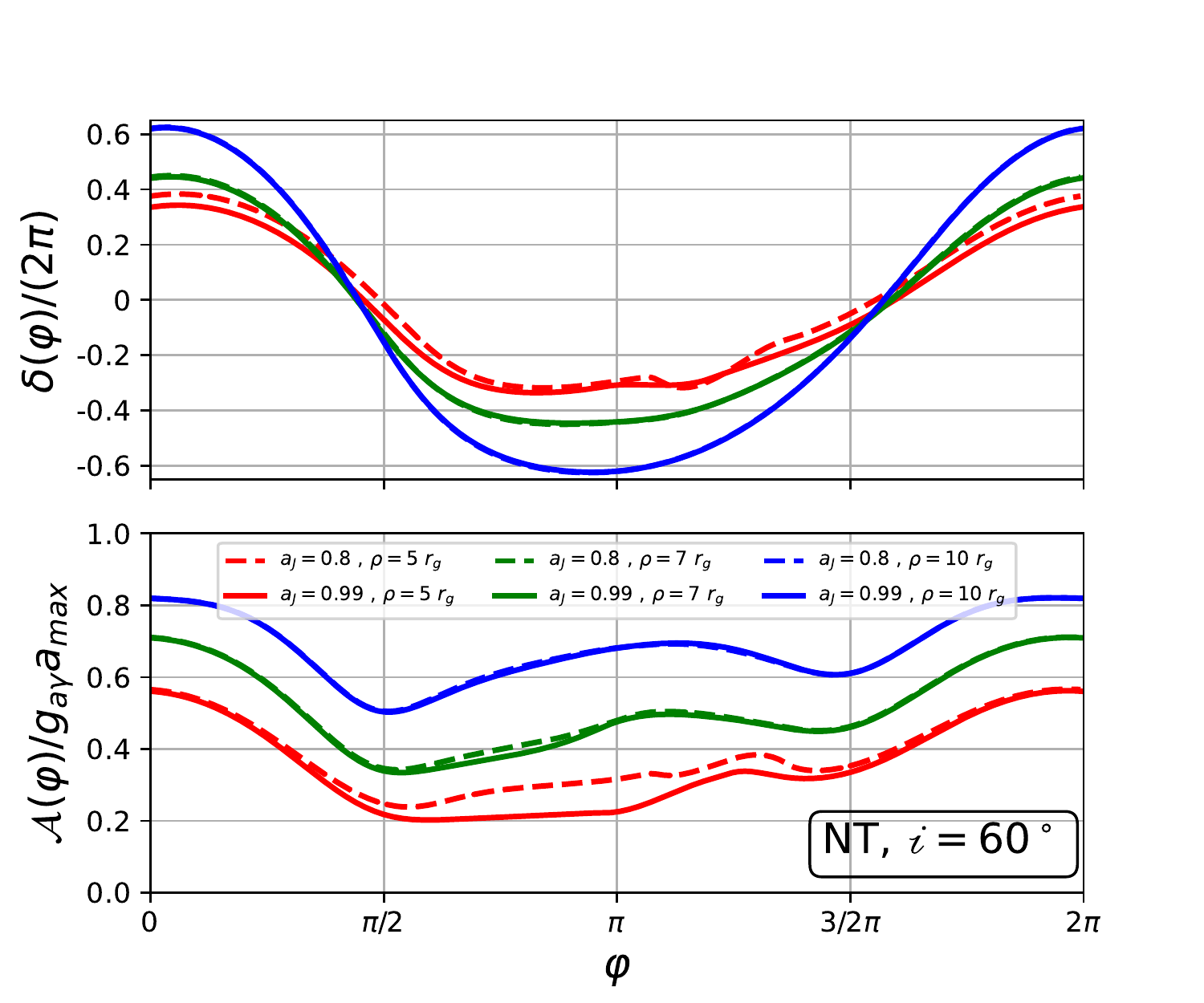}
\caption{\footnotesize{Left: We take the NT model and show a few benchmark results from the \texttt{IPOLE} simulation \cite{Moscibrodzka:2017lcu, Noble:2007zx}, with various choices of the inclination angle. We take $\alpha=0.25$, $g_{a\gamma\gamma}a_{\textrm{max}}=1$ rad and $a_j=0.99$. The black hole spin points along the $-x$ direction on the sky plane. 
The white quivers represent the astrophysical linear polarized radiations without the axion effect. From red to purple, the rainbow color is used to show the time variations of the EVPA in the presence of the axion cloud. Right: We show the relative phase, $\delta/2\pi$, and the amplitude, $\mathcal{A}/ g_{a\gamma\gamma} a_{\textrm{max}}$, as a function of the azimuthal angle $\varphi$. We take $\rho = 5\,r_g, 7\,r_g$ and $10\,r_g$ as benchmarks. The results for $a_J = 0.8$ are also shown for comparison. The deviations from the results of $a_J = 0.99$ are due to the difference in photons' geodesics. We note that $\varphi = 0$ corresponds to $+x$ direction on the left panel.  \vspace{3mm}}}
\label{inclination_a}
\end{figure*}

We substitute the axion cloud induced birefringence contribution, i.e., given in Eq. (\ref{ipole-mod}), into the radiative transfer code \texttt{IPOLE} \cite{Moscibrodzka:2017lcu, Noble:2007zx} and we specify the NT model as the emission source around the SMBH. The birefringence signals, with various choices of the inclination angle $i$ and the black spin $a_J$, are shown in Fig.\,\ref{inclination_a}.

\clearpage

\clearpage

\begin{figure*}[htb]
\begin{center}
\includegraphics[width=0.45\textwidth]{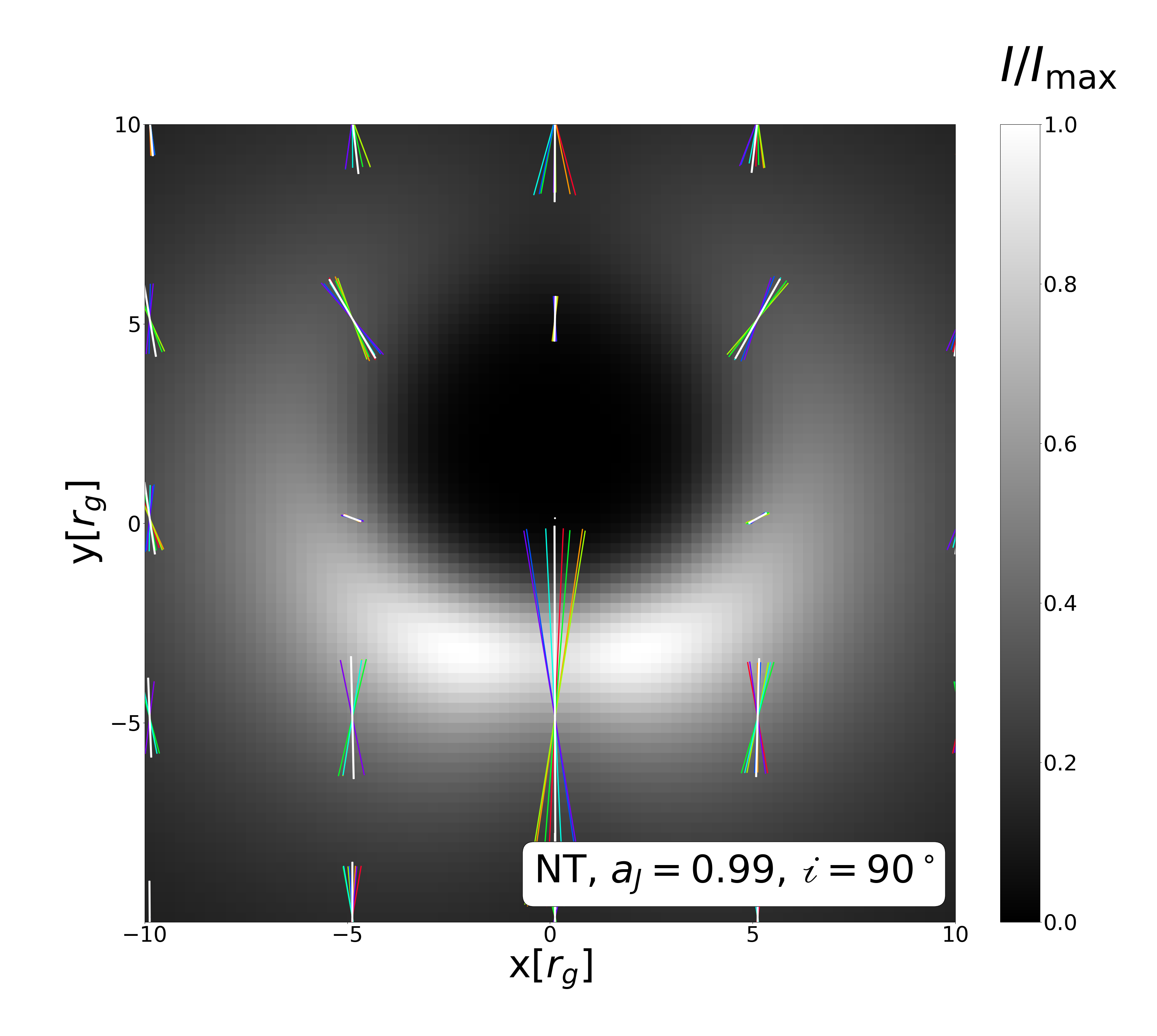}
\includegraphics[width=0.45\textwidth]{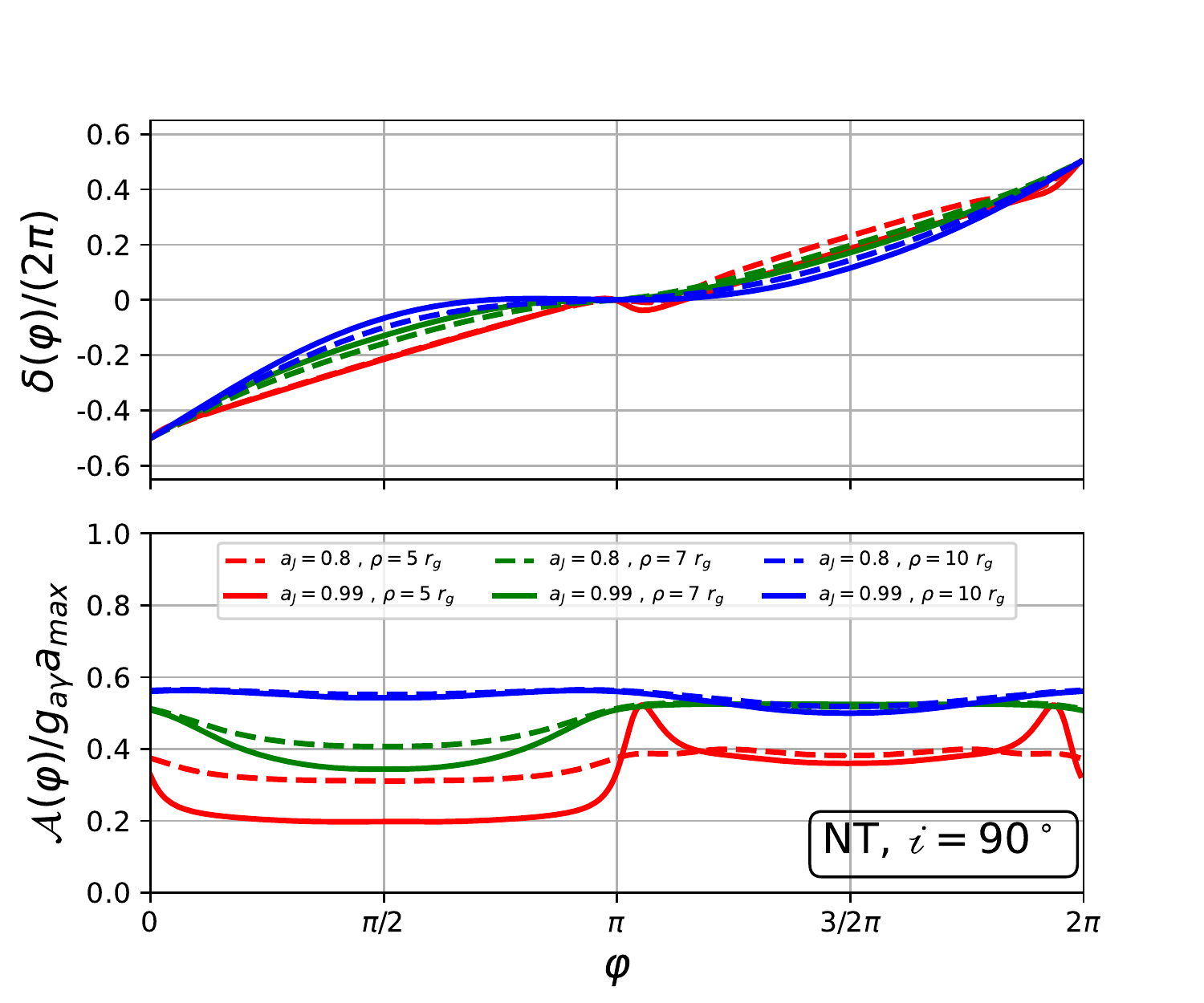}\\
\end{center}
\textbf{Figure 2}: \footnotesize{(Cont.)\vspace{3mm}}
\end{figure*}

On the left panel, each plot contains an intensity map. On top of it, the quivers with different colors provide the information about the linear polarization. The length of each quiver is proportional to the intensity of the linear polarization  $I_L = \sqrt{Q^2 + U^2}$, and the direction represents the EVPA. The white quiver lines show the EVPA without the axion. One oscillation period of the axion cloud is equally divided into eight segments, and the color of each quiver, from red to purple in the rainbow order, represents the time evolution. As expected, the birefringence signals from the axion cloud behave as a propagating wave along the azimuthal angle $\varphi$ of the sky plane.
On the right panel, we use the ansatz in Eq. (\ref{ansatz}) to fit the relative phase $\delta$ and the amplitude $\mathcal{A}$ of the EVPA oscillation along the $\varphi$ direction. We choose the radial coordinate as $\rho = 5\,r_g, 7\,r_g$ and $10\,r_g$ respectively, for black hole spin $a_J = 0.99$ and $0.8$. The axion mass is taken to satisfy $\alpha = 0.25$.

\subsubsection{Relative Phase of Azimuthal EVPA Oscillation}

\begin{figure*}[htb] 
\centering
\includegraphics[width=0.8\textwidth]{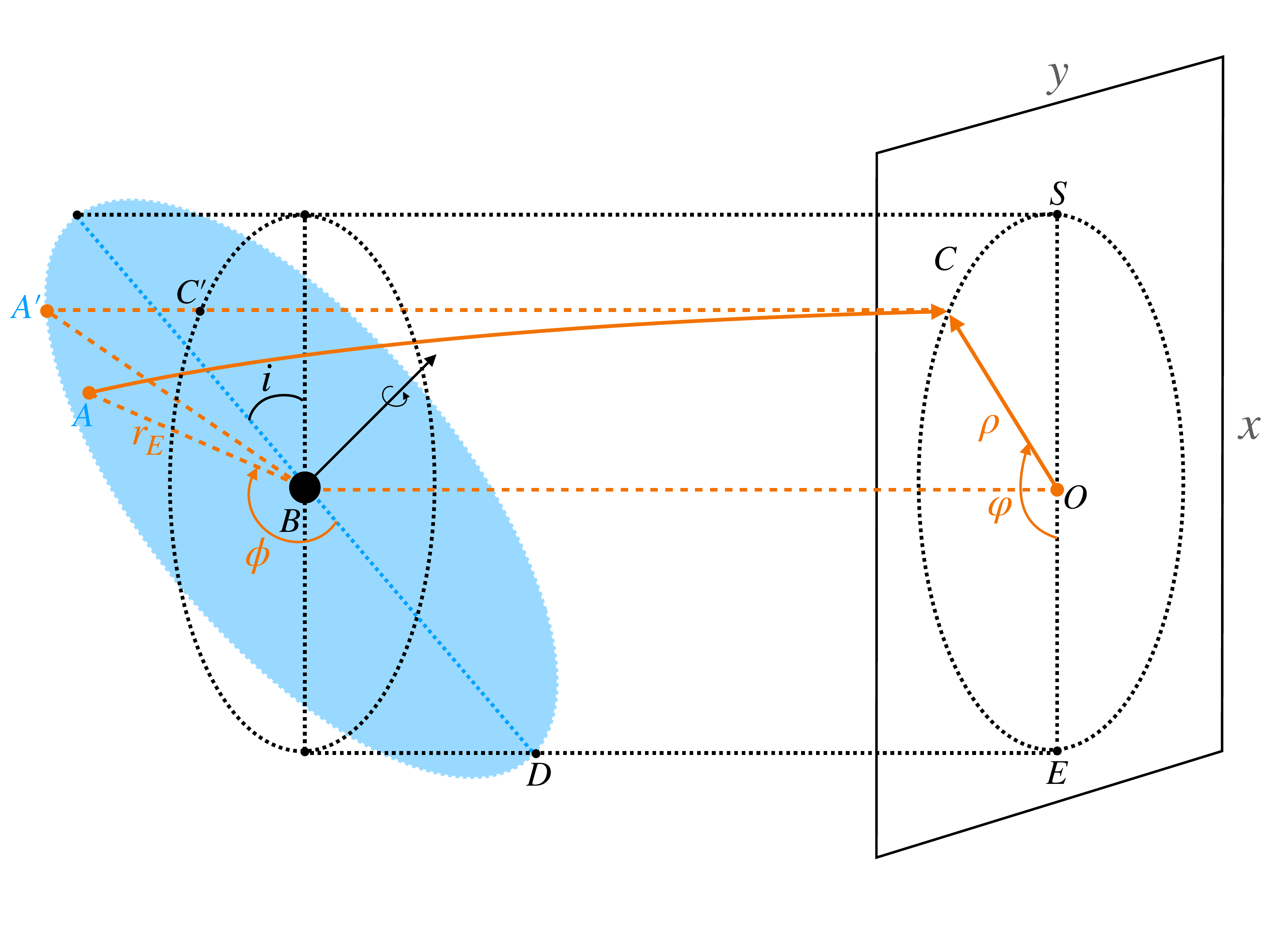}
\caption{\footnotesize{This figure shows the geometrical projection of a reference circle, at a fixed radius $\rho$ on the sky plane, onto the equatorial plane of the accretion disk marked in blue. The photons emitted at the point A on the equatorial plane reach the point C on the sky plane, with the polar angle $\varphi$. On the equatorial plane under the black hole coordinate, the point A is labeled as $(r_E,\phi)$. The direction of $OS$ is chosen to be along the projection of the black hole spin on the sky plane, and $i$ is the inclination angle of the black hole.
}}
\label{example}
\end{figure*}

The relative phase $\delta(\varphi, \rho)$ can be precisely obtained by reading out the numerical results from \texttt{IPOLE}. However, in the almost face-on scenarios, i.e., $i\simeq 0^\circ$ or $180^\circ$, this may be calculated analytically with a good approximation. In Fig.\,\ref{example}, we show the trajectories of photons from the emission point A on the accretion disk. We assume this point is relatively far from the black hole horizon, so that the frame dragging effects are not important. Under these assumptions, the relative phase $\delta(\varphi, \rho)$ can be written as
\be
\delta(\varphi,\rho) \approx \alpha\ \tan{i}\  \cos{\varphi}\ \rho/r_g \label{phasedelay}.
\ee
This relation can be understood as the time delay of the equatorial plane emission, induced by the inclination angle $i$ \cite{Chen:2021lvo}.
More explicitly, the time delay caused by the travel distance for the light from point $A$ can be approximated as $A'C'$ \cite{Loktev:2021nhk}.

On the other hand, the edge-on scenario with $i = 90^\circ$ is subtle. The observer is located on the equatorial plane. Such a plane is singular in the NT model due to the assumption of the infinitely thin disk. Consequently, the emission from the edge of the accretion disk, which propagates on the equatorial plane, is not included artificially. When the frame dragging effects are negligible, there is no $\varphi$-dependence in the phase component of Eq.\,(\ref{ansatz}) due to the rotation symmetry on the sky plane. This is consistent with the  $\delta(\varphi)$  result shown in last pair of Fig.\,\ref{inclination_a}, where the $\varphi$ dependence in $\delta(\varphi)$ approximately cancels the $\varphi$ term in cosine function in Eq.\,(\ref{ansatz}).

\subsubsection{Amplitude of Azimuthal EVPA Oscillation}
We next turn to the amplitude $\mathcal{A}$ of the EVPA oscillation. This quantity is determined by the axion field value at the emission point, shown in Fig.\,\ref{RWFaf}. Under the thin disk approximation, we simply need a map from the sky plane coordinate $(\rho, \varphi)$ to the equatorial plane coordinate $(r_E, \phi_E)$.

For the face-on case with ${i}=0^\circ$ or $180^\circ$, the amplitude $\mathcal{A}$  only depends on the $\rho-r_E$ mapping. This is consistent with the results shown in the first row of Fig.\,\ref{inclination_a}. Particularly, the amplitude at a fixed radius has no $\varphi$ dependence.
For general cases, such as ${i}=30^{\circ}$ and $60^{\circ}$, the rotation symmetry on the sky plane is broken.
However the curve of $\mathcal{A}(\varphi)$ still preserves an approximate reflection symmetry with respect to $\varphi=\pi$. Such a feature can be understood using ray tracing that connects the equatorial plane and the sky plane through geodesics.  The small violation of the reflection symmetry is caused by the frame dragging under the Kerr metric.

\begin{figure*}[htb] 
\centering
\includegraphics[width=0.7\textwidth]{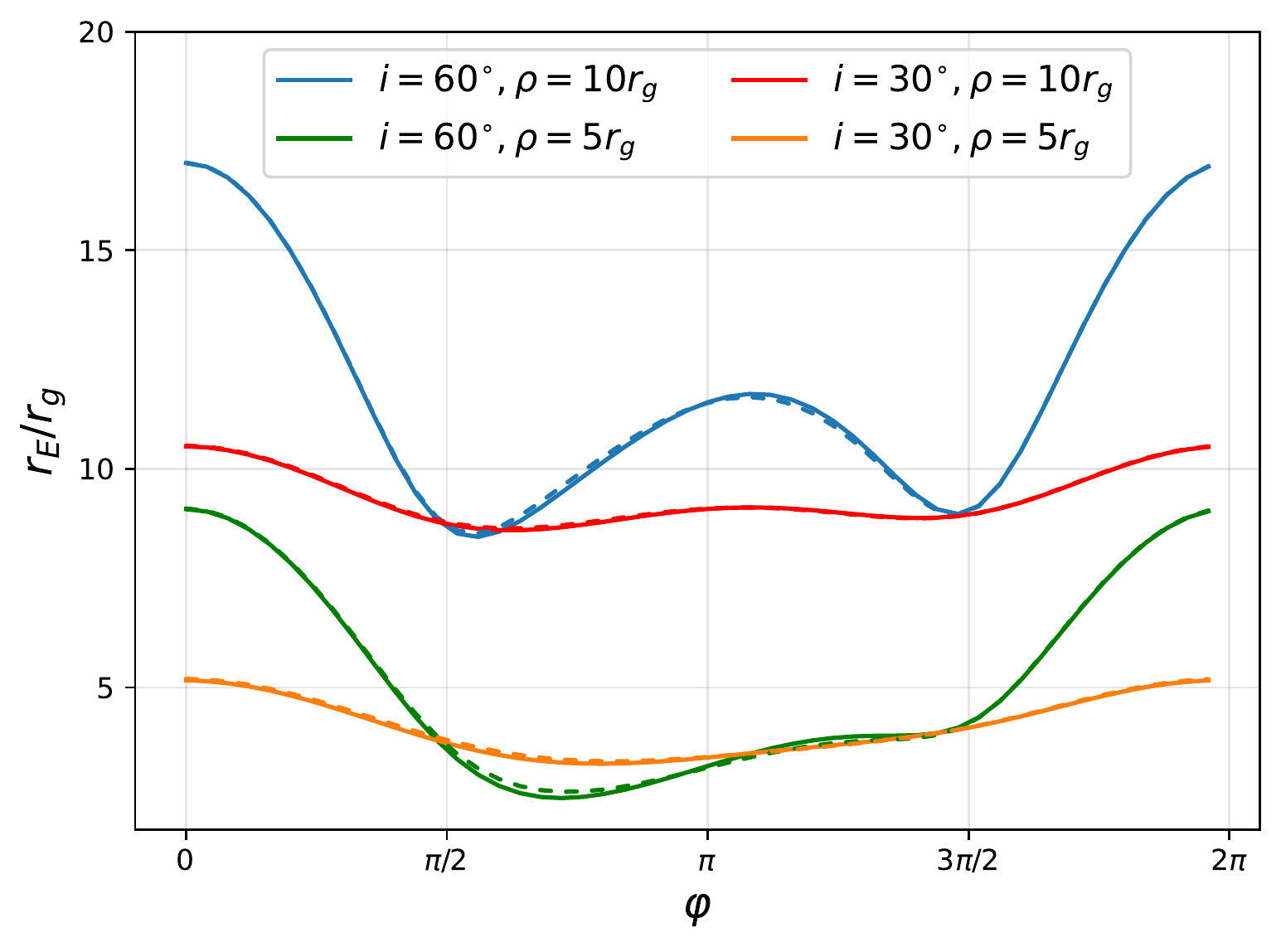}
\caption{\footnotesize{We study the mapping between the sky plane coordinate $(\rho, \varphi)$ to the polar coordinate on the equatorial plane. For fixed values of $\rho$, we show $r_E$ as a function of $\varphi$. The solid line and the dashed line represent the results with $a_J=0.99$ and $a_J=0.8$ respectively.}}
\label{analytic_A}
\end{figure*}

We calculate the photon geodesics according to the formalism developed in \cite{Gralla:2019ceu,Gelles:2021kti}. This constructs a map between the sky plane coordinate $(\rho, \varphi)$ and the equatorial plane coordinate $(r_E, \phi_E)$. The results are shown in Fig.\,\ref{analytic_A} for black holes with a spin of $a_J=0.99$ and $0.8$.

We show $r_E$ as a function of $\varphi$ at $\rho = 5\,r_g$ and $10\,r_g$ respectively, with different choices of the inclination angle ${i} = 30^\circ$ and $60^\circ$. Given the properties of the axion radial wave functions presented in Fig.\,\ref{RWFaf}, such a coordinate mapping explains the feature in Fig.\,\ref{inclination_a} nicely.
More explicitly, with ${i} = 60^{\circ}$ and $\rho = 10\,r_g$, the curve of $\mathcal{A}(\varphi)$ contains a double peak feature. This is caused by the geometric projection from a circle on the sky plane to an ellipse on the equatorial. As $\rho$ becomes smaller, the gravitational bending of a photon trajectory plays a more critical role. In this case, the mapping between two sets of coordinates becomes more subtle, and the approximation of the reflection symmetry with respect to $\varphi=\pi$ becomes worse.  

Particularly, for a given $\rho$, the photon emitted at the point $A$ with the sky plane angle $\varphi$ in Fig.\,\ref{example} experiences more gravitational bending than the photon from the opposite point with $\varphi-\pi$. Consequently, although both the photons from these two points reach the circle with the same $\rho$ on the sky plane, the point $A$ is more close to the black hole horizon. For the benchmark we consider here, this translates to a smaller value of the axion field, according to the radial wave function shown in Fig.\,\ref{RWFaf}. The explains the difference of $\mathcal{A}(\varphi)$ at $\varphi=0^\circ$ and $180^\circ$. Again, we emphasize that a slight asymmetry with respect to $\varphi=\pi$ is caused by the black hole spin.

On the right panel of Fig.\,\ref{inclination_a}, we show the results for both $a_J=0.99$ and $a_J=0.8$. Since the difference between the axion cloud wave functions for these two spin choices is negligible, the main difference of the birefringence signals comes from how spin modifies the geodesics. As shown on the right panel, the larger value of $a_J$ tends to decrease the signal amplitude $\mathcal{A}(\varphi)$. This effect is more pronounced for smaller radius $\rho$ and for $\varphi$ close to $\pi$, which is caused by the fact that photons reaching these regions are emitted at places closer to the black hole, where the black hole spin will have a stronger effect on geodesics.

Now let us consider the edge-on scenario where ${i} = 90^{\circ}$. In the limit of no black hole spin, $\mathcal{A}(\varphi)$ should be a constant, due to the rotation symmetry on the sky plane. The extra features, as shown in the right panel of the last pair in Fig.\,\ref{inclination_a}, are induced by the frame dragging. Such features become weaker when the black hole has a smaller spin or the distance to the black hole is larger.

\subsection{Global Feature: Angular Modes of the Azimuthal EVPA}

Without loss of generality, we focus on the cases with the inclination angle ${i} < 90^\circ$, and the axion cloud occupies the $m=1$ quantum state. In this case, the ansatz of the EVPA shift in Eq.\,(\ref{ansatz}) becomes 
\be 
\Delta \chi (t, \varphi, \rho)  = \frac{\mathcal{A} (\varphi, \rho)}{2} \left(  e^{i\omega t} e^{ -i \varphi + i \delta(\varphi, \rho)} + e^{-i \omega t} e^{ i \varphi - i \delta(\varphi, \rho)}  \right).
\label{ansatzE}
\ee

\begin{figure*}[htb] 
\centering
\includegraphics[width=0.8\textwidth]{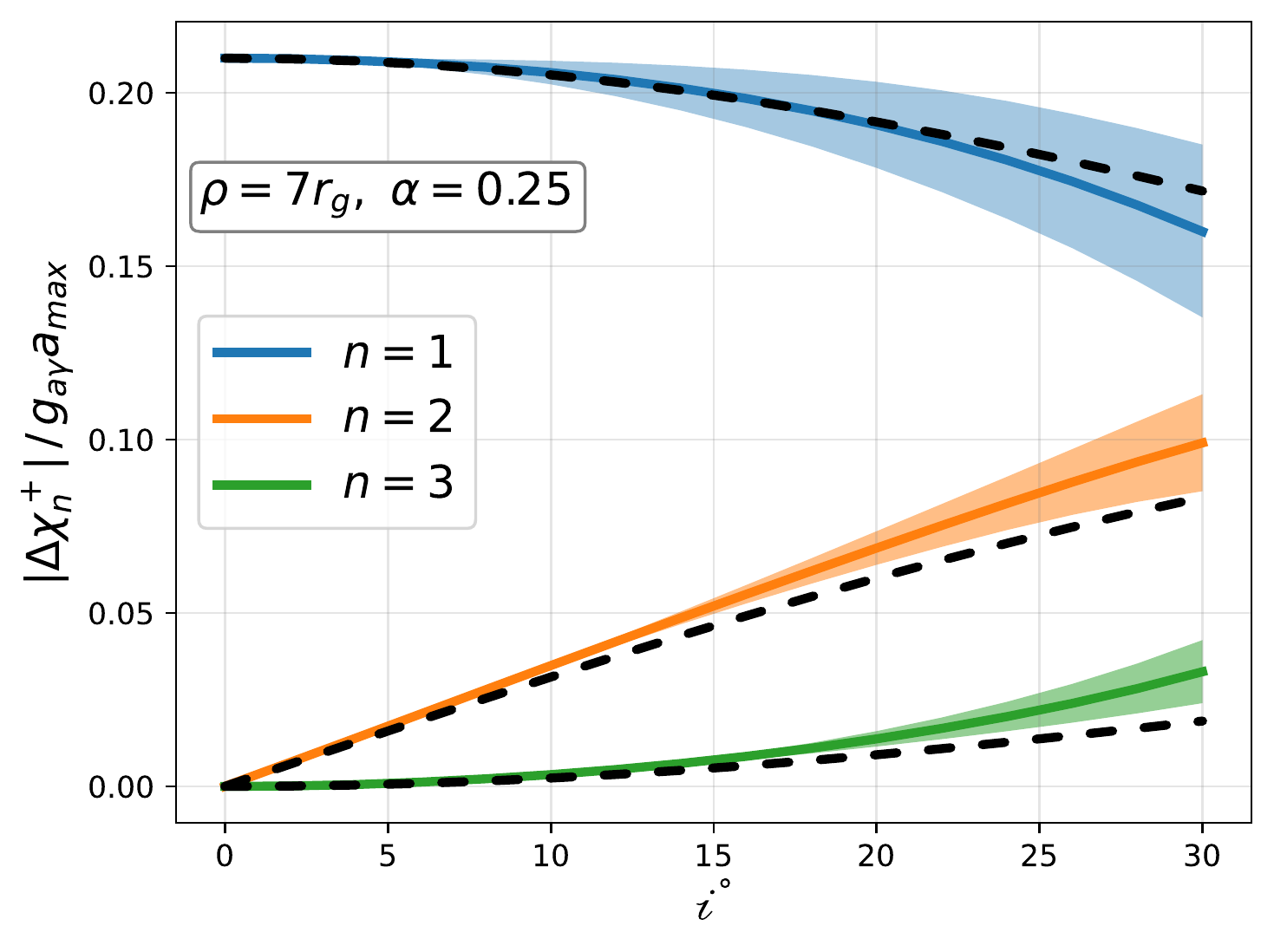}
\caption{\footnotesize{The magnitude of the Fourier coefficients defined in Eq.\,(\ref{AM}) are shown as blue, orange and green lines for $n=1,2$ and $3$ respectively. As a benchmark, we take $\rho=7\,r_g$ on the sky plane. The black hole spin is set as $a_J = 0.99$, and axion mass is chosen as $\alpha = 0.25$. The dashed lines show the results from the analytical approximation, derived in Eq.\,(\ref{AF}). 
\vspace{3mm}}}
\label{analytic_A_Fourier}
\end{figure*}

The features in the EVPA variation can be nicely captured by performing a Fourier transformation on Eq.\,(\ref{ansatzE}). To demonstrate that, we consider two scenarios. In the first scenario, we assume the observations are long enough to cover the whole period of the axion oscillation. In this case, the time dependence in Eq.\,(\ref{ansatzE}) can be properly extracted and we only need to focus on the angular dependence when we perform the Fourier transform.  Let us define $\Delta\chi_{n}^+$ as
\be \Delta\chi_{n}^+ = \frac{1}{4\pi}\int_{0}^{2\pi}\ \mathcal{A} (\varphi, \rho) \ e^{ -i \varphi + i \delta(\varphi, \rho)} \ e^{in\varphi}\ d\varphi. \label{AM}\ee
In Fig.\,\ref{analytic_A_Fourier}, we show the results of $|\Delta\chi_{n}^+|$ for $n = 1, 2, 3$ as a function of the inclination angle ${i}$ in solid lines.

For the face-on case, with a negligible relative phase $\delta(\varphi, \rho)$, only the mode with $n= 1$ is non-zero, as expected. When the inclination angle gradually increases, as shown Eq.\,(\ref{phasedelay}), we have $\delta(\varphi,\rho) \propto \sin {{i}}\ \cos \varphi$ for small $i$. This leads to mixtures among various Fourier modes. Approximately, one can take Eq.\,(\ref{phasedelay}) into Eq.\,(\ref{AM}) and ignore the $\varphi$ dependence in $\mathcal{A} (\varphi)$. This leads to 
\begin{equation}
    |\Delta\chi_n^+(\rho)| \simeq \frac{1}{2}\ \mathcal{A}(0, \rho)\ J_{n-1}\left(\alpha\ \sin {i} \ \rho/r_g\right),
\label{AF}
\end{equation}
where $J_n(x)$ is the first type Bessel function. Eq.\,(\ref{AF}) gives the dashed lines in Fig.\,\ref{analytic_A_Fourier}, which agree well with the numerical results for nearly face-on cases, e.g., with ${i} < 30^\circ$. We note that, for larger inclination angles ${i} > 30^\circ$, the mixtures among various angular modes become complicated and higher modes are also important. Consequently, the Fourier analysis suggested in Eq.\,(\ref{AM}) becomes less convenient to characterize the axion induced signal. One may simply perform a direct comparison between the ansatz in Eq.\,(\ref{ansatz}) with the data. We also highlight that, for the edge-on case with ${i} \simeq 90^\circ$, the Fourier mode with $n= 0$ is dominant, as consistent with the results shown in Fig.\,\ref{inclination_a}.

\section{Birefringence from Axion Cloud -- RIAF and  Washout}\label{RW}

In contrast to the geometrically thin and optically thick disk, such as the one discussed in the previous section, the emission of a RIAF has a larger spatial distribution along the line of sight. Also significant contributions may come from lensed photons that can propagate around the black holes for several times before reaching us \cite{Johannsen:2010ru, 
Gralla:2019xty, Johnson:2019ljv, Gralla:2019drh}, thanks to the optically thin disk. 
These lensed photons enhance the radiation intensity around $\rho \simeq 5\,r_g$ on the sky plane, forming the observed photon ring feature. Meanwhile, these photons contribute less than 10\% to the total intensity \cite{Johnson:2019ljv}. 
The RIAF is usually a good description for a low-luminosity active galactic nuclei (LLAGN), such as Sgr A$^\star$ and M87$^\star$ \cite{Narayan:1996wu,Yuan:2014gma}.
As we will see, the birefringence signals can be influenced by both the geometric thickness of the accretion flow and the lensed photons. 

Besides the axion-photon interaction, 
{the polarization state of the photon is significantly influenced by the medium}, such as by the Faraday conversion and rotation.  Thus one should use the differential radiative transfer Eq.\,(\ref{finalrte}) to properly describe the axion induced birefringence.

\begin{figure}[htb]
    \centering
    \includegraphics[width=0.7\columnwidth]{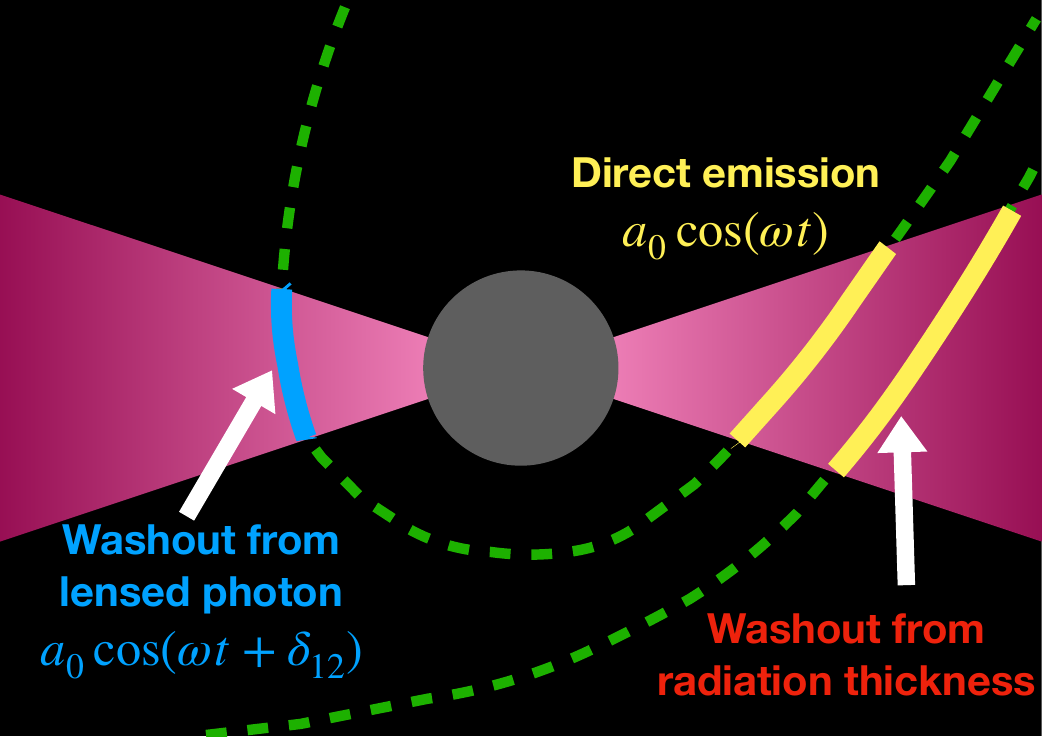}
    \caption{\footnotesize{Here we show the schematic diagram to demonstrate how the emissions are distributed along the line of sight. The green dashed lines are the geodesics along the line of sight. The yellow segments represent the contributions from the thickness of the accretion disk, and the blue ones lead to the lensed photon contribution. 
    \vspace{3mm}}}
    \label{SBL}
\end{figure}

In this section, we study in detail how the amplitude of the axion-induced EVPA oscillation can be influenced in various RIAFs. When the accretion flow is optically thin, photons that reach the Earth at the same time are emitted at different spatial points on the accretion flow and they experience different propagation time. Consequently, if there is an axion cloud, the axion oscillation causes different contributions to the EVPA variation. Adding these contributions together generically leads to a suppression factor to the amplitude of the EVPA oscillation in Eq.\,(\ref{ansatz}). 
One should expect a significant washout effect when a decent portion of photons are emitted from a large spatial region along the line of sight, especially when the size of such a region is comparable to the Compton wavelength of the axion, $2\pi \lambda_c$.
This indicates that such a washout effect becomes less important for lighter axion, due to a longer Compton wavelength.

In the following, we discuss two simple cases where one can study the washout effects quantitatively. One is a constant emission from a continuous and finite length. This mimics the photon emission from the finite thickness of the accretion flow. The other is the emission from two largely separate points. This is a good representation to describe the contribution from lensed photons. The washout effects in various accretion flows should be approximately described by a mixture of these two extreme scenarios. For illustration, we provide a schematic diagram in Fig.\,\ref{SBL} to demonstrate the two possible origins of the washout effects.

\subsection{Washout From Finite Radiation Length}

\begin{figure}[htb]
\centering
\includegraphics[width=0.7\textwidth]{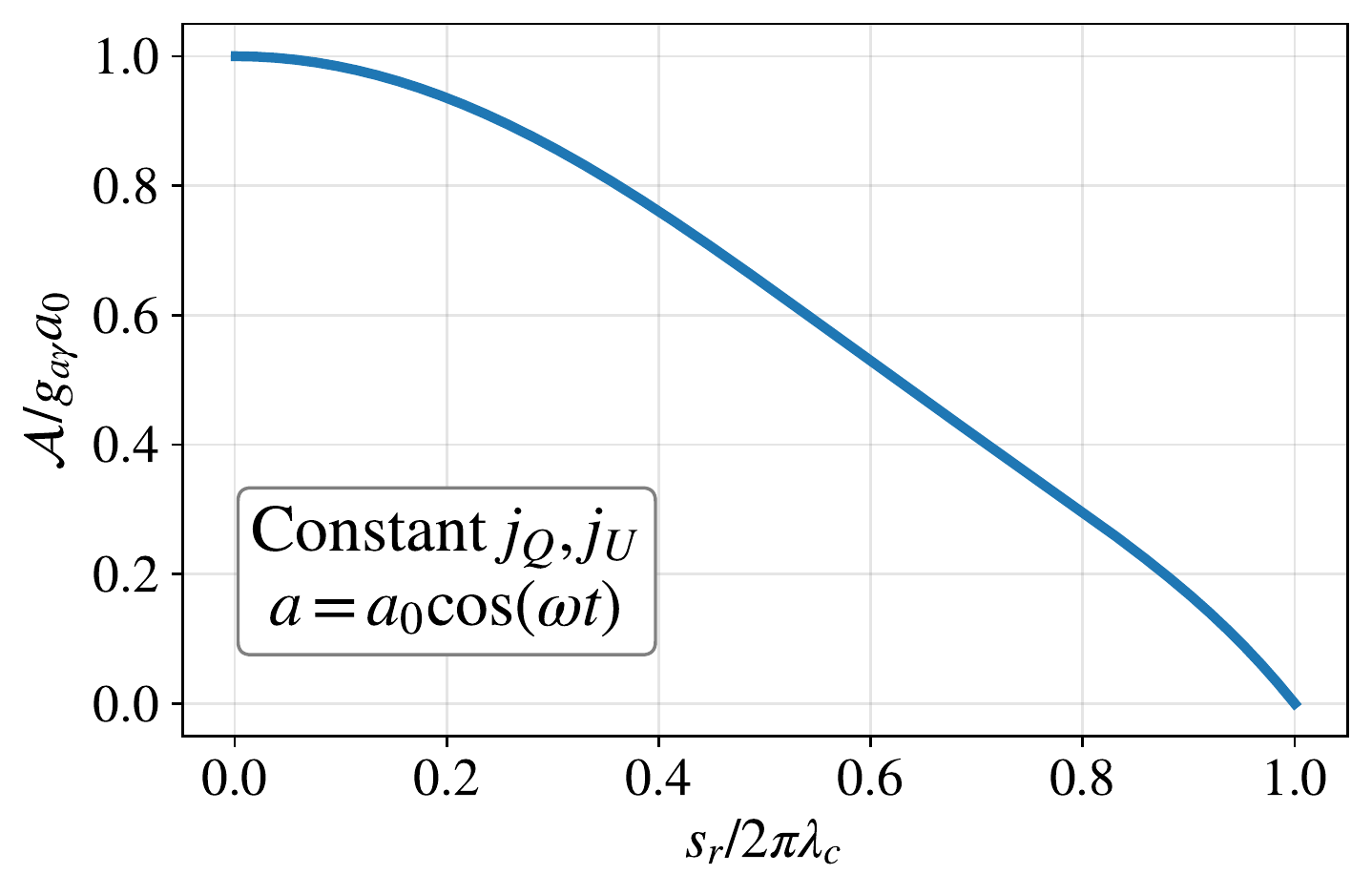}
\caption{\footnotesize{To demonstrate the washout effect from a finite radiation length, here we show the axion induced EVPA oscillation amplitude, i.e., $\mathcal{A}$, as a function of $s_r$. For simplicity, we assume $j_Q$ and $j_U$ are constant long the finite line of sight. The radiation length is normalized by the axion Compton wavelength $2\pi \lambda_c$.}}
\label{washout}
\end{figure}

For simplicity, we focus on the simple radiative transfer equation with only linearly polarized emissions and axion-induced birefringent terms
\be \frac{d\Big(Q + i\ U\Big)}{ds} = j_Q + i\ j_U - i 2 g_{a\gamma\gamma} \frac{d a}{d s} \Big(Q + i\ U\Big).\label{RTEQU}\ee
Here we neglect other contributions from the plasma in the radiative transfer equations since they are not relevant for the washout effects we consider here. The solution to Eq.\,(\ref{RTEQU}) is
\be Q (s_f) + i\ U (s_f) = \int_{s_i}^{s_f} e^{i 2 g_{a\gamma\gamma} \Big(a(s_f) - a(s)\Big)} \Big( j_Q (s) + i\ j_U (s) \Big) ds,\label{QUIaxion}\ee
where $s_i$ and $s_f$ are used to label the initial and final points along the line of sight respectively. We consider a simplified case in which the  linearly polarized emissions, $j_Q / j_U$, are constant in a finite length, $s_r$, along the line of sight. Without loss of generality, we take $j_U$ to be $0$ and the linearly polarized emissions in this case can be written as
\be j_Q^{\textrm{const}} (s) = j_Q^0 \ \Theta \left( |s| - \frac{s_r}{2} \right),\label{jQFL}\ee
where $\Theta$ is the heaviside function and $s = 0$ corresponds to the middle of the emission segment.
The axion field is taken to be a coherently oscillating background whose amplitude $a_0$ stays constant in the same region of emission but approaches to zero at the observer's location.  In this case, the washout effect on the amplitude $\mathcal{A}$ of Eq.\,(\ref{ansatz}) can be solved explicitly. We show the result in Fig.\,\ref{washout},
as a function of the radiation length $s_r$, normalized to the axion Compton wavelength $2\pi \lambda_c$. As expected, the amplitude approaches to $0$ when $s_r$ becomes comparable with $\lambda_c$.

In the analytic RIAF model \cite{Pu:2018ute}, there is a dimensionless parameter $H \equiv h/R$, defined as the ratio between the height $h$ and the horizontal scale $R$ of the accretion flow. This parameter is used to characterize the geometric thickness of the accretion flow. The linearly polarized radiation is proportional to the electron number density, which is exponentially suppressed respect to the distance from the equatorial plane. For a nearly face-on disk, neglecting the background metric when the emissions are far away from the horizon, the emission length in Eq.\,(\ref{jQFL}) can be approximated as
\be s_r \simeq \rho H.\ee
It is reasonable to expect that the washout effect induced by the finite radiation length is not important if the thickness of the accretion flow satisfies
\be \rho H \ll 2\pi \lambda_c.\ee
For example, if we take $\alpha \equiv r_g/\lambda_c = 0.4$ and $\rho \simeq 5\,r_g$, this condition leads $H \ll 3$, which applies to most kinds of accretion flows. For smaller $\alpha$, the finite length washout becomes even more negligible. 

\subsection{Washout From Lensed Photon}
We next consider the case that the linearly polarized radiation received on the sky plane comes from two discrete emission points. The separation in distance of these two points leads to a phase difference due to the axion cloud oscillation as well as the axion cloud spatial profile. We assume the linearly polarized emissions at the two points are independent, and the source of emission can be characterized as
 \be j_Q (s) + i\ j_U (s) = \sum_{p = 1, 2}\ e^{i 2\chi_p} I_L^p \ \delta (s - s_p).\ee
Here $I_L^p$ is the linear polarization intensity at each emission point, and the EVPA of the emission is $\chi_p$. 
Substituting this ansatz into Eq.\,(\ref{QUIaxion}), one gets
 \be Q (s_f) + i\ U (s_f) = \sum_{p = 1, 2} e^{- i 2 g_{a\gamma\gamma} a(s_p) + i 2\chi_p} I_L^p.\ee
We take the axion field value at the first point as $a(s_1) = a_0 \cos \left( \omega t \right)$. We further set the axion field amplitude to be the same at these two emission points for simplicity. The axion field at the second emission point can then be parametrized as  $a(s_2) = a_0 \cos \left( \omega t + \delta_{12} \right)$, with $\delta_{12}$ being the phase delay between these two points.  In this case, the oscillation amplitude of the EVPA can be written as
 \be \frac{\mathcal{A}}{g_{a\gamma\gamma} a_0} = \sqrt{\cos^2 \left( \frac{\delta_{12}}{2} \right) + \sin^2 \left( \frac{\delta_{12}}{2} \right) \left| \frac{I_L^1 - I_L^2 e^{i 2 \Delta \chi_{12}}}{I_L^1 + I_L^2 e^{i 2 \Delta \chi_{12}}} \right|^2},\label{EVPAA2E}\ee
with $\Delta \chi_{12} = \chi_2 - \chi_1$.

For optically thin RIAFs, some photons are nearly in bound states around the SMBH. These photons can propagate around the BH for several times before exiting, and they make a significant contribution to the photon ring observed on the sky plane \cite{Johannsen:2010ru, Gralla:2019xty, Johnson:2019ljv, Gralla:2019drh}. If the emissions happen dominantly around the equatorial plane, one gets a discrete sum of the radiation that differs with each other by the times that propagates around the black hole. Since the emission points of both the direct radiation and lensed photons have comparable radii from the black hole, the axion field values are comparable. Thus Eq.\,(\ref{EVPAA2E}) serves a good approximation for studying the EVPA oscillation amplitude on the photon ring.
The relative phase of the axion oscillation $\delta_{12}$ in Eq.\,(\ref{EVPAA2E}) is
\be \delta_{12} = \omega \Delta t - \Delta \phi . \label{d12}\ee
The time delay $\Delta t$ and the azimuthal angle difference $\Delta \phi$ are the critical parameters to characterize the lensed photons, and these quantities can be properly calculated \cite{Gralla:2019drh}.

\subsection{Landscape of Accretion Flows}
Now let us adopt the analytic RIAF \cite{Pu:2018ute} as a benchmark model.
We vary parameters for several aspects, such as the magnetic field structure, velocity distribution, and thickness $H$, in order to see how the birefringence signals are influenced. Three types of magnetic field geometries, including a vertical field, a toroidal field, and a radial field, are considered \cite{EHTM}.
Notice that the EHT observation for M87$^\star$ favors the vertical one \cite{EHTM}. 
Velocity distributions are characterized by a Keplerian, a sub-Keplerian, and a free-falling flow, respectively \cite{Pu:2016qak}.

\begin{figure*}[ht]
\centering
\includegraphics[width=0.45\textwidth]{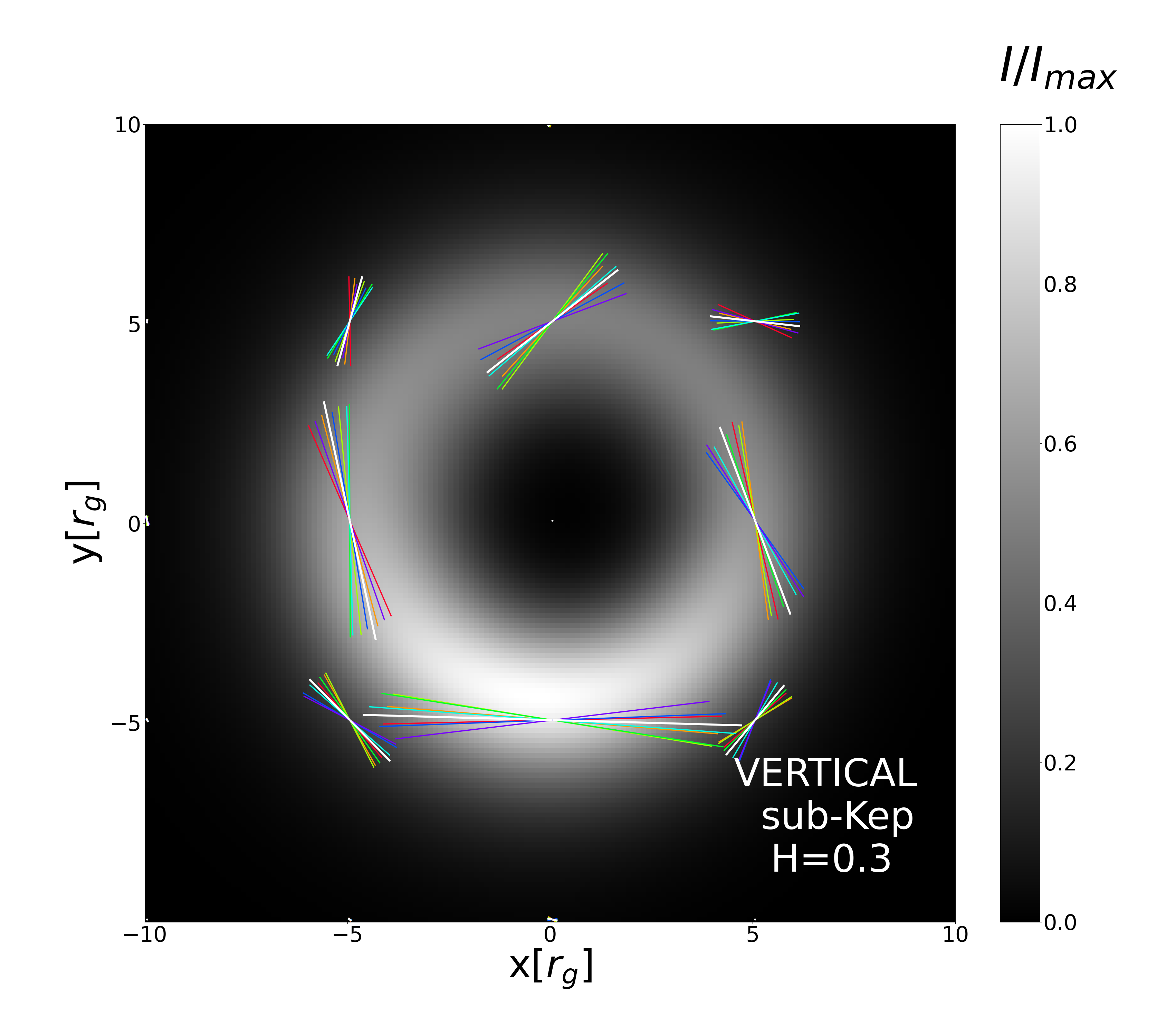}
\includegraphics[width=0.45\textwidth]{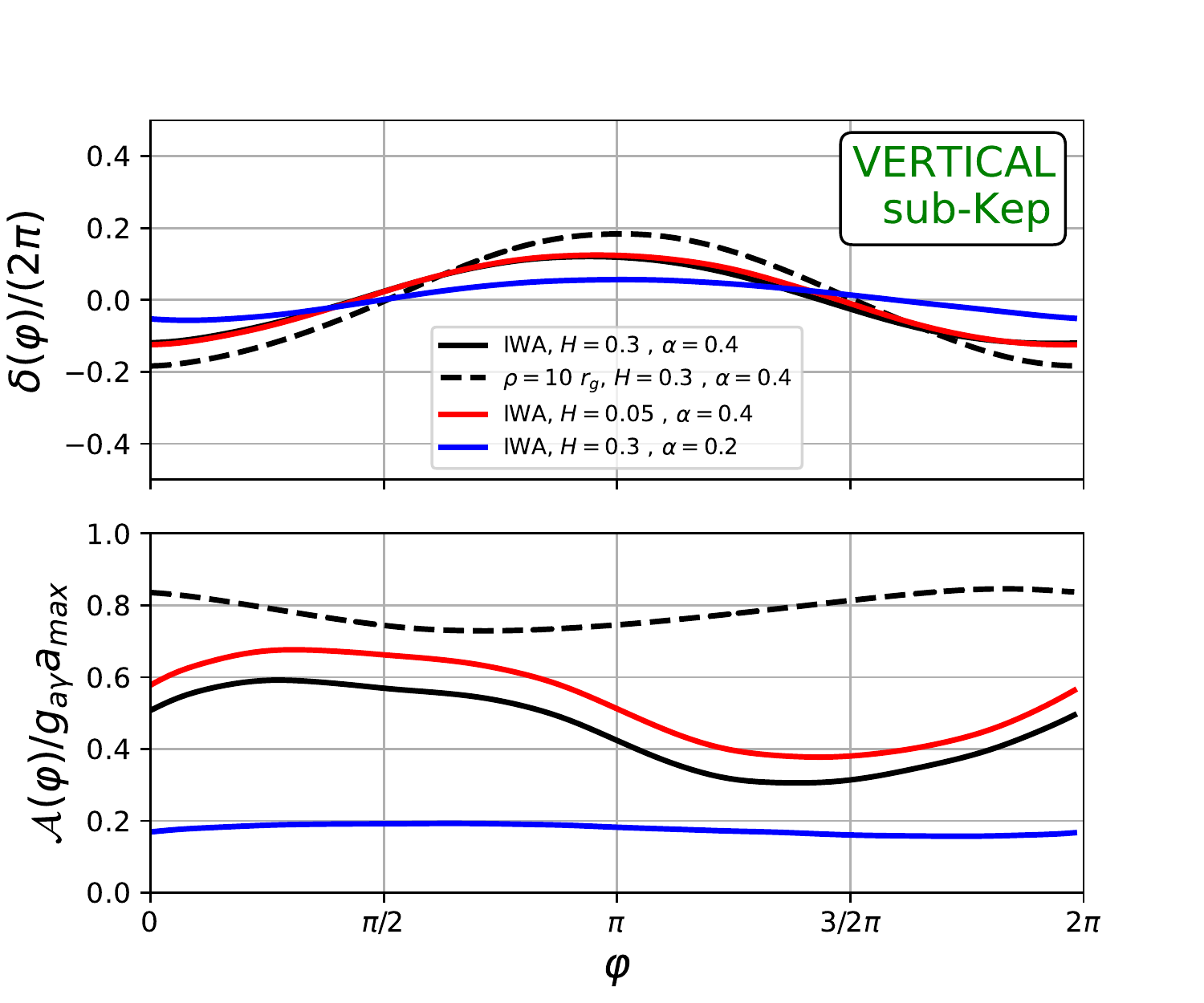}
\includegraphics[width=0.45\textwidth]{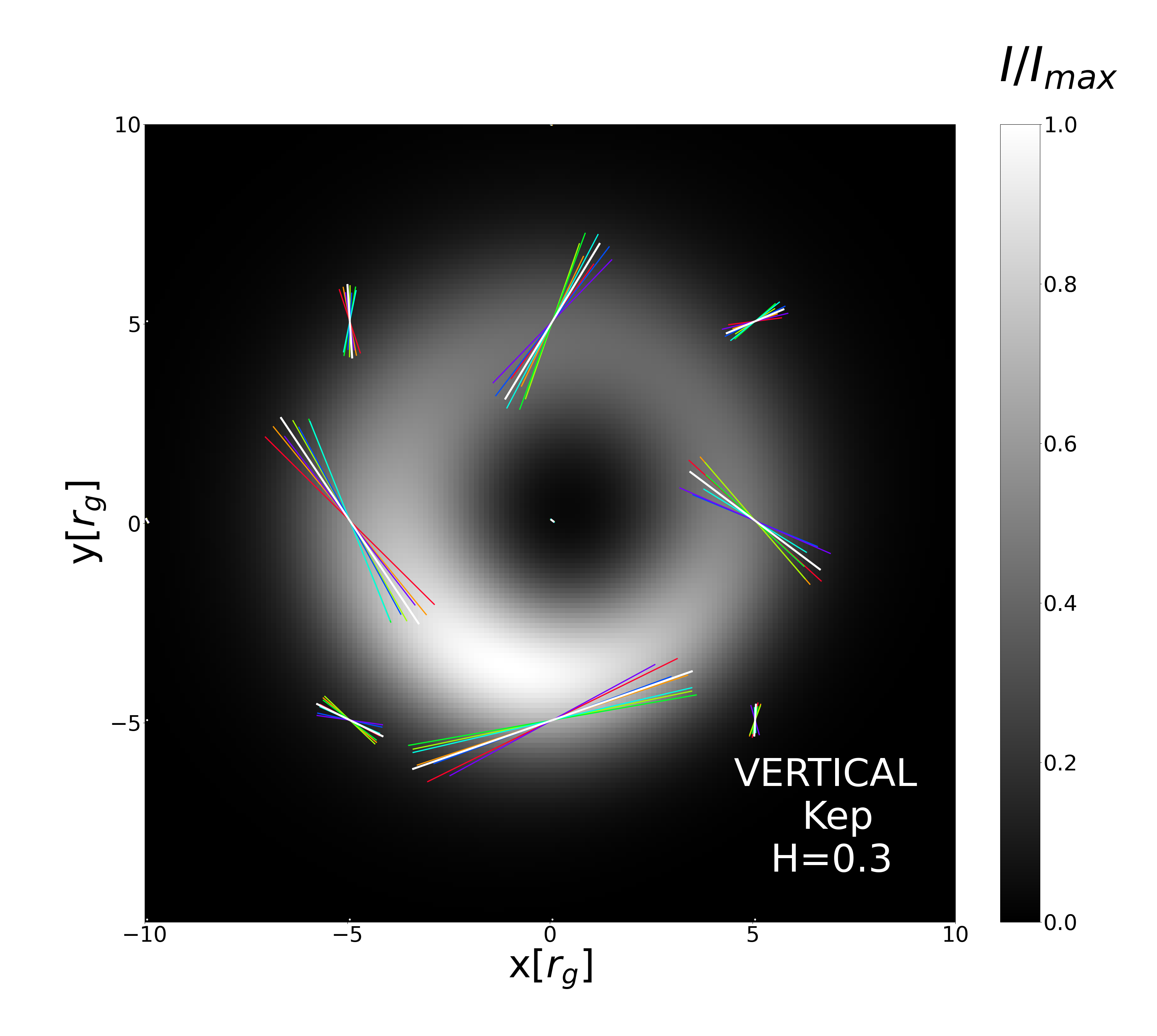}
\includegraphics[width=0.45\textwidth]{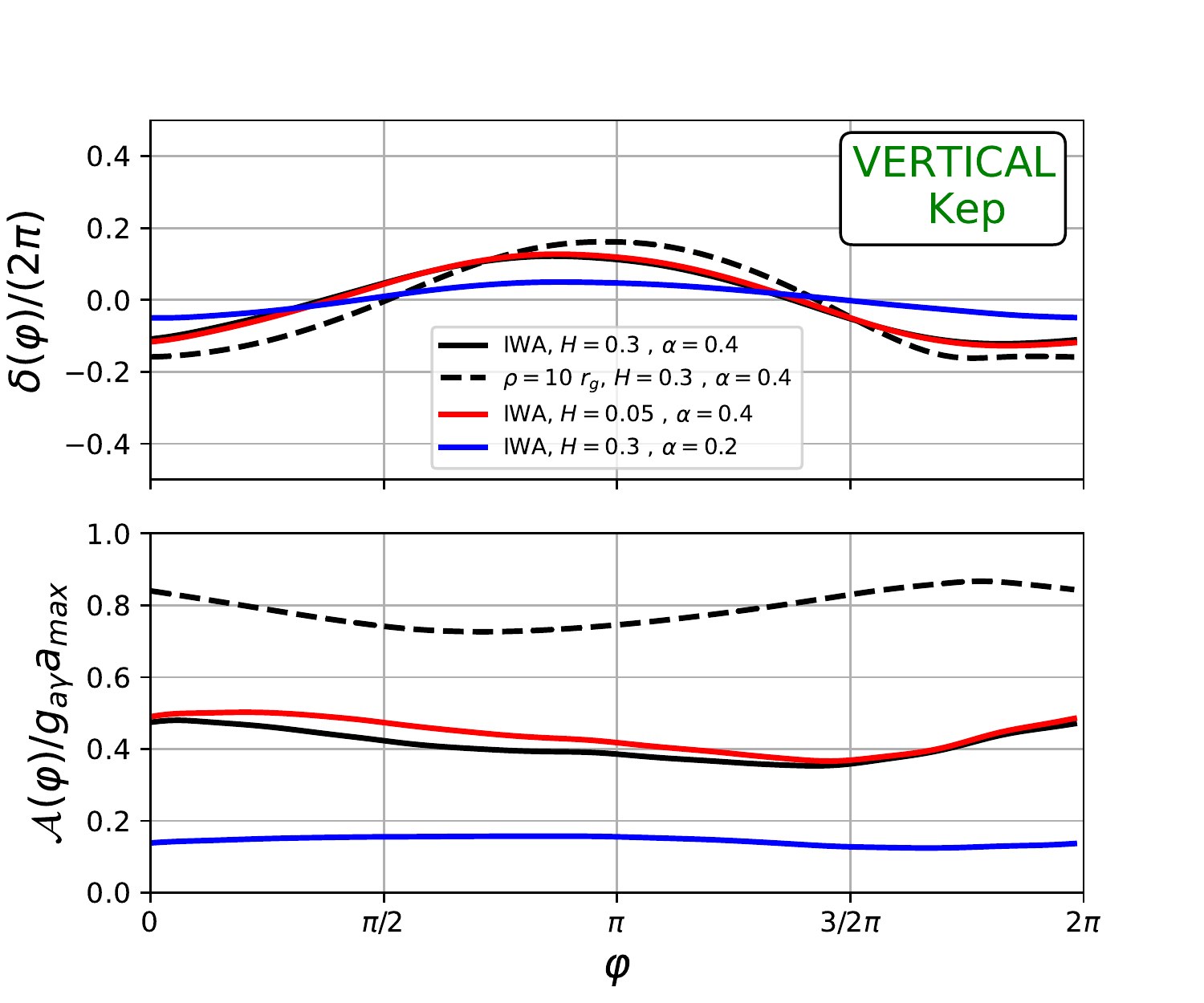}
\caption{\footnotesize{Left: Examples from the \texttt{IPOLE} simulation  \cite{Moscibrodzka:2017lcu, Noble:2007zx} with different types of RIAFs are shown.
The magnetic field structure, velocity distribution, and thickness $H$ are labelled in each panel.
We take $\alpha = 0.4$, $g_{a\gamma\gamma} a_{\textrm{max}} = 1$ rad  and $a_J = 0.99$ for the axion and black hole parameters. Inclination angle ${i}$ is set to $163^\circ$ motivated by M87$^\star$, and the spin points to $- x$ on the sky plane.
At each point, the white quiver represents the EVPA without the axion correction.
The rainbow color, from red to purple, shows the time variations of EVPA in the presence of the axion cloud. 
Right: We demonstrate the EVPA oscillation in terms of the relative phase, $\delta (\varphi)/2\pi$, and amplitude, $\mathcal{A} (\varphi) / g_{a\gamma\gamma} a_{\textrm{max}}$, as functions of the azimuthal angle on the sky plane. Intensity weighted average (IWA) EVPAs and EVPAs at $\rho = 10\,r_g$ are shown for comparison. Results with various choices of $H$ and $\alpha$ are also presented. We note that the direction of $\varphi = 0$ corresponds to the $+ x$ direction on the left panel.}}
\label{figRIAF0.3}
\end{figure*}

In Fig.\,\ref{figRIAF0.3}, we show several examples of RIAFs. The impacts induced by the inclination angle and the black hole spin are qualitatively the same as those of the NT thin disk model. For illustration, we fix these quantities as $163^\circ$ and $a_J = 0.99$, which are motivated by M87$^\star$. We consider the EVPA distribution as a function of the azimuthal angle with a given radius $\rho$ on the sky plane. In addition, motivated by the study in  \cite{EHTP}, we also calculate the intensity weighted average (IWA) EVPAs, 
\be \langle \chi (\varphi) \rangle \equiv \frac{1}{2} \textrm{arg}\Big{(} \langle Q \times I \rangle + i \langle U \times I \ \rangle \Big{)}.\label{defIEVPA}\ee
Here the IWA region covers the dominant emission on the sky plane \cite{Akiyama:2019bqs}.

\clearpage

\begin{figure*}[htb]
\begin{center}
\includegraphics[width=0.45\textwidth]{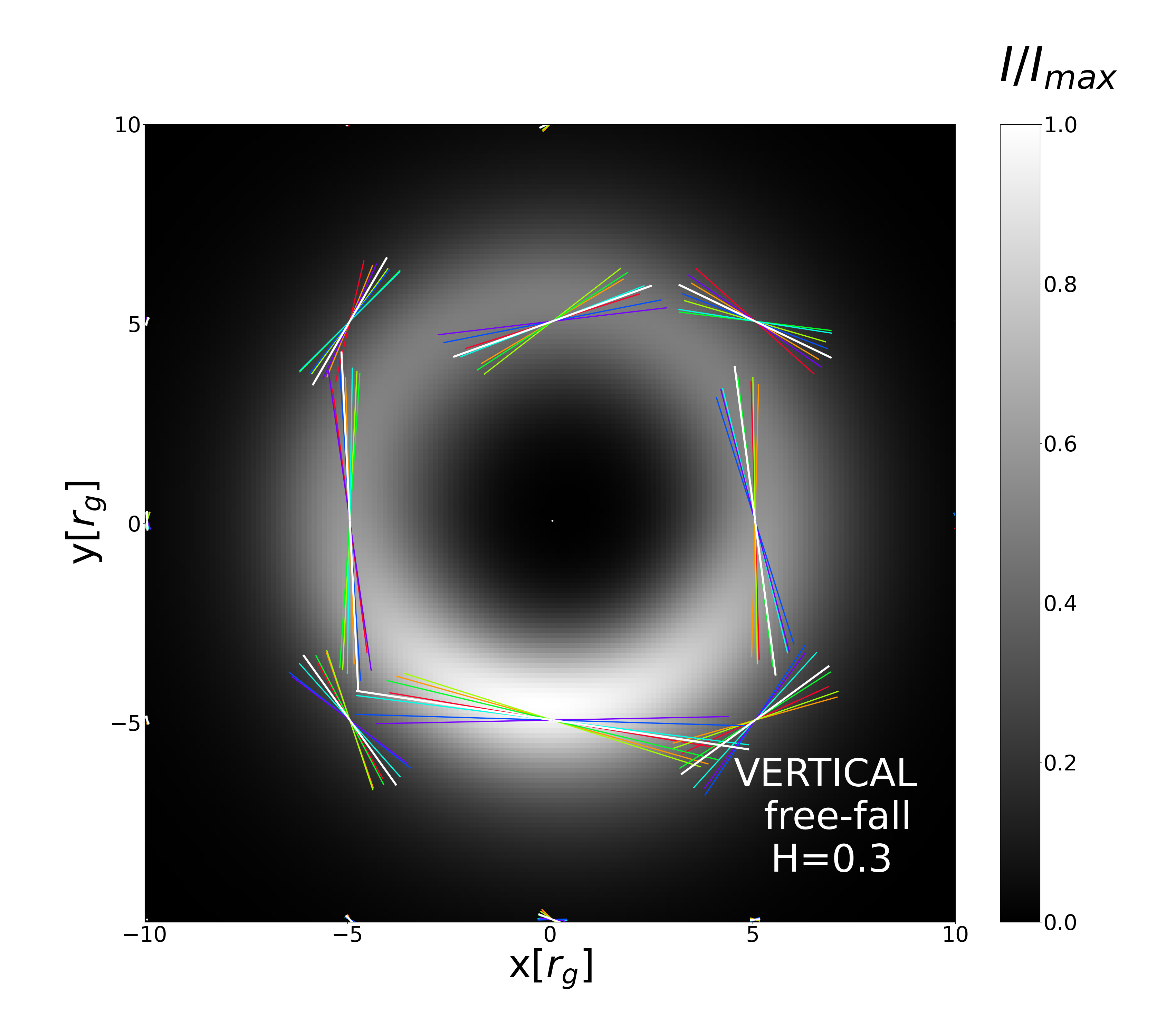}
\includegraphics[width=0.45\textwidth]{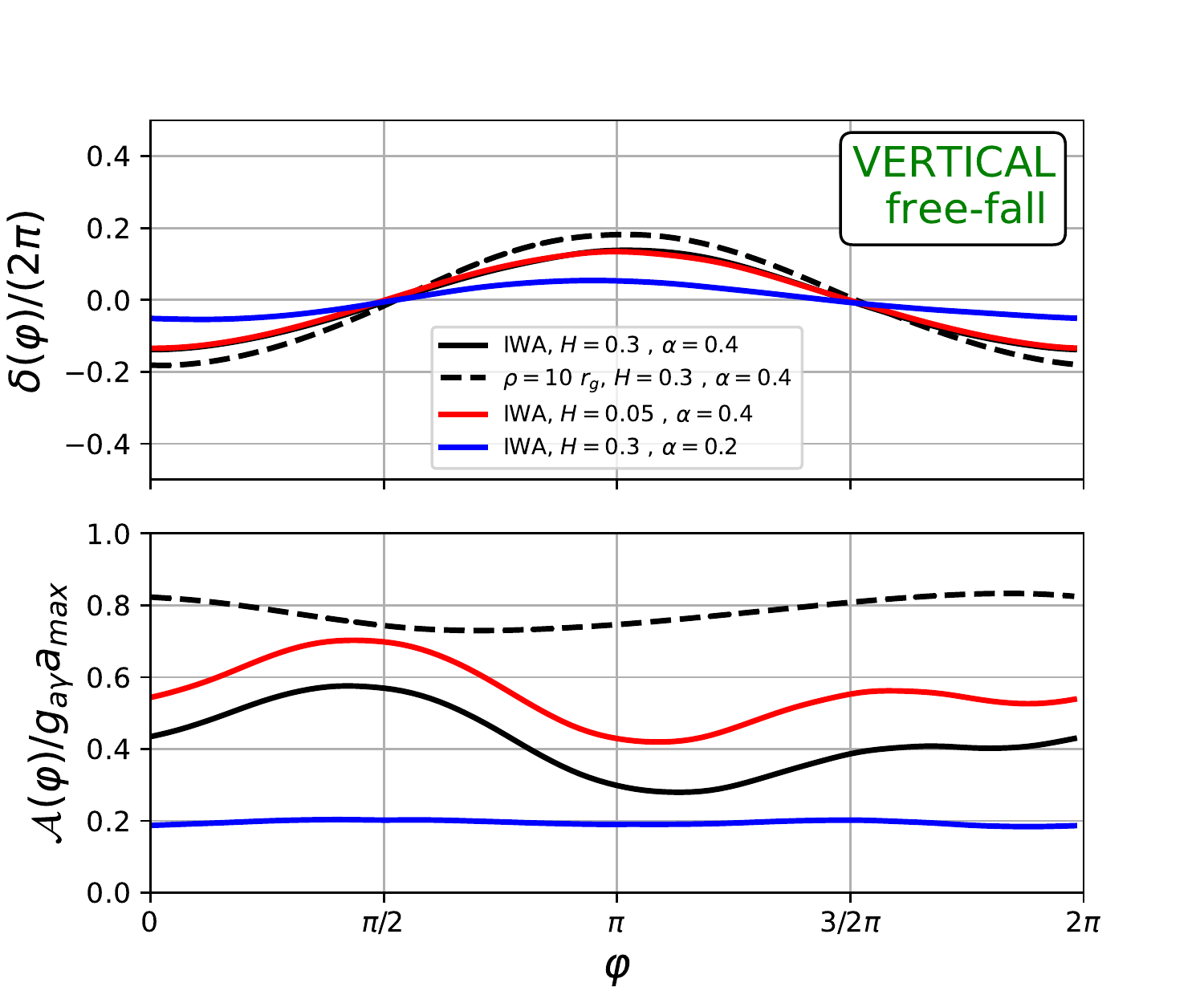}
\includegraphics[width=0.45\textwidth]{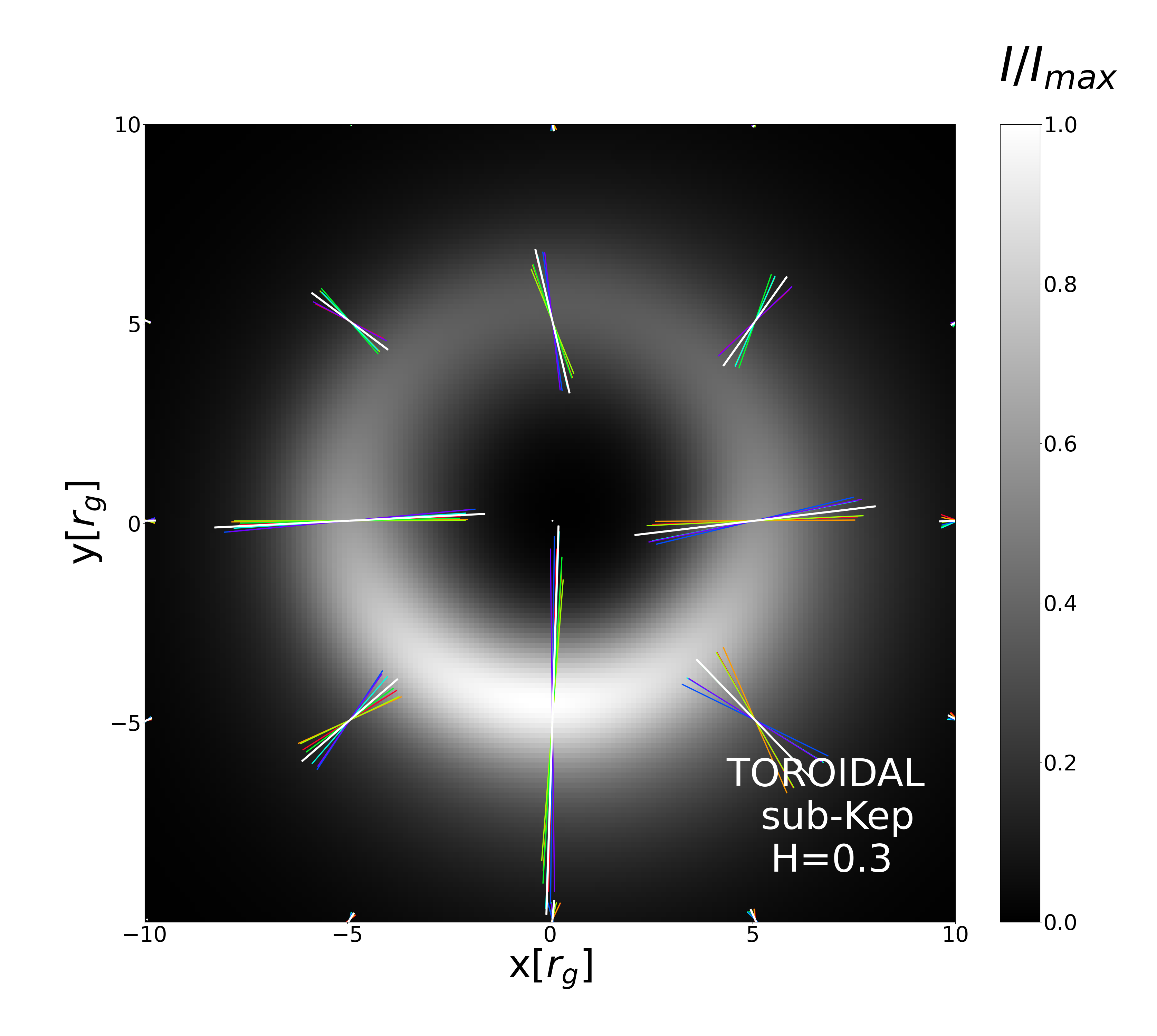}
\includegraphics[width=0.45\textwidth]{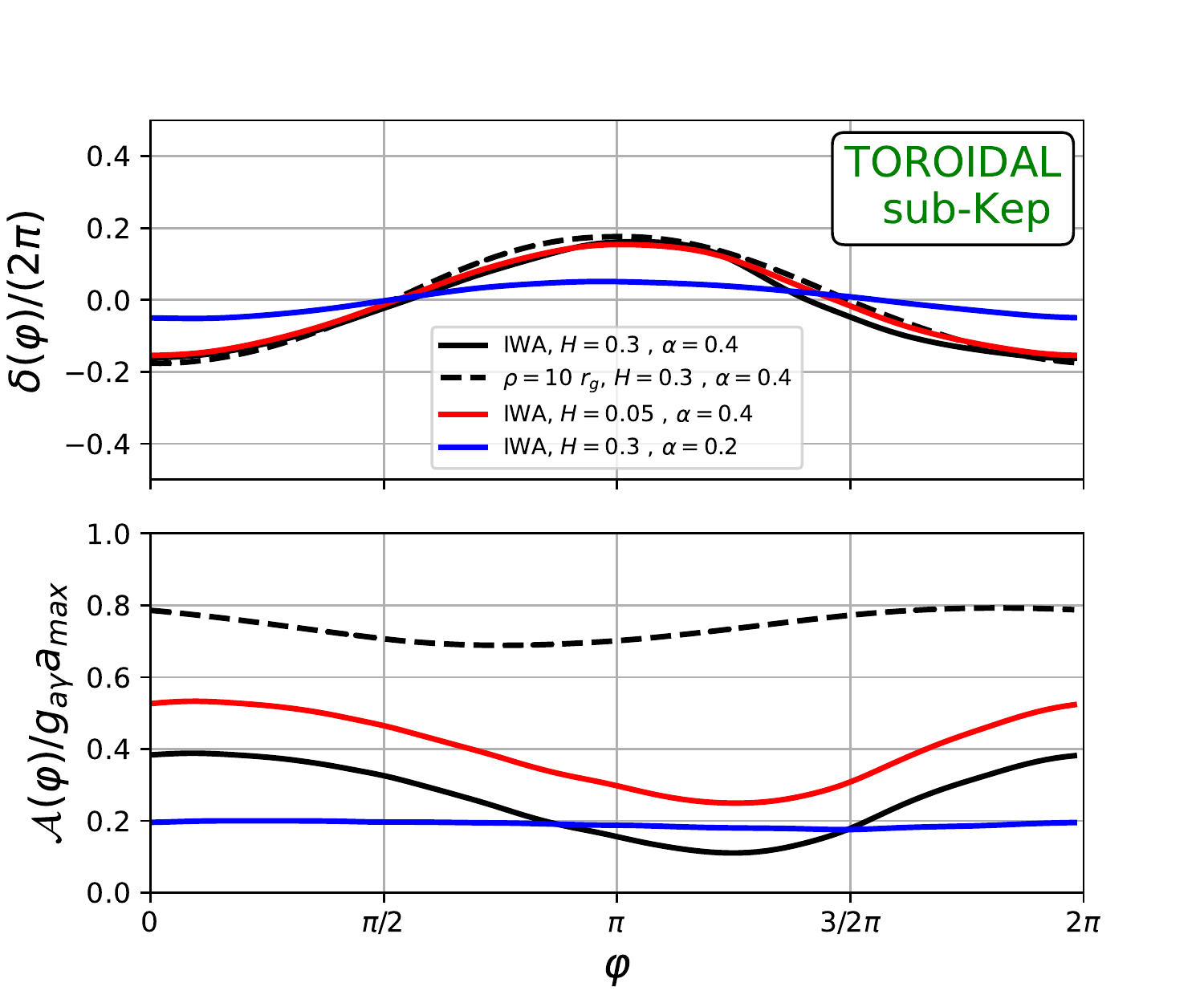}
\includegraphics[width=0.45\textwidth]{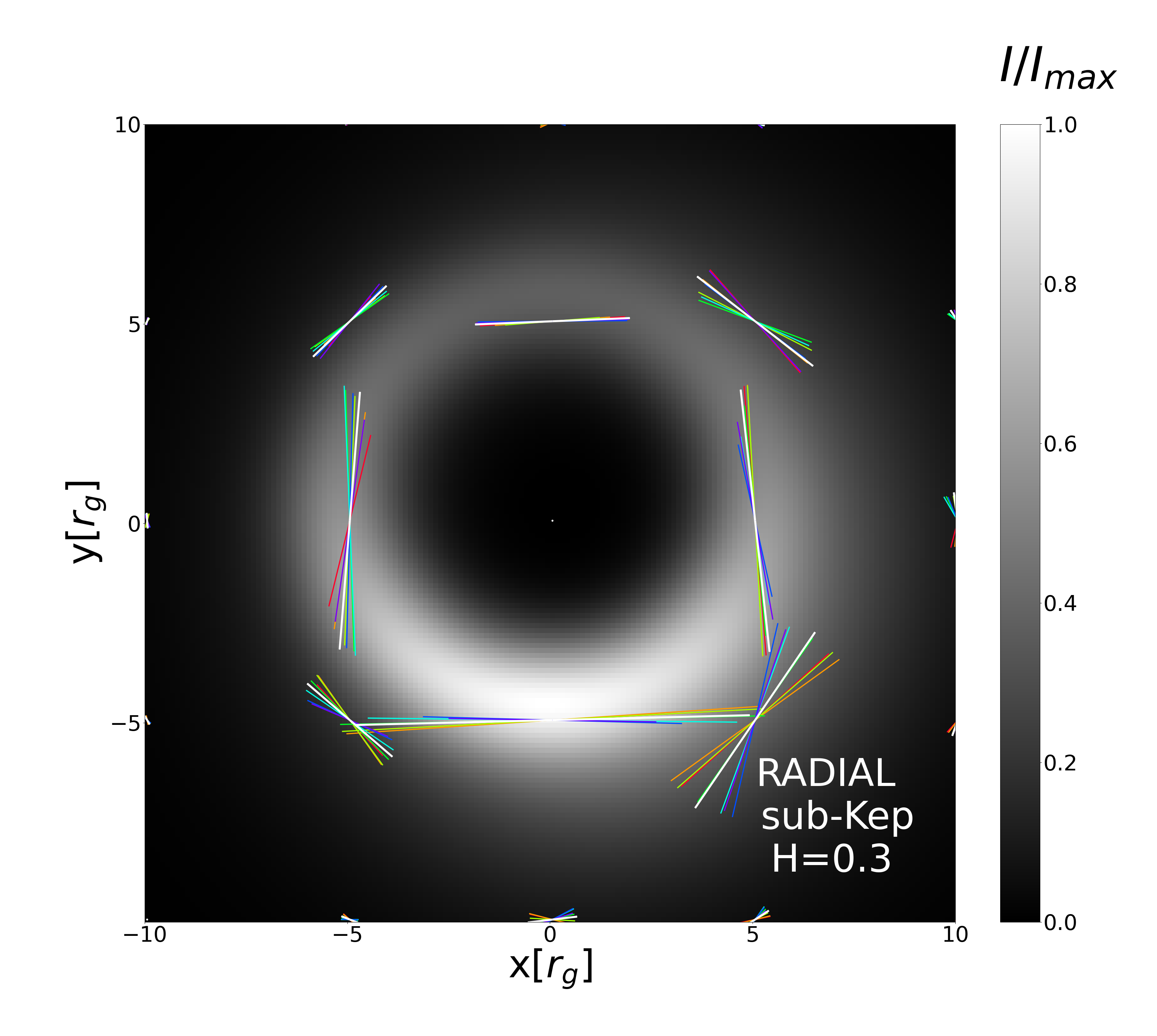}
\includegraphics[width=0.45\textwidth]{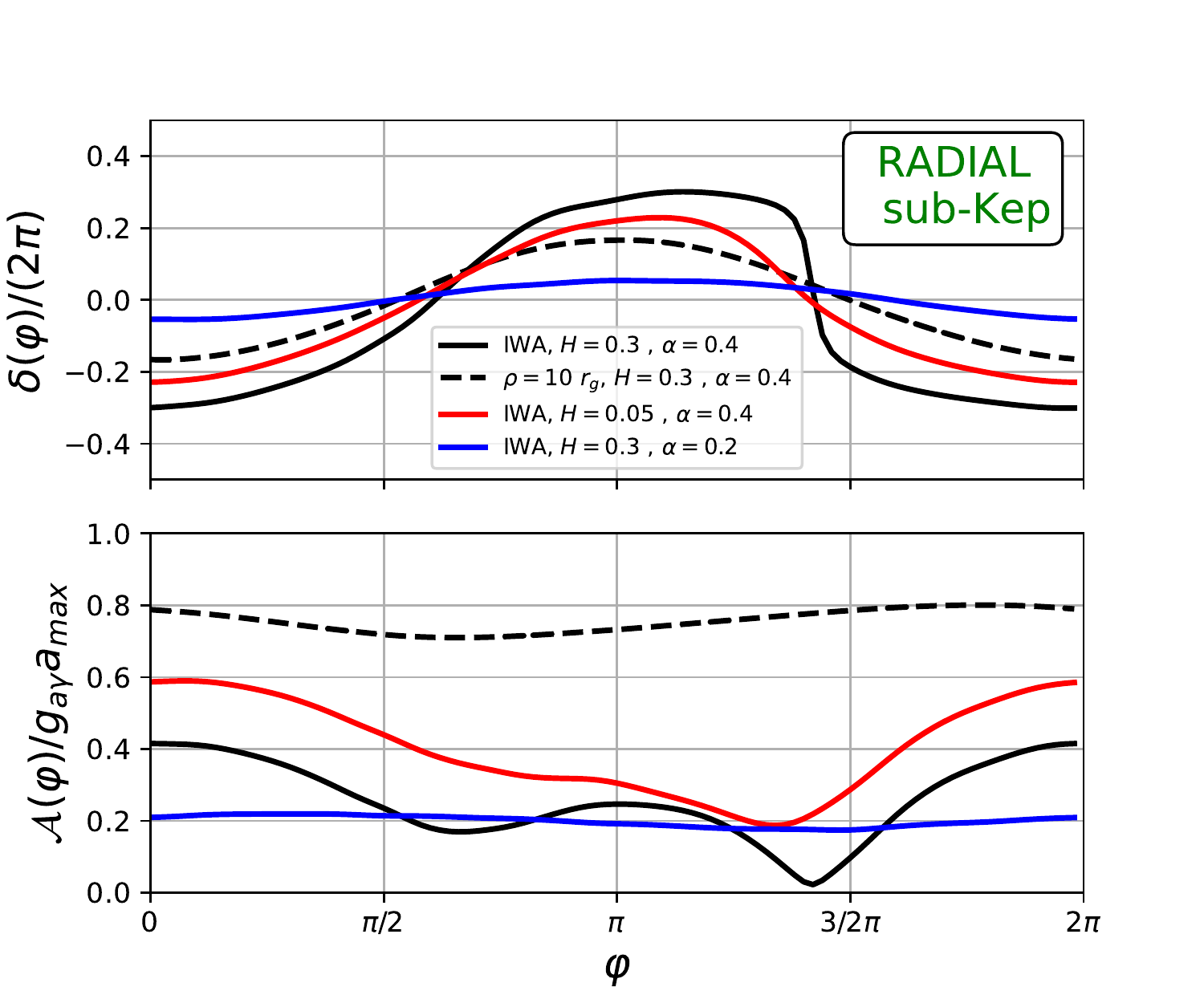}
\end{center}
\textbf{Figure 8}: \footnotesize{(Cont.)\vspace{3mm}}
\end{figure*}

There are several features to be explained. First, let us study the impact of the thickness of the accretion flow.  As discussed in \cite{EHTP}, a magnetically arrested disk (MAD) \cite{EHTM} has a strong magnetic field which compresses the thickness parameter of the RIAF, $H$, to $0.05$ in the inner region, and extends it to about $0.3$ in the outer region  \cite{Igumenshchev:2003rt, Narayan:2003by, McKinney:2012vh, Tchekhovskoy2015}. We take $H = 0.3$ and $0.05$ for comparison as benchmarks in this study.
As demonstrated in Fig.\,\ref{figRIAF0.3}, the oscillation amplitude, $\mathcal{A}$, for $H = 0.3$ is typically smaller than that for $H = 0.05$ by a simple scaling factor. This is consistent with the washout effect induced by the finite thickness of the accretion flow, discussed previously.

\begin{figure*}[ht] 
\centering 
\includegraphics[width=0.45\textwidth]{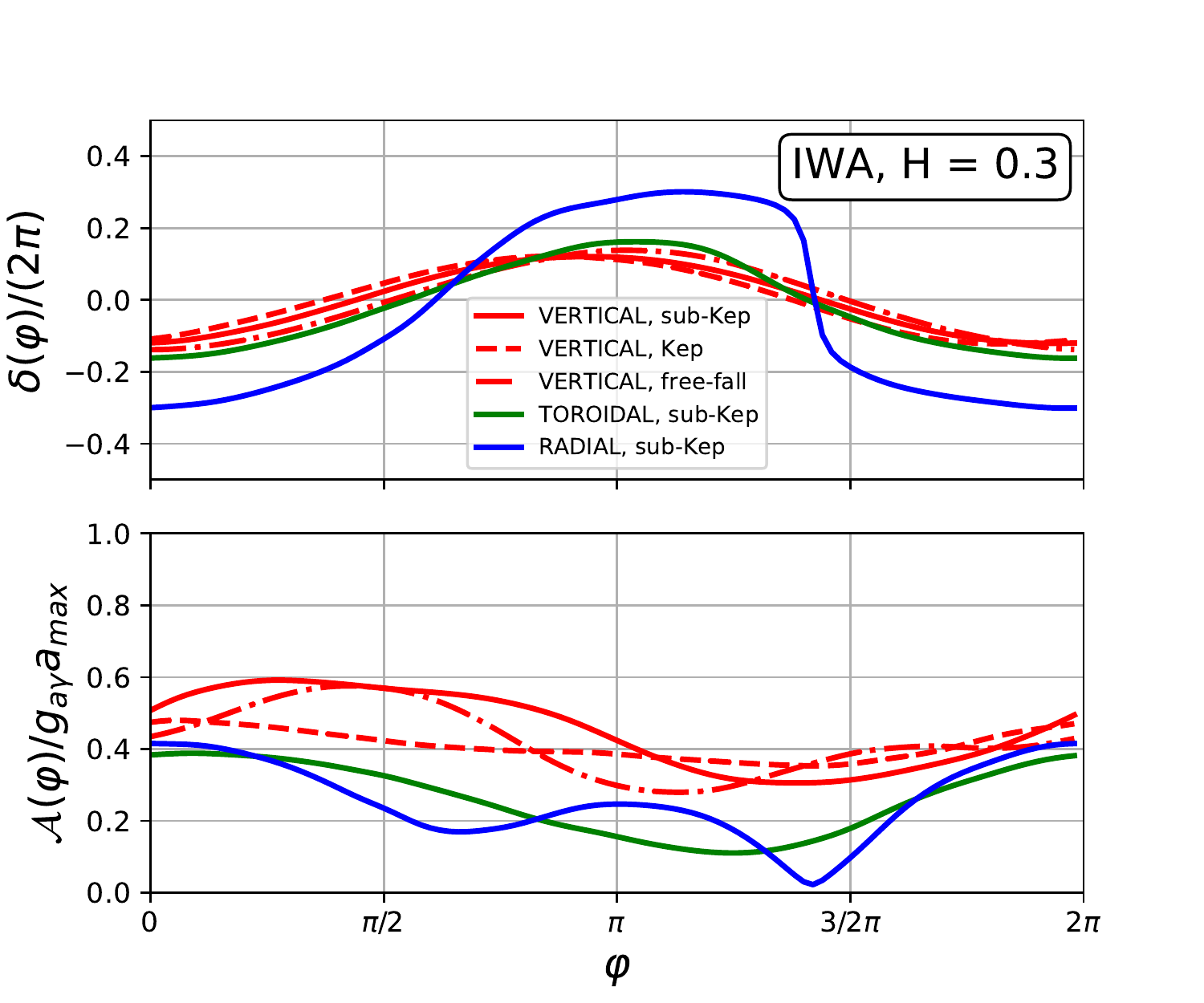}
\includegraphics[width=0.45\textwidth]{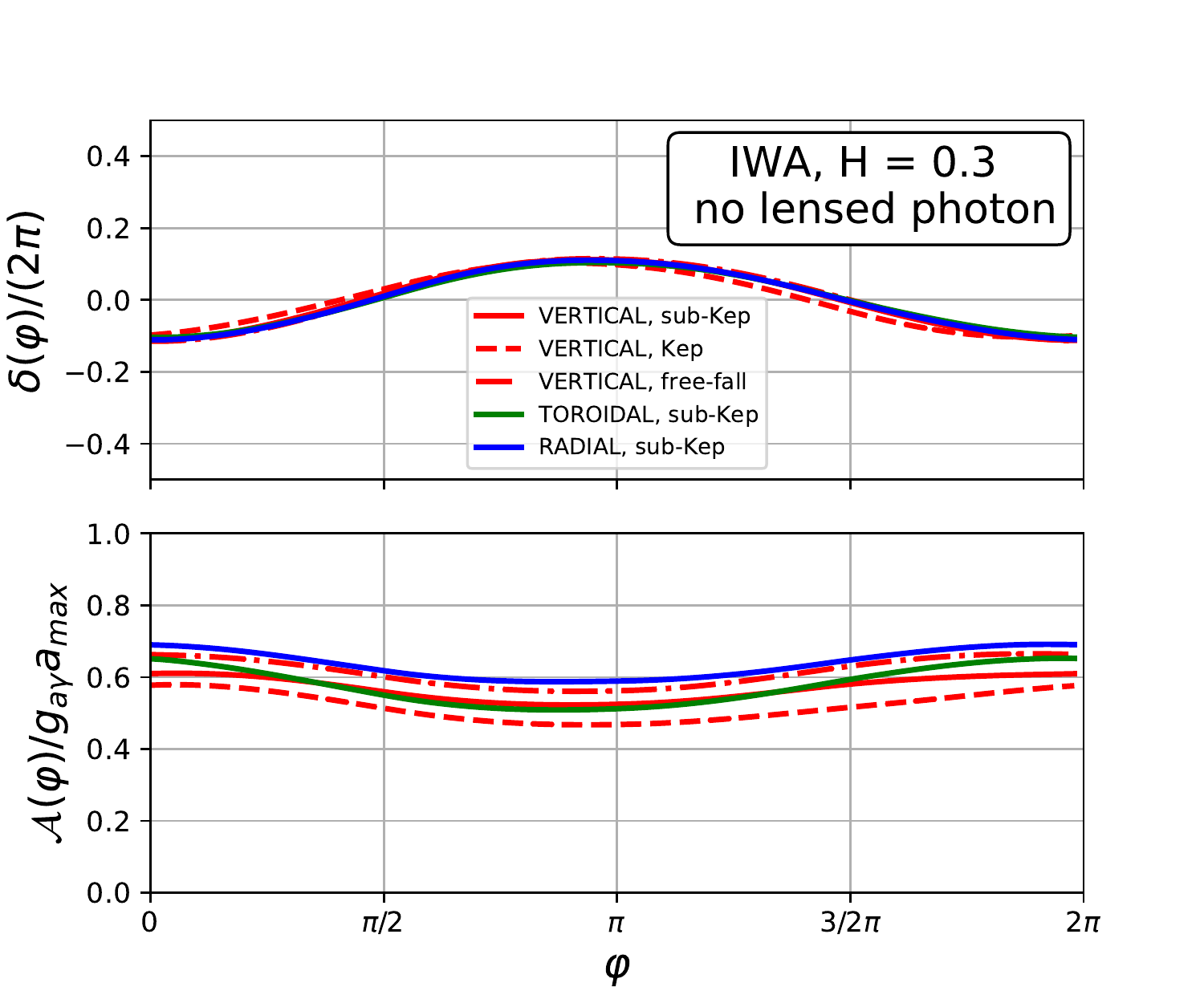}
\caption{\footnotesize{Here we show the relative phase, $\delta (\varphi)/2\pi$, and amplitude, $\mathcal{A} (\varphi) / g_{a\gamma\gamma} a_{\textrm{max}}$ of the EVPA oscillation as functions of the azimuthal angle, with (left) and without (right) lensed photons. In the latter case, the IWA EVPAs present universal features, in terms of various choices of the magnetic fields and the velocity distributions. }}
\label{AEVPA5}
\end{figure*}

Furthermore, lensed photons also contribute significantly to the washout effects. Such a washout effect, led by lensed photons, can be reduced if one focuses on EVPAs away from the neighbourhood of the black hole, e.g., $\rho \gg 5\,r_g$ for M87$^\star$. Particularly, the EVPA variations at $\rho = 10\,r_g$ are shown as the black dashed lines in the right panel of Fig.\,\ref{figRIAF0.3}. On the other hand, when we consider IWA EVPAs, lensed photons may lead to a substantial impact. In order to disentangle the washout effect from the finite thickness and that from the lensed photons, we artificially remove the lensed photon and recalculate the EVPA variations. This is done by a simple manipulation in \texttt{IPOLE} \cite{Moscibrodzka:2017lcu, Noble:2007zx}. We show the new results in Fig.\,\ref{AEVPA5}.  
As we can see, after artificially removing lensed photons, the variations of the IWA EVPA show universal structures in both the relative phase $\delta$ and the amplitude $\mathcal{A}$ for various choices of the accretion flow parameters. This indicates that more astrophysical model independent analyses can be carried out if the future VLBI measurements provide detailed information about the EVPA variations in the regions away from the SMBH, where lensed photons are not important.

It is also worth to mention that when we increase the axion Compton wavelength, i.e., decrease $\alpha$ from 0.4 to 0.2, both washout effects become less important. In this case, and the amplitudes are mainly influenced by the radial wave-function of the axion cloud.

\begin{figure}[ht] 
\centering
  \begin{subfigure}[b]{0.45\columnwidth}
    \includegraphics[width=\linewidth]{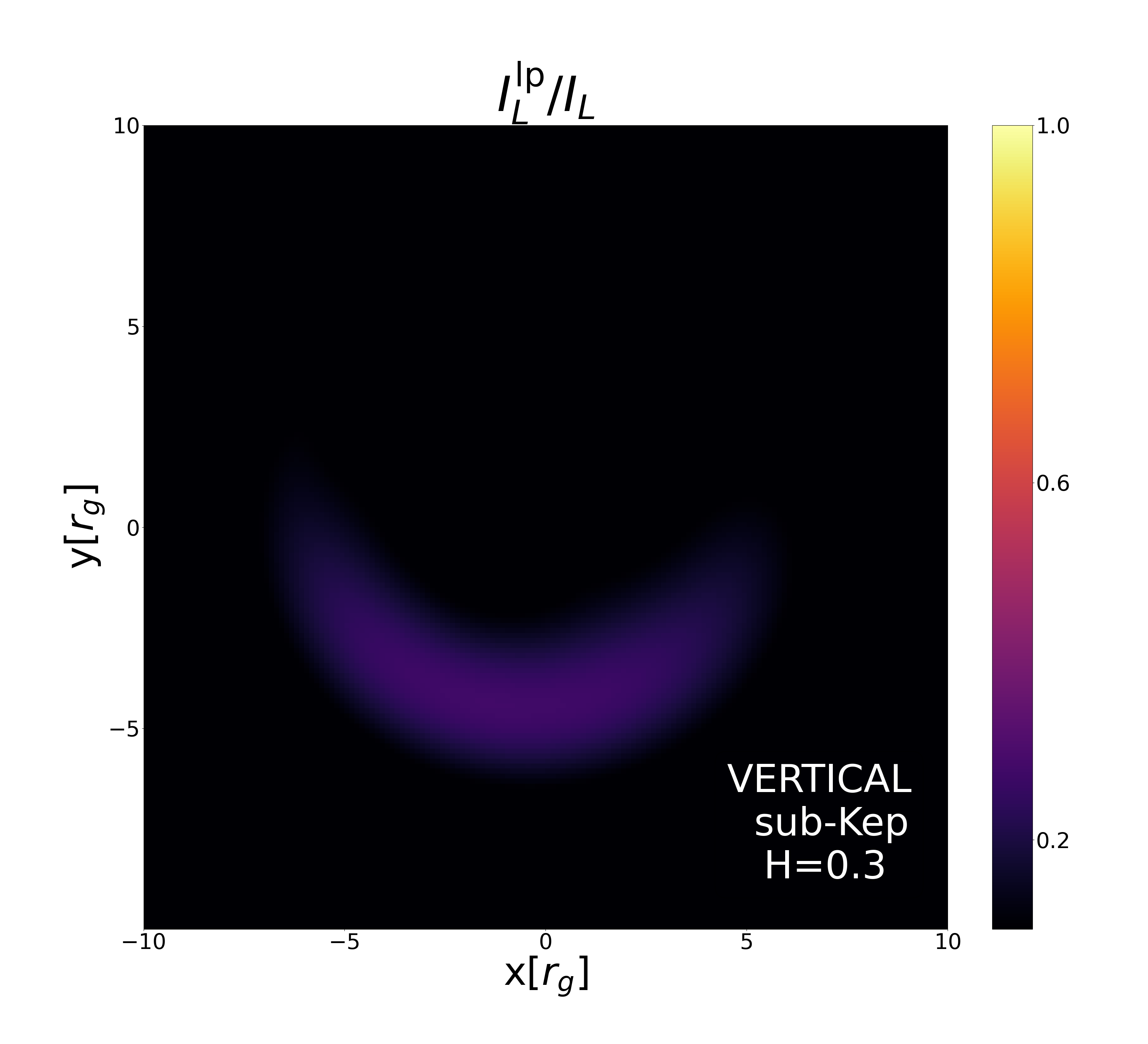}
  \end{subfigure} 
  \begin{subfigure}[b]{0.45\columnwidth}
    \includegraphics[width=\linewidth]{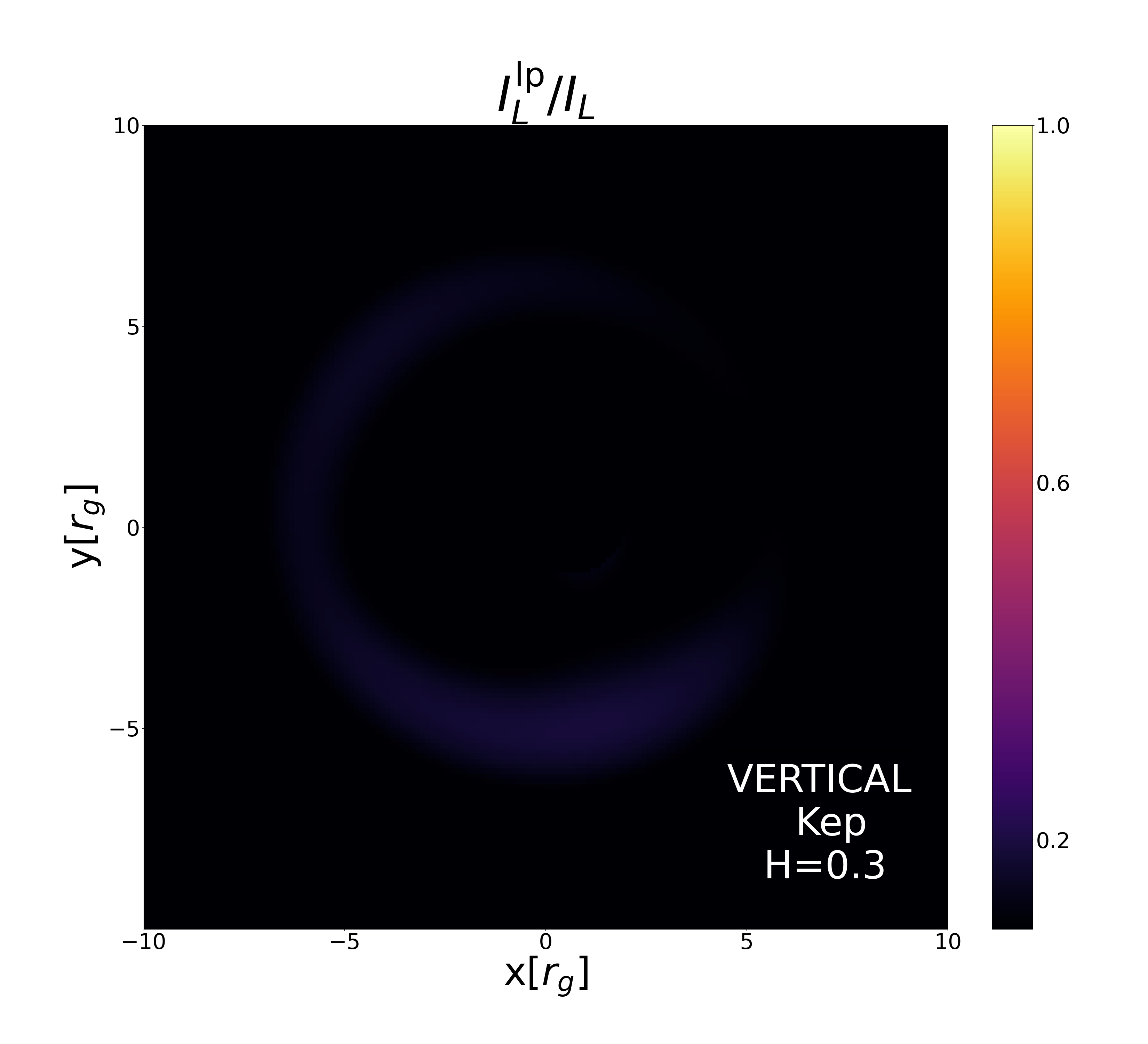}
     \end{subfigure} 
  \begin{subfigure}[b]{0.45\columnwidth}
    \includegraphics[width=\linewidth]{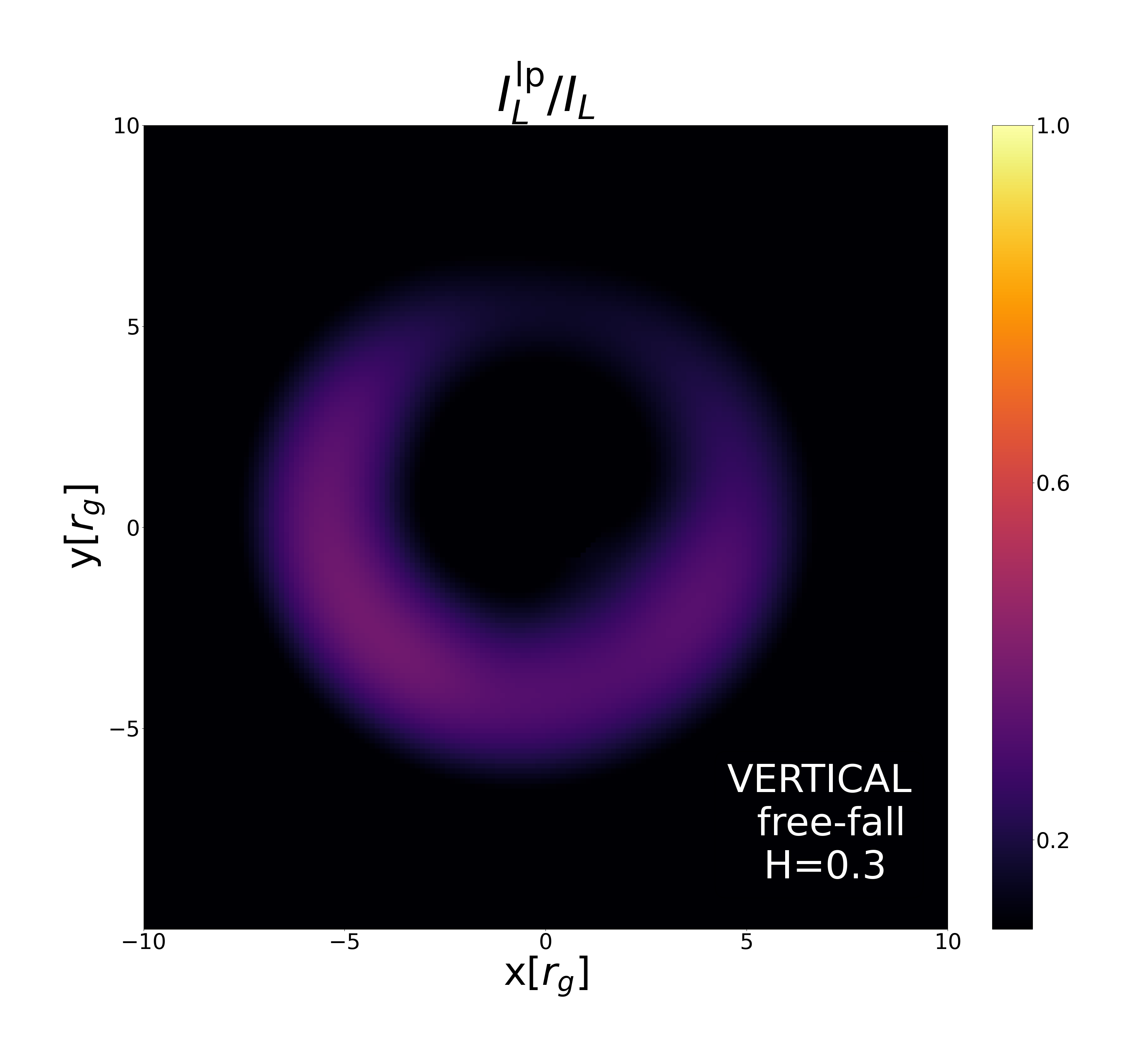}
     \end{subfigure} 
  \begin{subfigure}[b]{0.45\columnwidth}
    \includegraphics[width=\linewidth]{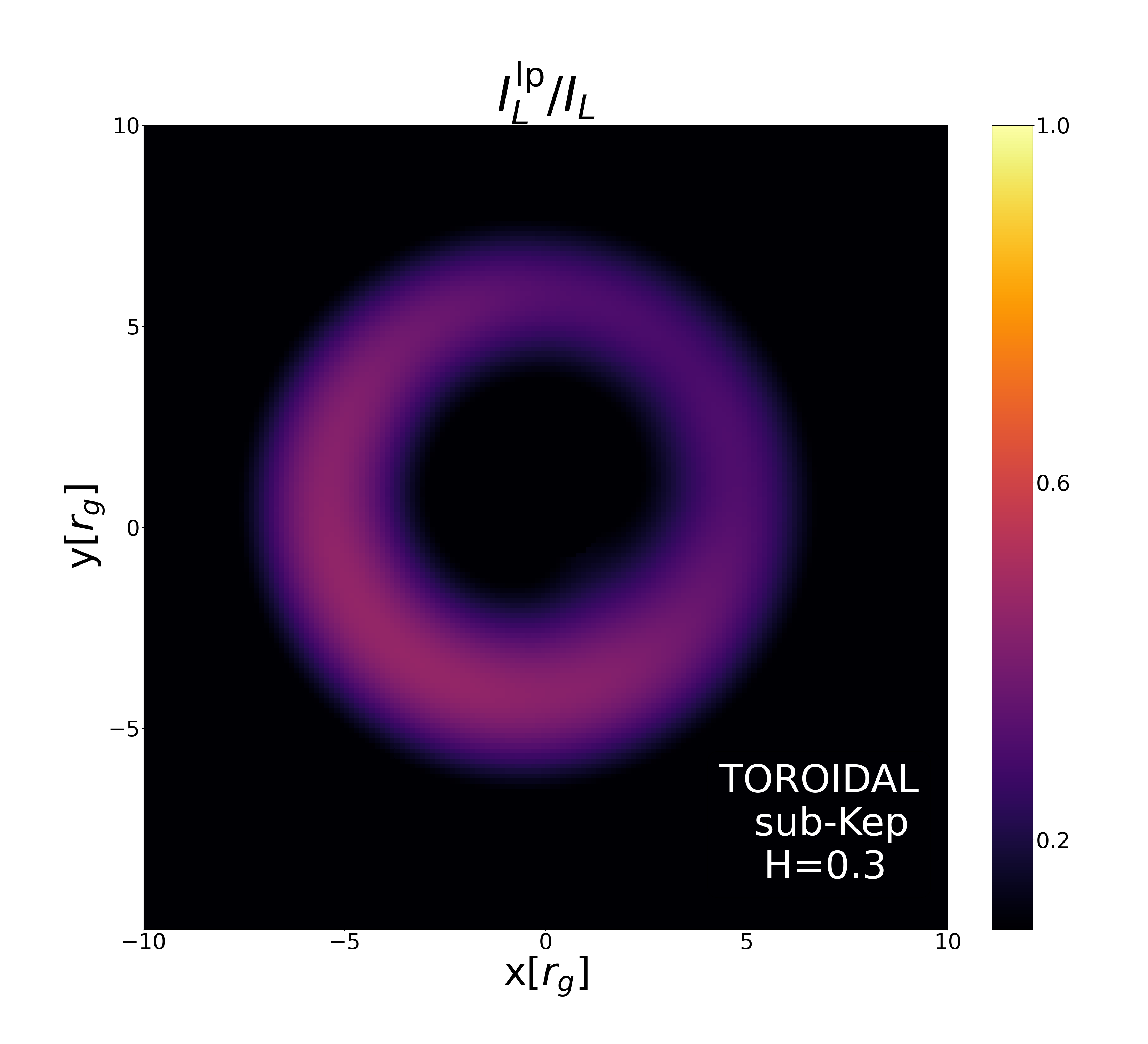}
      \end{subfigure} 
  \begin{subfigure}[b]{0.45\columnwidth}
    \includegraphics[width=\linewidth]{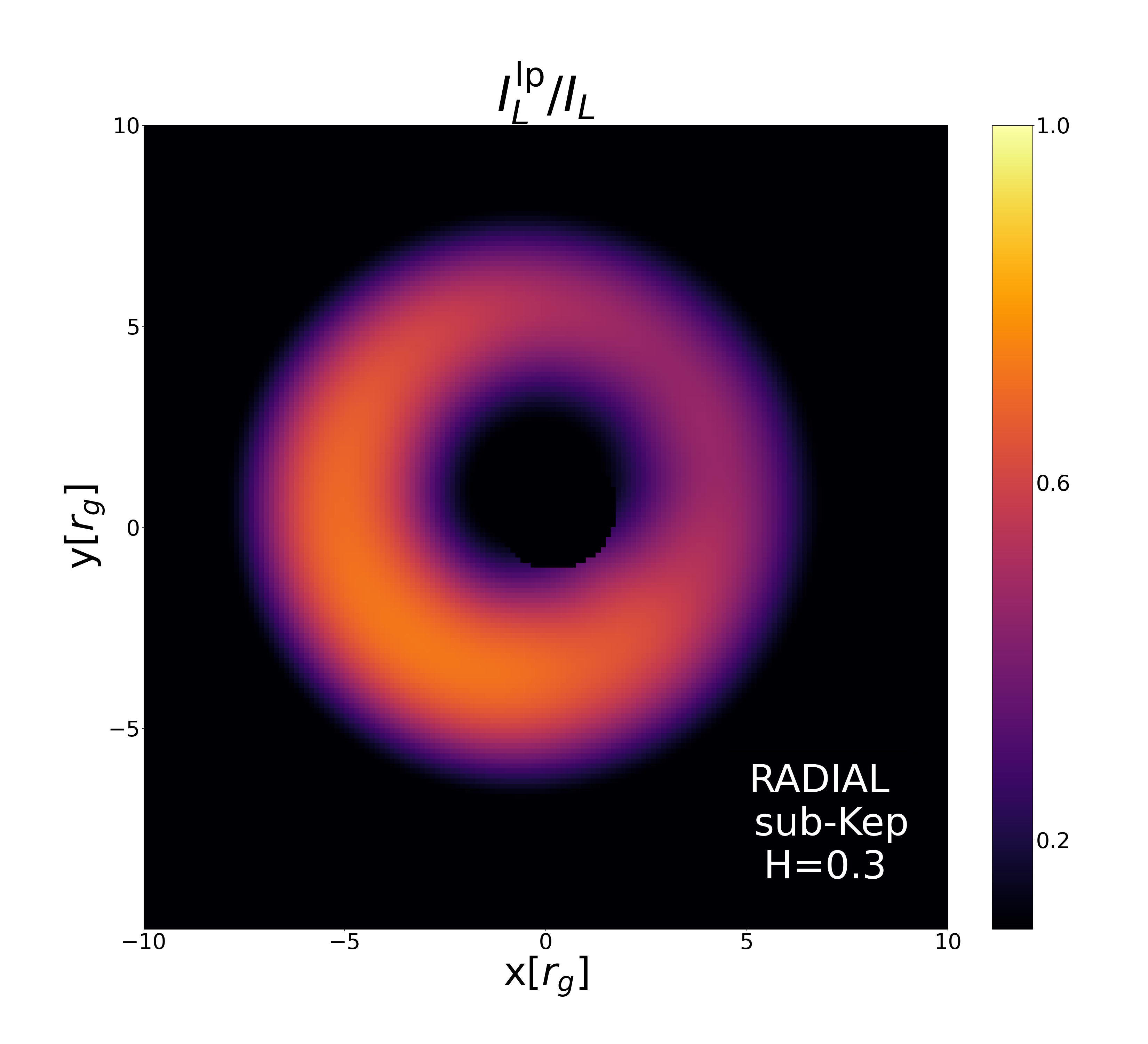}
      \end{subfigure} 
      \ \
  \begin{subfigure}[b]{0.43\columnwidth}
    \includegraphics[width=\linewidth]{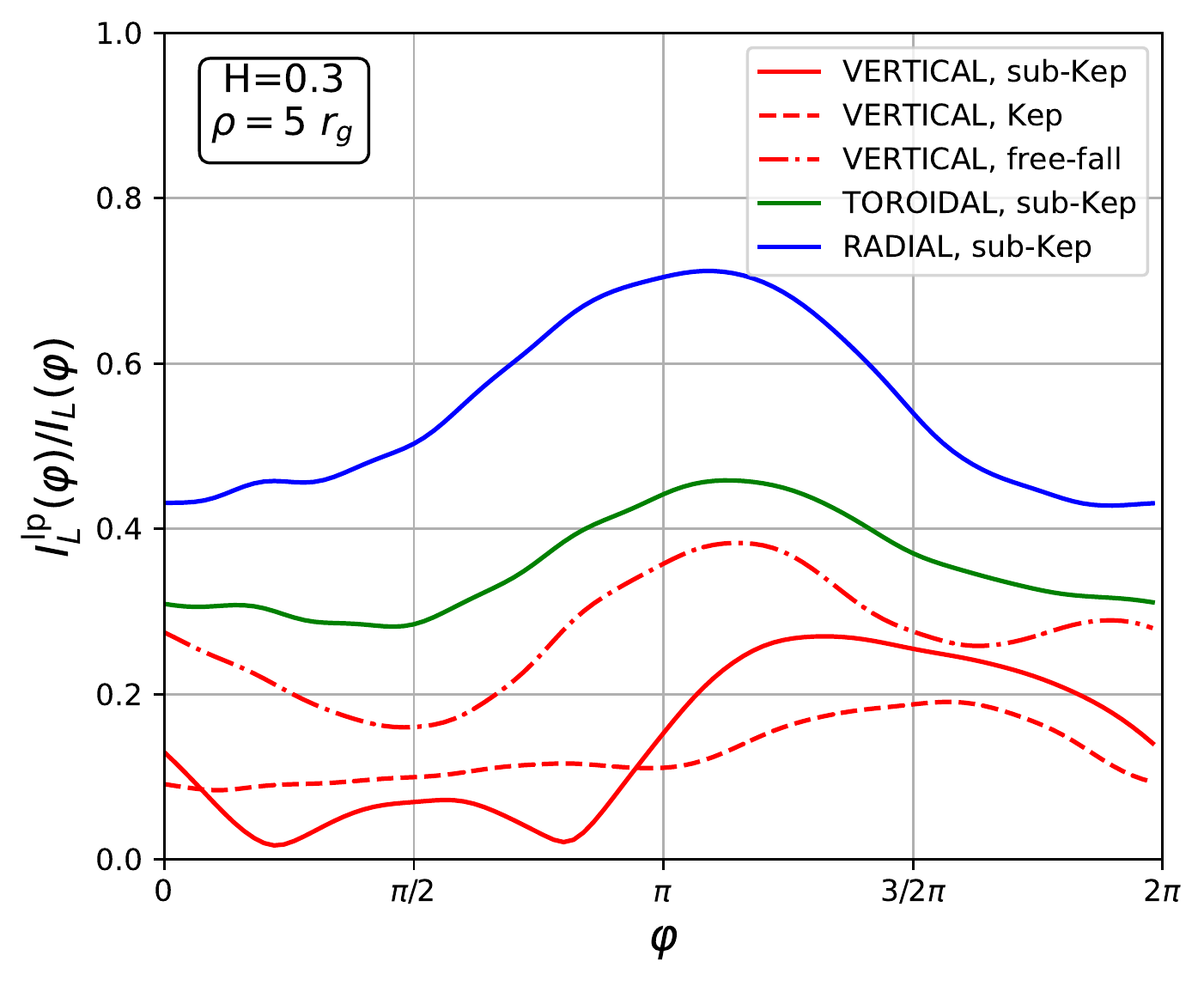}
      \end{subfigure}
      \caption{\footnotesize{Here we present the ratio between the linear polarization intensity from lensed photons and that of the total emission. We choose various magnetic field configurations and velocity distributions of RIAF. The last plot shows the ratio at a fixed radius, i.e., $\rho = 5\,r_g$, as a function of the azimuthal angle $\varphi$.}} 
\label{ILRf}
\end{figure}

By comparing $\delta (\varphi)/2\pi$ distributions in Fig.\,\ref{AEVPA5}, we find that most of them are fit well by Eq.\,(\ref{phasedelay}), except for the one with a radial magnetic field. It turns out that such a deviation is caused by a significant contribution from lensed photons. In Fig.\,\ref{ILRf}, we compare the linear polarization intensity from lensed photons, $I_L^{\textrm{lp}}$, with the total linear polarization intensity $I_L$. It is clear that, with a radial magnetic field, the lensed photons give a much larger contribution, and they dominate in the region near $\varphi\simeq \pi$ where the largest deviation from Eq.\,(\ref{phasedelay}) appears.

Notice that in more realistic cases, such as the accretion flows described by general relativistic magnetohydrodynamic (GRMHD) simulations, lensed photons are typically less polarized than the ones from direct emissions, due to the magnetic turbulence \cite{Jim_nez_Rosales_2021, Palumbo:2022pzj}. Consequently, our study in this section, based on the analytic RIAF, tends to overestimate the washout effect from lensed photons.

\begin{figure*}[ht] 
\centering 
\includegraphics[width=0.45\textwidth]{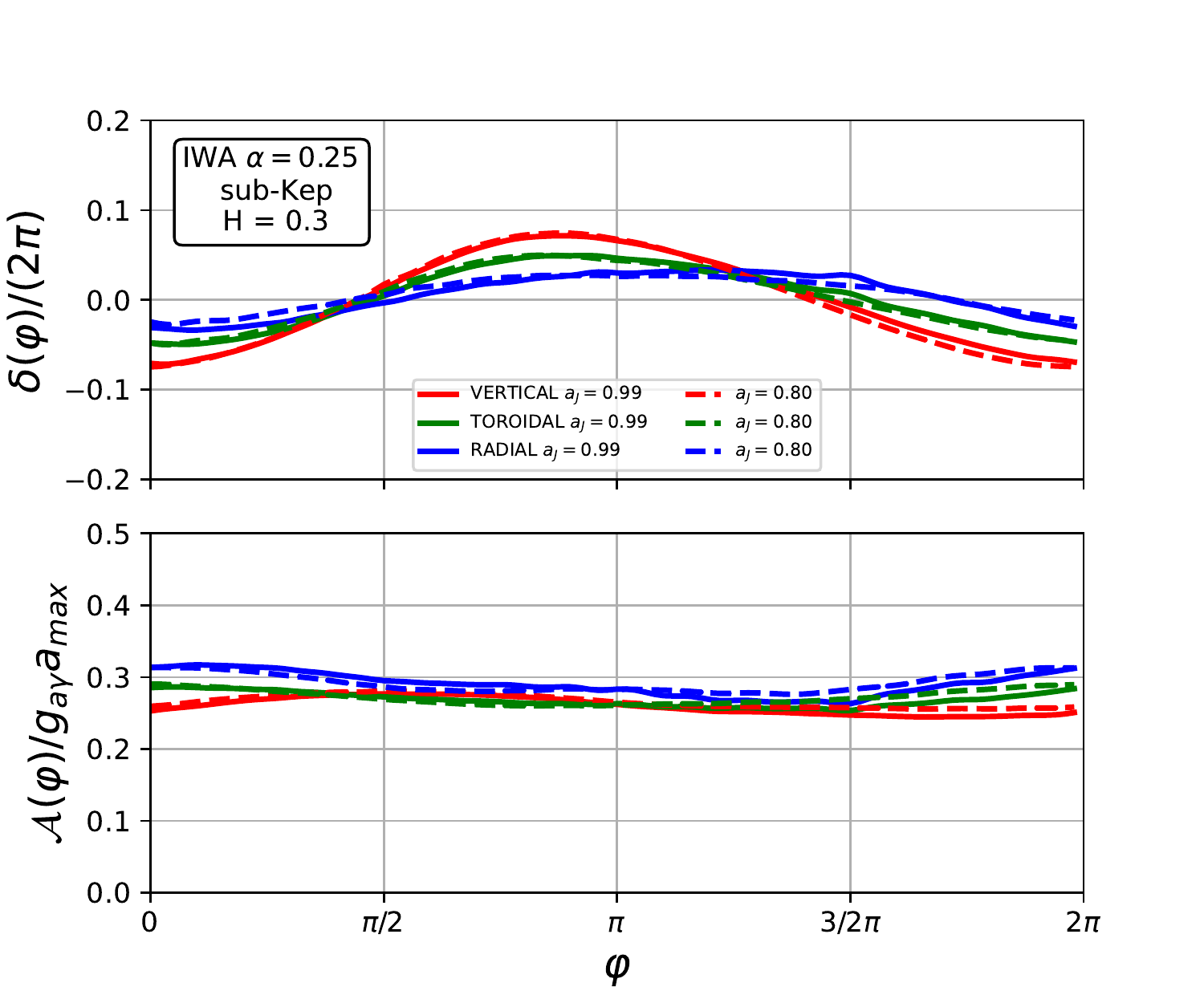}
\includegraphics[width=0.45\textwidth]{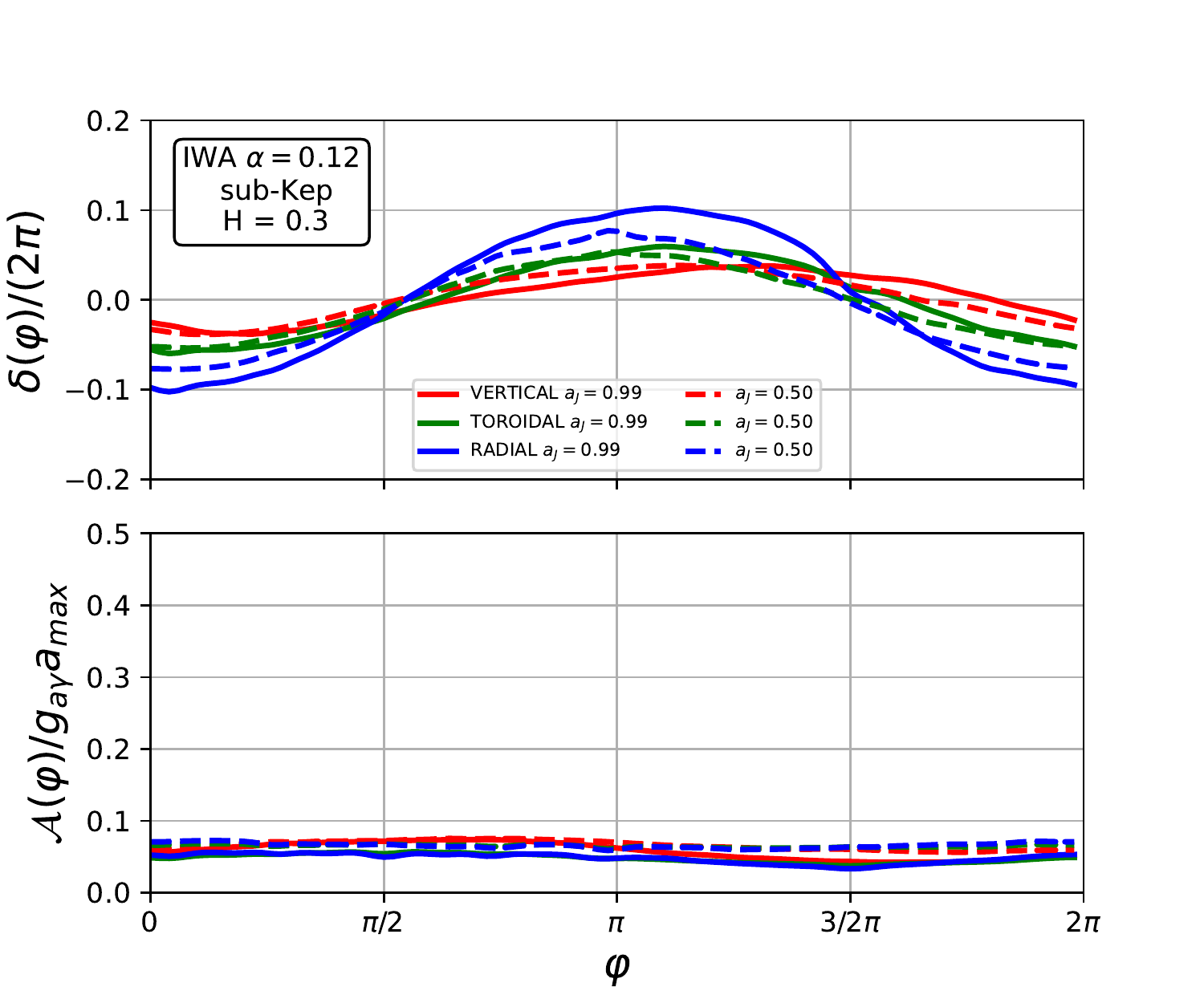}
\caption{\footnotesize{Here we compare IWA EVPAs for different spins, with $\alpha = 0.25$ in the left panel and $\alpha = 0.12$ in the right panel. The EVPA variations have slight difference with the case of $a_J = 0.99$. }}
\label{AEVPAspin}
\end{figure*}

In addition, we also study the effect of the black hole spin $a_J$. 
In Fig.\,\ref{AEVPAspin}, we show the comparison of IWA EVPAs with various choices of spins.
We find that, as long as the superradiance can happen, the EVPA variations remain qualitatively the same as the ones of $a_J = 0.99$.

\section{Prospect for future VLBI observations}\label{Pn}

\subsection{Statistics}

\subsubsection{Search for EVPA variations}

In this section, we characterize the statistics method for the axion-induced birefringence search. The EVPA data from observation can be parametrized as 
\begin{align}
&\chi_{D}=\chi^{\rm astro}_{D}+\chi^a_{D}(\boldsymbol{\vartheta}^a)+n_{D}.\label{chiD}
\end{align}
Here $\chi^{\rm astro}$ is the EVPA variation with an astrophysical origin, and  $\chi^{a}(\boldsymbol{\vartheta}^a)$ is the EVPA variation induced by the axion cloud. Further, $\boldsymbol{\vartheta}^a$ represents the axion related parameters, such as its mass and its coupling to photons, and $n_{D}$ is the measurement noise. 
The subscript $D$ labels the properties of the observation data, including the time of a measurement, the coordinates on the sky plane and the photon frequency.
For simplicity, we assume the measurement noise follows a Gaussian distribution, thus $n_{D}$ has a probability distribution as
\begin{align}
P(n_{D})= \frac{1}{\sqrt{2\pi}\sigma_{D}}\exp\left[-\frac{1}{2}\frac{n_{D}^2}{\sigma_{D}^2}\right].\label{white_noise}
\end{align}
We note that our following discussion can be easily generalized to include non-diagonal noise correlations.
Given a set of observation data $\chi_D$, the likelihood function can be written as
\begin{equation}
\mathcal{L} \left[ \chi_{D}| {\boldsymbol{\vartheta}}^a; \chi^{\rm astro}_D \right] = \prod_{D} \frac{1}{\sqrt{2\pi}\sigma_{D}}\exp\left[-\frac{\Big(\chi_{D}-\chi^{\rm astro}_D-{\chi}^a_{D}({\boldsymbol{\vartheta}}^a)\Big)^2}{2\sigma_{D}^2}\right].\label{likelihood_original}
\end{equation}

In order to estimate the likelihood, one needs to properly model the astrophysical contribution, $\chi_{D}^{\rm astro}$. The complexity of the accretion flow leads to the biggest technical obstacle. Now let us introduce a method in order to characterize the behavior of $\chi_{D}^{\rm astro}$.

The dynamics of the accretion flow can be modeled by numerical simulations, e.g., based on GRMHD. There are many parameters, such as electron density and temperature, velocity distribution, and magnetic field structure and strength, serving as the inputs. For a specific SMBH, one can determine the range of these input parameters by comparing the simulation results with the observation data, such as the photon ring morphology, the luminosity distribution, etc. We label the input parameters as $\{p_i\}$, and their allowed ranges to describe such a SMBH as $\{\Delta p_i\}$. Now one can perform the numerical simulations with a scan of these parameters within their allowed region $\{\Delta p_i\}$. For each choice of $\{p_i\}$, we obtain a distribution of EVPA, labeled as $\chi_{D}^{\rm astro}(\{p_i\})$. This forms an ensemble of EVPA with various choices on the astrophysical input parameters. 

First, let us define the ensemble average of the EVPA for each $D$, which can be written as
\begin{equation}
\chi^0_D=\frac{1}{N_{\rm ens}}\sum_{\{p_i\}}\chi_{D}^{\rm astro}(\{p_i\}),\label{ensemble-ave}
\end{equation}
with $N_{\rm ens}$ as the number of simulations carried out in this ensemble. 

In order to characterize the uncertainties from the accretion flow modeling, let us assume that $\chi_{D}^{\rm astro}$ for various choices of $\{p_i\}$ follows a Gaussian distribution. More explicitly, we define 
\begin{equation}
    M_{DD'} = \frac{1}{N_{\rm ens}}\sum_{\{p_i\}} (\chi_{D}^{\rm astro}(\{p_i\}) - \chi^0_D)(\chi_{D'}^{\rm astro}(\{p_i\}) - \chi^0_{D'}).
\end{equation}
Here we maintain the potential correlations in time, space and photon frequency. The Gaussian approximation is exact if the following two requirements are met. First, we need the parameters within $\{\Delta p_i\}$ follow a multivariate normal distribution. Further, $\chi_{D}^{\rm astro}(\{p_i\})$ needs to respond linearly to all $p_i$ within $\{\Delta p_i\}$, which can be approximately justified using Taylor expansion. In practice, whether the Gaussian approximation is valid should be examined in the GRMHD simulation. If this is not satisfied, a more complicated probability distribution of $\chi_{D}^{\rm astro}(\{p_i\})$ can be numerically introduced and our analysis method can be easily generalized. For now, let us stick with the Gaussian approximation for the simplicity.

After obtaining the probability distribution of $\chi_{D}^{\rm astro}$, one can convolute it with the likelihood calculation in Eq.\,(\ref{likelihood_original}) and integrate out $\chi_{D}^{\rm astro}$ as nuisance parameters. The likelihood distribution can be written as
\begin{align}
    \mathcal{L}\left[ \chi_{D} | {\boldsymbol{\vartheta}}^a\right] = \frac{1}{\sqrt{|2\pi M'|}}\exp\left[
-\frac{1}{2}\sum_{DD'} \left(\chi_{D} - \chi_{D}^0 - {\chi}_{D}^a({\boldsymbol{\vartheta}}^a)\right) M'^{-1}_{DD'} \left(\chi_{D'} - \chi_{D'}^0 - {\chi}_{D'}^a({\boldsymbol{\vartheta}}^a)\right)\right],\label{likelihoodAG} \end{align}
where $M'_{DD'} \equiv M_{DD'} + \sigma_{D}^2 \delta_{DD'}$. 
This gives a viable calculation to estimate the sensitivity on the axion related parameters $\boldsymbol{\vartheta}^a$.

In order to perform a back-of-envelop estimation on the sensitivity in the parameter space, let us assume that the uncertainties, for all values of $D$, are uncorrelated with each other and they are approximately at the same order of magnitude. Under this assumption, one can calculate the typical size of the uncertainty as \begin{equation}
    \overline{\sigma}^2 = \frac{1}{{\rm Tr}[M'^{-1}_{DD'}]}.\label{fix_chi0}
\end{equation}
Consequently, the signal to noise ratio (SNR) can be estimated by comparing the typical size of the axion-induced birefringence signal with  $\overline{\sigma}$. By this simple approximation, the sensitivity on the axion-photon coupling, i.e., $c$, scales as $1/\sqrt{N_D}$, where $N_D$ is the total number of data points.

\subsubsection{Search for differential EVPA variations}

The statistical method considered above requires a systematic study of the accretion flow, using GRMHD for example. Now let us consider an alternative analysis using the differential EVPA. This method has been introduced in \cite{Chen:2021lvo} to perform an axion search using the EHT observation on M87$^\star$. Although we pay the price of a suppression factor, one does not need a very comprehensive understanding on the accretion flow dynamics.

Remember that the index $D$ labels the observation time, the coordinates on the sky plane and the photon frequency. Let us single out the time information, using the index $i$, and the other information is labeled by $d$, i.e., $D\equiv \{i, d\}$. We will compare the EVPA at different times, with fixed coordinates and frequency. Let us define the differential EVPA as 
\be \Delta\tilde{\chi}_{i} \equiv \frac{{\chi}_{i+1}}{\sigma_{i+1}} - \frac{{\chi}_{i}}{\sigma_{i}},\label{diffEVPA} \ee
where all the indices $d$ are dropped for convenience.

Let us focus on the axion parameter space where the following condition is met, 
\begin{equation}
    \left| \Delta\tilde{\chi}_i^{\rm astro} \right| \ll \left| \Delta\tilde{\chi}_i^{a} ({\boldsymbol{\vartheta}}^a) \right|. \label{small_variation_limit}
\end{equation}
This condition implies that the change of the EVPA between two observation times is dominated by the axion birefringence effect rather than an astrophysical origin.

Under this assumption, one can easily calculate the likelihood function for $\Delta\tilde{\chi}_i^a$ as
\begin{align}
    \mathcal{L}\left[ \Delta\tilde{\chi}_{i} | {\boldsymbol{\vartheta}}^a \right] = \frac{1}{\sqrt{|2\pi \tilde{M} |}}\exp\left[
-\frac{1}{2}\sum_{ii'} \Big( \Delta\tilde{\chi}_{i} -  \Delta\tilde{\chi}_i^a ({\boldsymbol{\vartheta}}^a) \Big) \tilde{M}^{-1}_{ii'} \Big( \Delta\tilde{\chi}_{i'} - \Delta\tilde{\chi}_{i'}^a ({\boldsymbol{\vartheta}}^a) \Big) \right].\label{likelihoodDEVPA} \end{align}
Here $\tilde{M}$ characterizes the measurement noise for the differential EVPA. Using the notation in Eq.\,(\ref{chiD}), \begin{align}\tilde{M}_{ii'}=\Big\langle\left(\frac{n_{i+1}}{\sigma_{i+1}}-\frac{n_{i}}{\sigma_{i}}\right)\left(\frac{n_{i'+1}}{\sigma_{i'+1}}-\frac{n_{i'}}{\sigma_{i'}}\right)\Big\rangle. \end{align}
Assuming the measurement noise is uncorrelated, we obtain  $\tilde{M}_{ii} = 2$, $\tilde{M}_{i\,(i\pm1)} = 1$ and 0 for all other matrix elements.

Here we see that the benefit of using differential EVPA is to remove the non-trivial dependence on $\chi_i^{\rm astro}$ in the likelihood function. In order to justify the condition in Eq.\,(\ref{small_variation_limit}), one only need to understand the accretion flow at the level of orders of magnitude. This is much easier to achieve than a comprehensive understanding required in the previous analysis method.

On the other hand, the analysis based on the differential EVPA needs to pay the price of a suppression factor. To demonstrate that, let us use the axion signal ansatz, presented in Eq.\,(\ref{ansatz}), to calculate the differential EVPA. 

Assuming $\sigma_i \simeq \sigma_j$, the axion contribution to the differential EVPA, defined in Eq.\,(\ref{diffEVPA}), can be written as
\begin{equation}
   \Delta\tilde{\chi}_i^a = 2{\mathcal{A}}\sin\left[\frac{{\omega}\Delta t}{2}\right]\,\cos\left[\frac{{\omega}(t_i+t_j)}{2}+{\delta}\right]/\sigma_i,
\end{equation}
where $\Delta t \equiv t_j - t_i$ is the time interval of the sequential observations. Thus we see that the axion signal in terms of the differential EVPA suffers from a suppression factor as $2 \sin\left[\frac{{\omega}\Delta t}{2}\right]$. This suppression is more severe for a smaller axion mass.

\subsection{Data sets increase}

In this subsection, we discuss the prospective improvements that can be achieved using the EVPA data from the future VLBI observations, e.g., ngEHT \cite{Raymond_2021,Lngeht}. We focus on the correlations among the axion induced birefringence signals, which can potentially increase the sensitivity as well as discriminate against astrophysical background.

\begin{figure}[htb]
\centering
\includegraphics[width=0.45\textwidth]{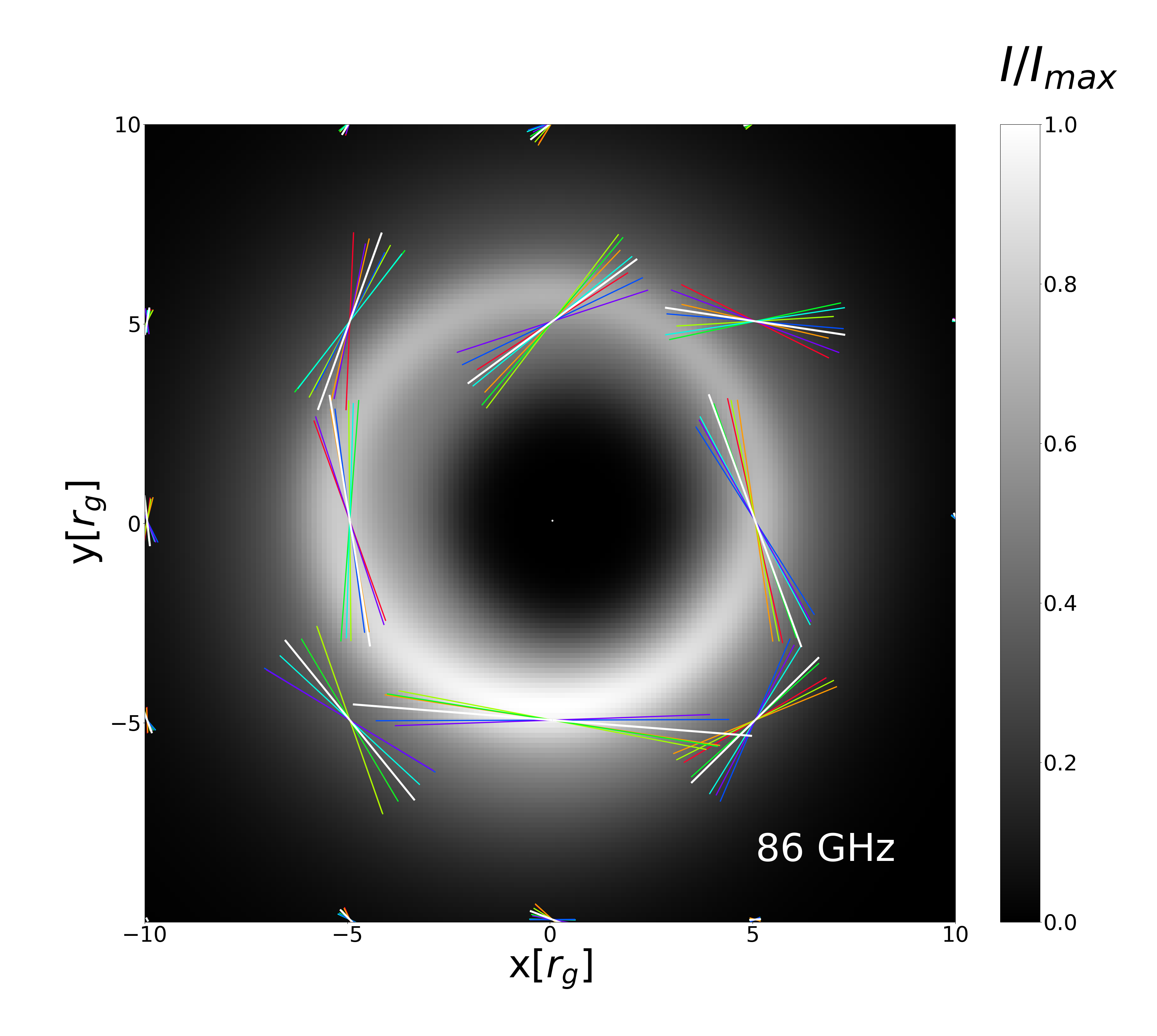}
\includegraphics[width=0.45\textwidth]{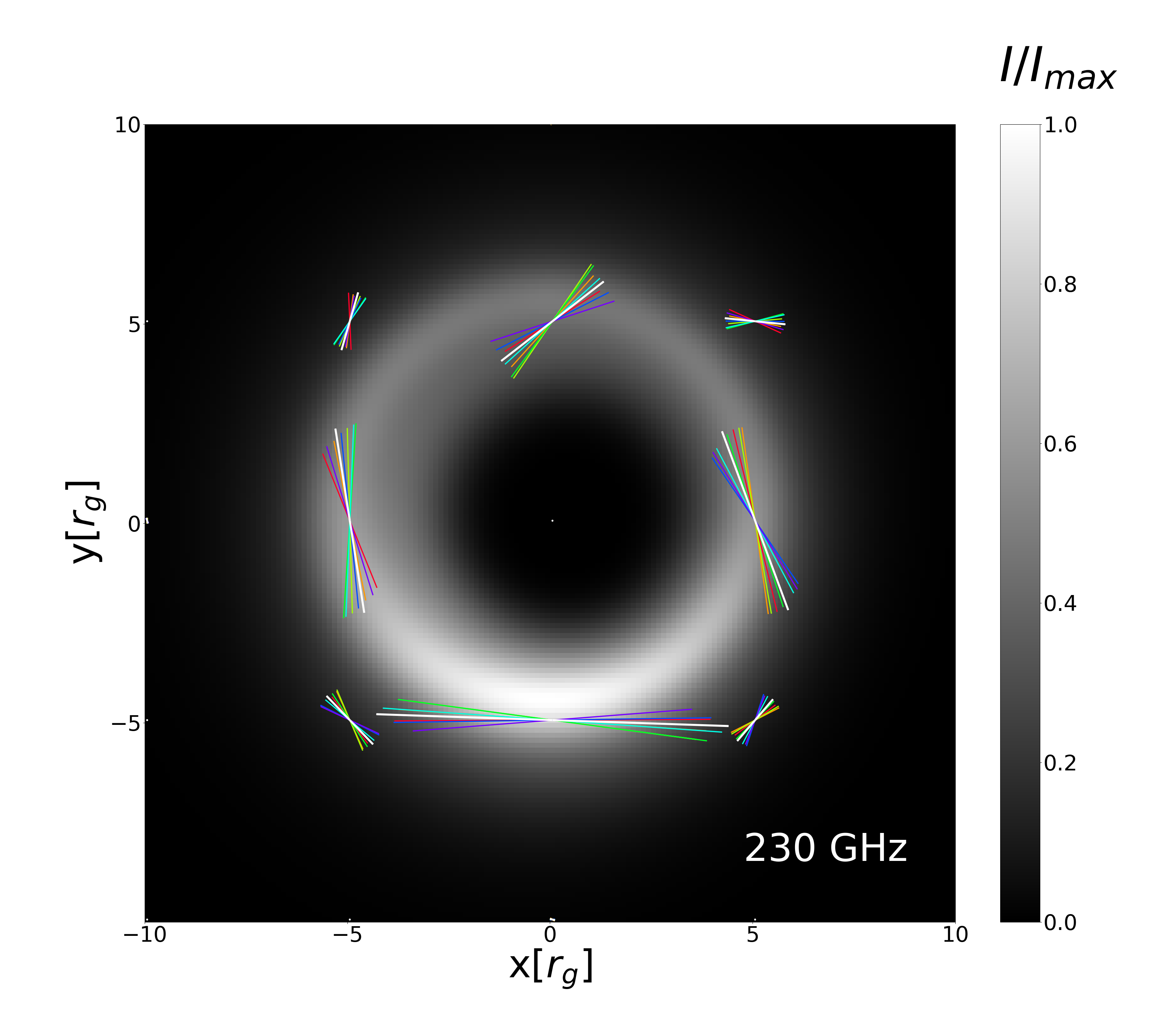}
\includegraphics[width=0.45\textwidth]{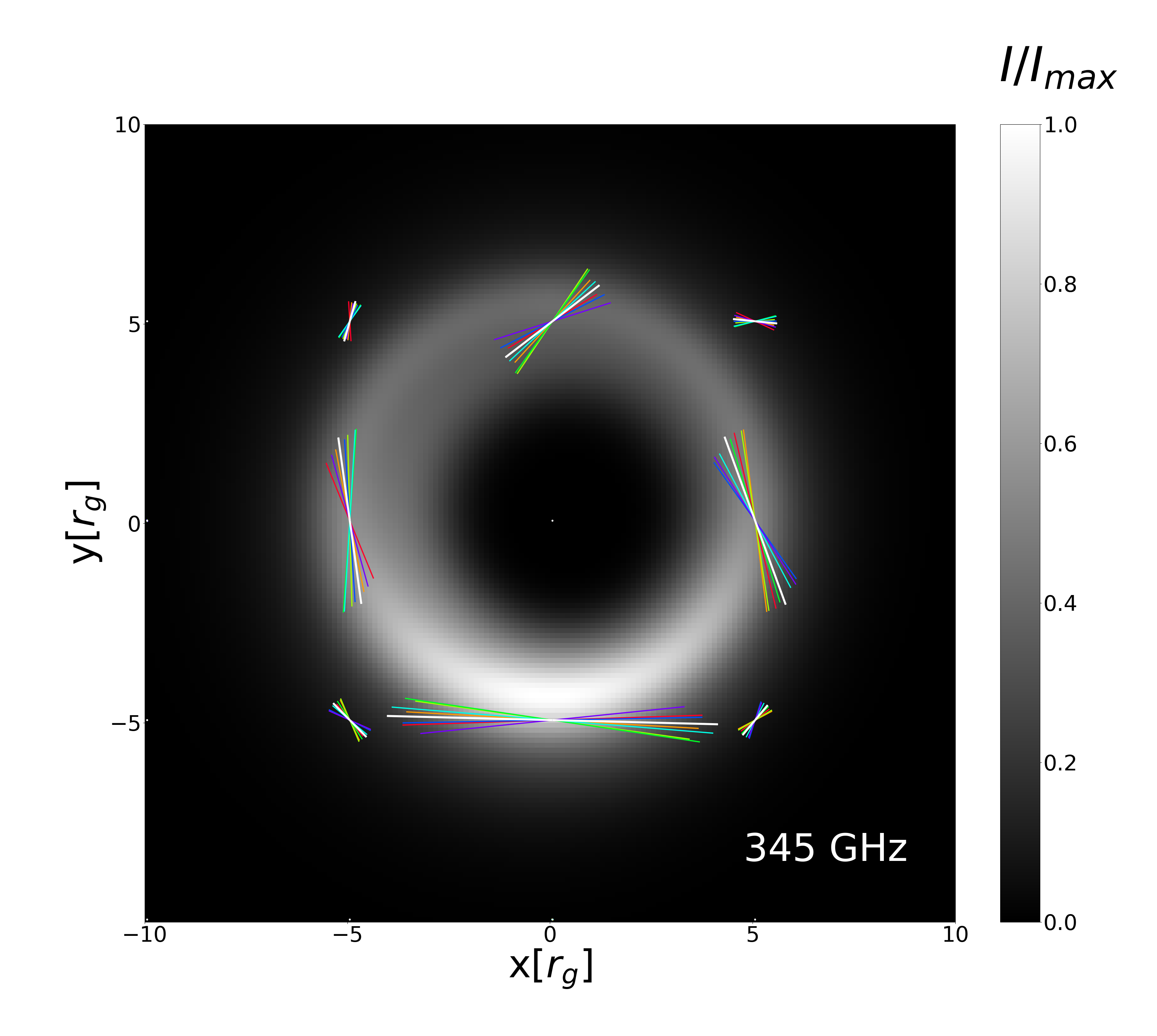}
\includegraphics[width=0.45\textwidth]{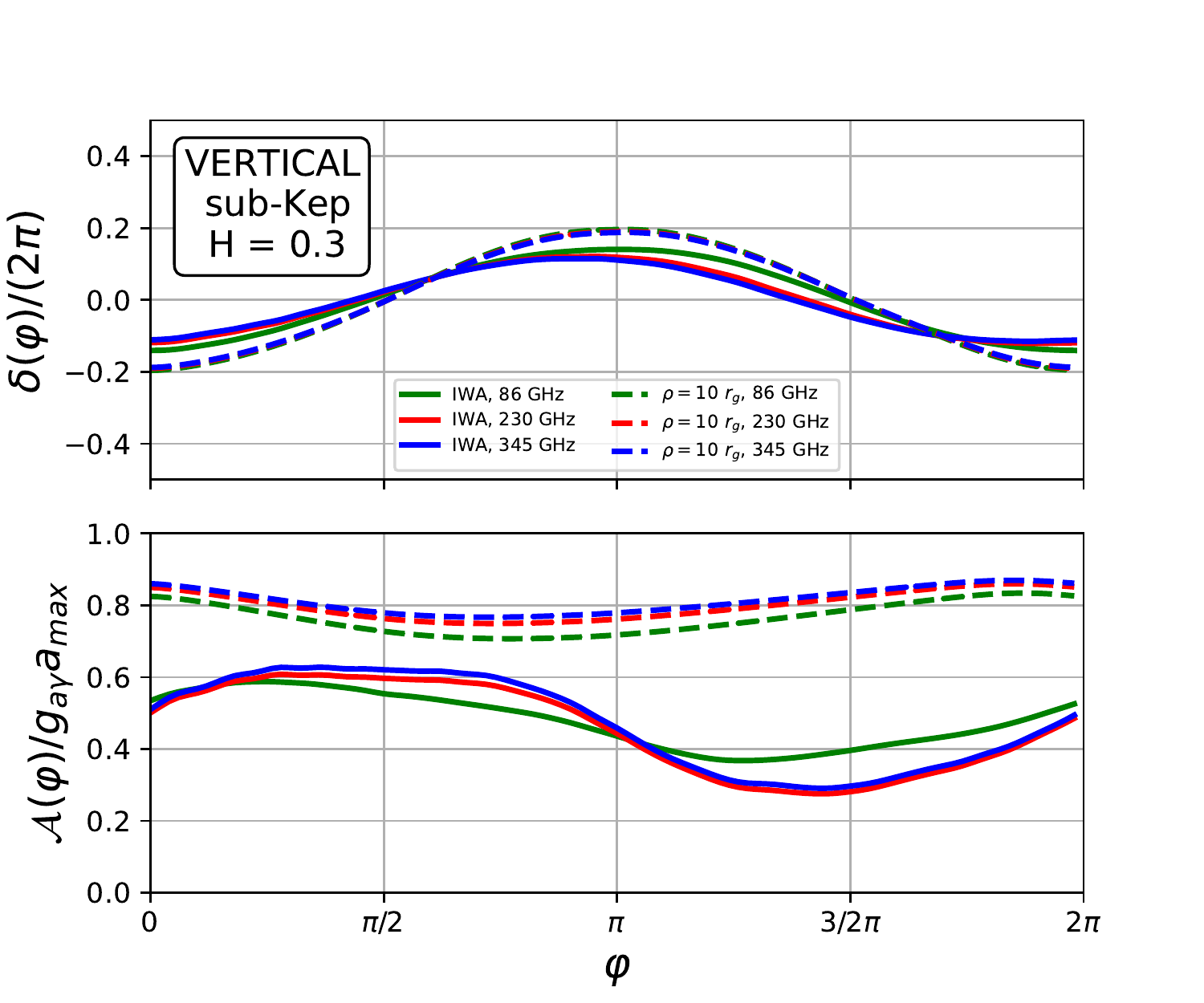}
\caption{\footnotesize{ Here we show the results from the \texttt{IPOLE} simulation based on RIAF model. We choose three frequencies as benchmarks, 86 GHz, 230 GHz and 345 GHz. They are corresponding to the ones that ngEHT \cite{Raymond_2021,Lngeht} plans to observe. 
The white quiver represents the EVPA without the axion contribution at each point.
The rainbow color, from red to purple, indicates the time variation of the EVPA in the presence of the axion cloud. 
The inclination angle ${i} = 163^\circ$ is taken to be consistent with M87$^\star$, and the spin points to $- x$ on the sky plane.
The last plot shows the relative phase, $\delta (\varphi)/2\pi$, and the amplitude, $\mathcal{A} (\varphi) / g_{a\gamma\gamma} a_{\textrm{max}}$, of the EVPAs oscillation. We show the results for IWA and those at $\rho = 10\,r_g$. The features in signals are similar with each other with various choices on the frequency, indicating a strong correlation among them. 
In this analysis, we take $\alpha = 0.4$, $g_{a\gamma\gamma} a_{\textrm{max}} = 1$ rad and $a_J = 0.99$ for the axion and the black hole. Further, we choose vertical magnetic field, sub-Keplerian velocity distribution and $H = 0.3$ as the benchmark parameters for the RIAF model.
}}
\label{3f}
\end{figure}

First, we study the axion signal correlations in various frequency bands. The ngEHT can potentially observe at three different frequencies simultaneously, i.e., $86$ GHz, $230$ GHz and $345$ GHz \cite{Raymond_2021,Lngeht}. 
Since the axion-induced birefringence is achromatic, the EVPA variations at different frequencies are the same while propagating in the vacuum. After including the plasma effects based on RIAF models, we show the comparison on the IWA EVPA oscillations at these three frequencies in Fig.\,\ref{3f}.  There are slight differences, which are caused by the washout effects induced by the finite thickness of the accretion flow and the lensed photons. 
Notice that, for $86$ GHz, the accretion flow is optically thicker compared to that at higher frequencies. Thus the contribution from lensed photons is less important, which makes the green solid line (IWA at $86$ GHz) in Fig.\,\ref{3f} less asymmetric respect to $\varphi=\pi$. 
The correlations among EVPA variations at different frequencies appear to be quite strong for this benchmark. The Faraday rotation modifications on the EVPA, characterized by $\rho_{V}$ in Eq.\,(\ref{ipole-mod}), has a square dependence on the photon wave-length, while the axion-induced term is universal for all frequencies. Thus this provides a powerful way to subtract the astrophysical contributions.

Furthermore, the future {VLBI experiments} have potentials to increase the spatial resolution and improve the dynamic range. The EVPA variations at different radii from the black hole can be measured.
As mentioned in previous sections, the lensed photons contribute significantly to the washout effects. However, since these lensed photons contribute dominantly at small radii, such as $\sim 5.5 r_g$ for M87$^\star$, EVPA variations at the outer region are almost free from such a washout, as demonstrated previously. In addition, for the parameter space we are interested in this study, the axion wave-functions generically peak at a larger radius, e.g., $\sim10\,r_g$ on the sky plane. Therefore, correlating the EVPA variations at different radii on the sky plane can be a powerful handle to reduce the lensed photon contamination.

\begin{figure}[htb]
\centering
\includegraphics[width=0.7\textwidth]{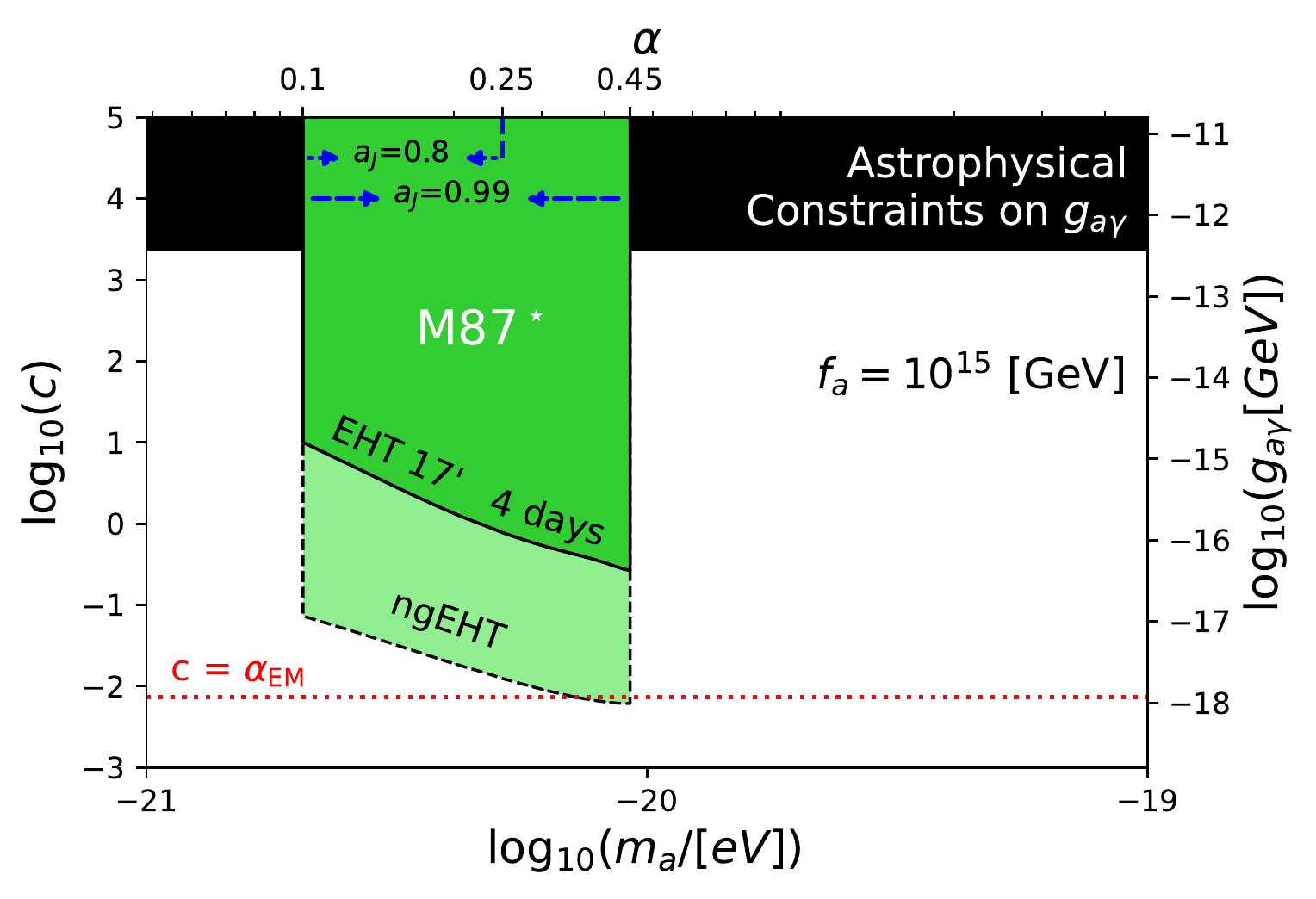}
\caption{\footnotesize{ The comparison between constraints on dimensionless axion-photon 
coupling  $c \equiv 2 \pi g_{a\gamma\gamma} f_a$ in \cite{Chen:2021lvo}, other astrophysical constraints assuming $f_a = 10^{15}$ GeV, and prospect for ngEHT is shown. The red dashed line corresponds to the minimal possible value of $c = \alpha_{\rm EM}$.
}}
\label{pngEHT}
\end{figure}

In order to demonstrate the prospects of the axion search at ngEHT, we perform a back-of-envelop estimation based on several potential improvements to be achieved on future observation of M87$^\star$. The results are shown in Fig.\,\ref{pngEHT}. As a comparison, we also present the existing constraint in the axion parameter space, using the recently published EHT results \cite{Chen:2021lvo}.
The improvements on the sensitivity mainly come from the following aspects:
\begin{itemize}
    \item  three different frequencies;
\item five different radii between $\rho = 5.5\,r_g$ and $\rho = 9.5\,r_g$;
\item ten times the observation time ($\sim 40$ days span);
\item the axion field values at different radii, according to the axion cloud wavefunction, are taken into consideration;
\item compared with the differential EVPA analysis carried in \cite{Chen:2021lvo}, we remove of the suppression factor $\sin{[\omega t_{\textrm{int}}/2]}$ assuming a better understanding on the accretion disk dynamics.
\end{itemize}
We note that, in this estimation, we assume that the uncertainty in each measurement is at the same order of magnitude as that in \cite{EHTP}, for simplicity. 
Here we emphasize that the ngEHT observation on M87$^\star$ can potentially probe $c_{\rm min} \sim  \mathcal{O}(1) \alpha_{\rm EM}$. This serves as a well-motivated theoretical benchmark, in which the axion-photon coupling is induced by $\mathcal{O}(1)$ numbers of chiral fermions with $\mathcal{O}(1)$ units of the electric charge.

Finally, the future {VLBI experiments} \cite{Gurvits:2022wgm} have the potential to observe more SMBHs at the horizon scale \cite{SMBHVLBI,Pesce:2021adg}. In this case, one can perform an axion search at a broader mass window,  potentially covering from $10^{-22}$ eV to $10^{-17}$ eV. 
In Table\,\ref{TSMBH}, we list some candidates of SMBHs in \cite{SMBHVLBI} that can be observed by the future {VLBI experiments}. Such observations require the photon rings of these SMBHs to have open angles larger than $2 \mu$as, and enough flux at the radio frequency band can be received. In the table, we provide the axion mass range corresponding to $\alpha$ between $0.1$ and $0.5$, which should be eventually determined by the individual spin of each SMBH.

\begin{table}[htb]
\begin{center}
\begin{tabular}{ | c | c | c | c | c | c |}		
\hline	
SMBH & $M/M_\odot$ & $\theta_{\rm ring} / \mu$as & $\mu/$eV range &  $T_a/s$ at $\alpha = 0.3$ \\
\hline
Sgr A$^\star$ &$4.3\times10^6$ & 53 & $3.1 \times 10^{-18}\sim 1.6 \times 10^{-17}$  &  $4.4\times 10^2$    \\
M87$^\star$ & $6.5\times10^9$ & 42 & $2.1 \times 10^{-21}\sim 1.0\times 10^{-20}$  & $6.7\times 10^5$    \\
\hline
IC 1459 & $2.8\times10^9$ & 9.2 &  $4.9 \times 10^{-21}\sim 2.4\times 10^{-20}$  &  $2.8\times 10^5$    \\
NGC 4374 & $1.5\times10^9$ & 9.1 & $8.8 \times 10^{-21}\sim 4.4\times 10^{-20}$  & $1.6\times 10^5$    \\
NGC 4594 & $5.8\times10^8$ & 5.7 & $2.3 \times 10^{-20}\sim 1.2\times 10^{-19}$  & $6.0\times 10^4$    \\
IC 4296 & $1.3\times10^9$ & 2.5 & $9.9 \times 10^{-21}\sim 5.0\times 10^{-20}$  & $1.4\times 10^5$    \\
NGC 3031 & $7.9\times10^7$ & 2.0 & $1.7 \times 10^{-19}\sim 8.4\times 10^{-19}$  & $8.2\times 10^3$    \\
\hline
\end{tabular}
\caption{\footnotesize{Here we provide a list of SMBHs. Two of them are already by measured by EHT, M87$^\star$ and Sgr A$^\star$. The rest are potential candidates to be resolved in the future \cite{SMBHVLBI}. We also provide the typical axion mass window, corresponding to $\alpha$ between $0.1$ and $0.5$, as well as the typical axion oscillation timescale $T_a$ for each SMBH. }}
\label{TSMBH}
\end{center}
\end{table}

\section{Conclusion}
The polarimetric measurements of the horizon scale emissions from SMBHs open a new window to probe the existence of ultralight axion fields \cite{Chen:2019fsq,Chen:2021lvo}. An axion cloud can be generated through the superradiance process, and the axion field can potentially reach the highest possible value. On the other hand, accretion flows generate large amount of linearly polarized radiations from the neighborhood of rotating black holes, overlapping with the densest region of the axion cloud. Consequently the EVPA of these photons will oscillate periodically due to the axion photon coupling. The current and next-generation VLBI polarimetric measurements \cite{EHTP,Lngeht} are powerful ways to search for axion clouds around SMBHs.

The strong gravity and medium effects highly influence the horizon scale observations. Both the axion cloud and the accretion flow dynamics can lead to EVPA variations. In our study, we show explicitly how the axion photon coupling can be embedded into the polarized covariant radiative transfer equations where both the curved spacetime and plasma effects are taken into consideration. The axion effect can be included by a simple modification on the numerical radiative transfer simulation, such as \texttt{IPOLE} \cite{Moscibrodzka:2017lcu, Noble:2007zx}.

The mapping from the SMBH coordinates to the sky plane for observation is non-trivial.  For a geometrically thin and optically thick disk, such as the NT model, one needs to follow the photon geodesics which connects the sky plane to the surface of the accretion disk. We study in detail on how such a mapping depends on the size of the black hole spin and its inclination angle. This mapping is further used to generate the amplitude and the relative phase of the axion induced EVPA signal on the sky plane.

For a more realistic model of the accretion flow, photons being observed at each point on the sky plane may have different spatial and temporal origins along the line of sight. The sum of these photons generically leads to a suppression on the EVPA oscillation amplitude. We study such washout effects in two simple toy models. One is the constant radiation source along a continuous and finite length, representing the thickness of the accretion flow. The other one is the radiation from two spatially separated point sources, mocking the contributions from lensed photons.

The future {VLBI experiments}, such as the next-generation Event Horizon Telescope \cite{Raymond_2021,Lngeht}, will be able to perform better measurements and provide more detailed information about the EVPA variations. The sensitivities of the axion searches can therefore be significantly improved, especially by correlating the EVPA oscillations at different radii and frequencies. In addition, a much larger axion mass window is expected to be explored since more SMBHs will be observed.

\acknowledgments
We are grateful to useful discussions with Richard Brito, Vitor Cardoso, Horng Sheng Chia, Ru-Sen Lu, Alexandru Lupsasca, Elias Most, Chen Sun, George N. Wong, Ziri Younsi and Yunlong Zhang.
This work is supported by the National Key Research and Development Program of China under Grant No. 2020YFC2201501. 
Y.C. is supported by the China Postdoctoral Science Foundation under Grant No. 2020T130661, No. 2020M680688, the International Postdoctoral Exchange Fellowship Program, by the National Natural Science Foundation of China
(NSFC) under Grants No. 12047557, by VILLUM FONDEN (grant no. 37766), by the Danish Research Foundation, and under the European Union’s H2020 ERC Advanced Grant “Black holes: gravitational engines of discovery” grant agreement no. Gravitas–101052587. 
Y.M. is supported by the ERC Synergy Grant ``BlackHoleCam: Imaging the Event Horizon of Black Holes'' (Grant No. 610058) and the National Natural Science Foundation of China (Grant No. 12273022). 
J.S. is supported by the National Natural Science Foundation of China under Grants No. 12025507, No. 12150015, No.12047503; and is supported by the Strategic Priority Research Program and Key Research Program of Frontier Science of the Chinese Academy of Sciences under Grants No. XDB21010200, No. XDB23010000, and No. ZDBS-LY-7003 and CAS project for Young Scientists in Basic Research YSBR-006.
X.X. is supported by  Deutsche Forschungsgemeinschaft under Germany’s Excellence Strategy EXC2121 “Quantum Universe” - 390833306.
Q.Y. is supported by the Key Research Program of CAS under Grant No. XDPB15, and by the Program for Innovative Talents and Entrepreneur in Jiangsu. 
Y.Z. is supported by U.S. Department of Energy under Award No. DESC0009959.
Y.C. would like to thank the SHAO and TDLI for their kind hospitality.
Y.Z. would like to thank the ITP-CAS for their kind hospitality.
The simulation codes used in this study are a modified version of publicly available code \texttt{IPOLE} \cite{Moscibrodzka:2017lcu,Noble:2007zx}.

\providecommand{\href}[2]{#2}\begingroup\raggedright\endgroup

\end{document}